\documentclass[reprint, superscriptaddress, amsmath,amssymb, prx, floatfix]{revtex4-2}

\usepackage{graphicx}
\usepackage{dcolumn}
\usepackage{bm}
\usepackage{placeins}
\usepackage{mathtools}
\usepackage{hyperref}
\usepackage[dvipsnames]{xcolor}
\usepackage{color,soul}
\hypersetup{
    colorlinks=true,
    citecolor={blue!50!black},
    linkcolor={blue!50!black},
    urlcolor={blue!50!black}
}

\newcommand{\textss}[1]{\scriptsize \mbox{#1}}
\usepackage{upgreek}

\begin{document}

\title{Momentum-Resolved Exciton Coupling and Valley Polarization Dynamics in Monolayer WS$_2$}

\author{Alice Kunin}
\author{Sergey Chernov}
\author{Jin Bakalis}
\affiliation{Department of Chemistry, Stony Brook University, Stony Brook, New York 11794, USA.}

\author{Ziling Li}
\author{Shuyu Cheng}
\affiliation{Department of Physics, The Ohio State University, Columbus, Ohio 43210, USA.}

\author{Zachary H. Withers}
\affiliation{Department of Physics and Astronomy, Stony Brook University, Stony Brook, New York 11794, USA.}

\author{Michael G. White}
\affiliation{Department of Chemistry, Stony Brook University, Stony Brook, New York 11794, USA.}
\affiliation{Chemistry Division, Brookhaven National Laboratory, Upton 11973 New York, USA.}

\author{Gerd Sch\"onhense}
\affiliation{Johannes Gutenberg-Universit\"at, Institut f\"ur Physik, D-55099 Mainz, Germany.}

\author{Xu Du}
\affiliation{Department of Physics and Astronomy, Stony Brook University, Stony Brook, New York 11794, USA.}

\author{Roland K. Kawakami}
\affiliation{Department of Physics, The Ohio State University, Columbus, Ohio 43210, USA.}

\author{Thomas K. Allison}
 \email{thomas.allison@stonybrook.edu}
\affiliation{Department of Chemistry, Stony Brook University, Stony Brook, New York 11794, USA.}
\affiliation{Department of Physics and Astronomy, Stony Brook University, Stony Brook, New York 11794, USA.}

\date{\today}

\begin{abstract}

Coupling between exciton states across the Brillouin zone in monolayer transition metal dichalcogenides can lead to ultrafast valley depolarization.
Using time- and angle-resolved photoemission, we present momentum- and energy-resolved measurements of exciton coupling in monolayer WS$_2$.
By comparing full 4D ($k_x, k_y, E, t$) data sets after both linearly and circularly polarized excitation, we are able to disentangle intervalley and intravalley exciton coupling dynamics.
Recording in the exciton binding energy basis instead of excitation energy, we observe strong mixing between the B$_{1s}$ exciton and A$_{n>1}$ states. The photoelectron energy and momentum distributions observed from excitons populated via intervalley coupling (e.g. K$^-$ $\rightarrow$ K$^+$) indicate that the dominant valley depolarization mechanism conserves the exciton binding energy and center-of-mass momentum, consistent with intervalley Coulomb exchange. On longer timescales, exciton relaxation is accompanied by contraction of the momentum space distribution.

\end{abstract}

\maketitle

Monolayer transition metal dichalcogenides (TMDs) have garnered significant interest in the last 10 years following the discovery of valley-selective circular dichroism in these novel, atomically thin, direct band gap semiconductors \cite{Xiao2007, Yao2008, Cao2012, Xiao2012, Mak2012, Zeng2012, Kioseoglou2012, Wang2018}.
Right ($\sigma^+$) and left ($\sigma^-$) circularly polarized light selectively excites interband transitions in the inequivalent K$^+$ and K$^-$ valleys, respectively, where the band extrema are located \cite{Sallen2012, Liu2013}. 
Strong Coulomb forces and spin-orbit coupling in these materials yield two series of tightly bound exciton states of opposite spin character in each K valley, the A and B excitons, giving rise to the potential for long-lived, spin-valley locked excitons \cite{Jones2013, Qiu2013, Xu2014, Glazov2015, Koperski2017, Li2019b}. 
These unique excitons provide a promising platform for novel optoelectronic and valleytronic device applications \cite{Jariwala2014, Schaibley2016, Mak2016, Xiao2017, Mueller2018, Vitale2018}.

\begin{figure*}[t]
	\includegraphics{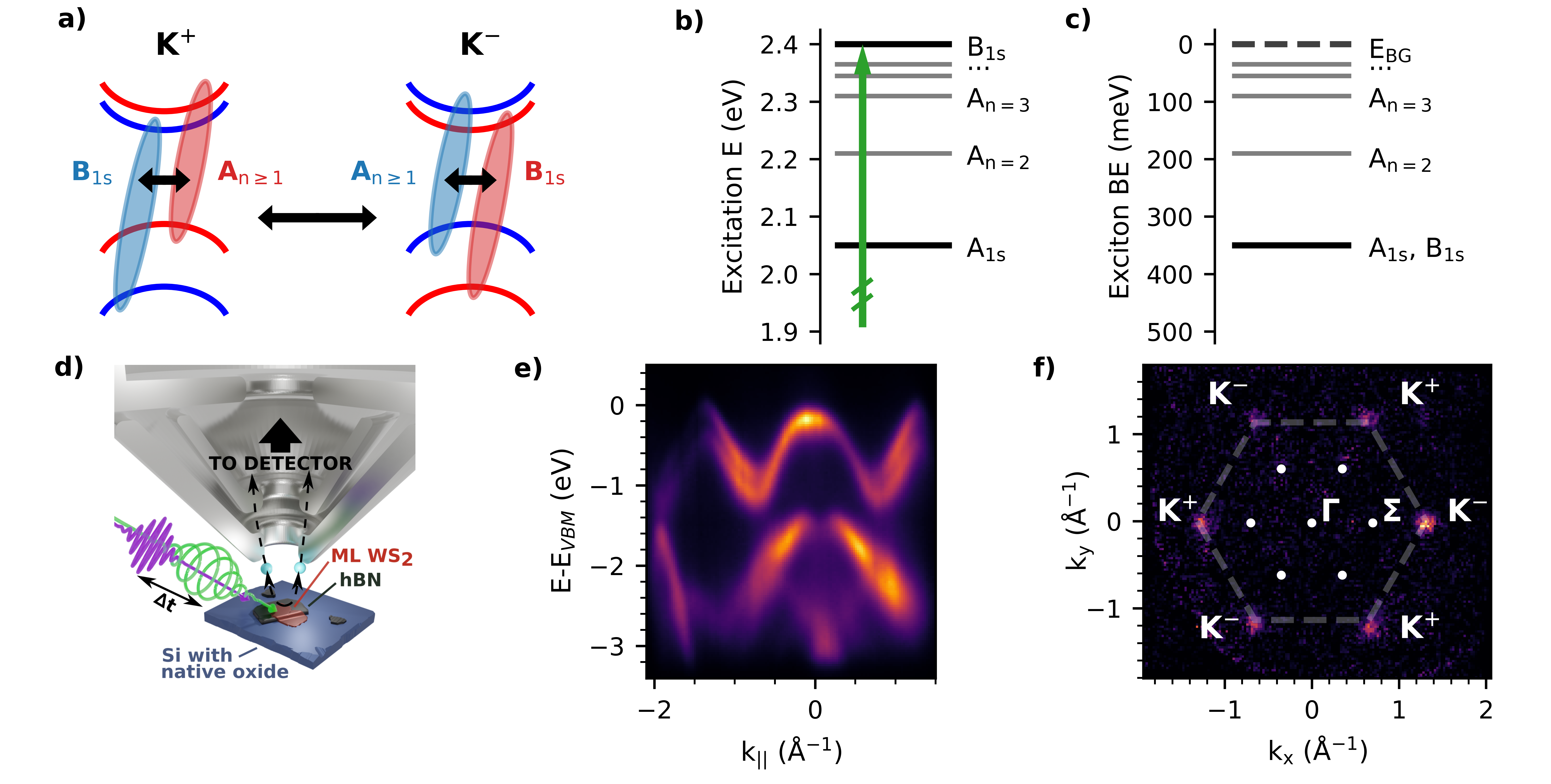}
    \caption{\textbf{Time-resolved ARPES of excitons in monolayer WS$_2$.}
        \textbf{a)} The Coulomb exchange interaction couples both intravalley and intervalley exciton states of opposite spin character. 
\textbf{b)} The 2.4 eV pump pulses employed here are resonant with the B exciton of WS$_2$.
 \textbf{c)} In our tr-ARPES spectra, the exciton signals are separated by exciton binding energy. 
Thus, exciton states with lower binding energies appear at higher energies above the VBM.
E$_{\textss{BG}}$ denotes the electronic band gap.
Excitation and binding energies are derived from \cite{Li2014a, Chernikov2014, Hill2015}.
 \textbf{d)} Linearly or circularly polarized pump pulses excite the sample, and a time-delayed XUV probe pulse photoejects electrons that are extracted into the momentum microscope column. 
    \textbf{e)} Cut along the WS$_2$ K$-\Gamma-$K valence band structure, collected with 		 
$h\nu_{\textss{probe}} = 27.6$ eV. 
    \textbf{f)} Raw exciton signal at 210 fs delay ($h\nu_{\textss{probe}} = 25.2$ eV).
The K$^+$, K$^-$, $\Sigma$, and $\Gamma$ valley locations are indicated. }
    \label{fig:intro1}
\end{figure*}

In TMD monolayers, the same strong Coulomb forces that give exciton binding energies on the order of $\sim$0.5 eV \cite{Chernikov2014, Hill2015}  
can also give rise to substantial interactions between exciton states, both within the same valley (\emph{intra}-valley coupling) and between different valleys (\emph{inter}-valley coupling). In particular, the Coulomb exchange interaction couples bright excitons of opposite spin character, coupling A and B excitons within the same valley (A$^+$ $\xleftrightarrow{}$ B$^+$) or degenerate excitons in opposite valleys (A$^+$ $\xleftrightarrow{}$ A$^-$, B$^+$ $\xleftrightarrow{}$ B$^-$), as illustrated in Fig. \ref{fig:intro1}a) \cite{Pikus1971, Yu2014a, Glazov2014, Qiu2015}. 
Due to this strong coupling, the exciton eigenstates are, in general, a combination of exciton states with mixed spin and valley characters \cite{Glazov2014, Yu2014a, Zhu2014a, Hao2016, Glazov2017, Guo2018}. 
Optical excitation addresses only the bright states, in which the electron and hole occupy the same valley with small total momentum $\mathbf{Q} = \mathbf{k}_e - \mathbf{k}_h$ and have net spin zero. 
Photoexcitation thus creates a superposition of eigenstates which then rapidly evolves in time, leading effectively to relaxation of the initial excitation and valley depolarization.
The strength of the eigenstate splitting due to Coulomb exchange, and thus its contribution to valley depolarization, is disputed among theoretical models
\cite{Yu2014a, Yu2014, Glazov2014, Kormanyos2015, Hao2016, Moody2016}. 
The additional role of exciton-phonon interactions in both intervalley and intravalley exciton dynamics is also non-negligible \cite{Wang2018a, Selig2018, He2020, Jiang2021}.
   
Many optical spectroscopy techniques have been employed to investigate depolarization lifetimes in monolayer
 TMDs, including photoluminescence \cite{Mak2012, Zhu2014, Wang2014, Lagarde2014, Yan2015}, 
differential transmission \cite{Mai2014, Schmidt2016}, 
time-resolved Kerr and Faraday rotation \cite{Zhu2014a, Conte2015, Hsu2015, Plechinger2016, Plechinger2017, McCormick2017, Schwemmer2017, Dey2017}, 
and multidimensional spectroscopies \cite{Hao2016, Smallwood2018, Guo2018, Lloyd2021, Purz2021}, among others \cite{Mahmood2017}. 
Valley polarization lifetimes ranging from a few picoseconds \cite{Wang2014, Lagarde2014, Schmidt2016} to hundreds \cite{Mai2014, Conte2015, Plechinger2016} or tens \cite{Lloyd2021} of femtoseconds have been reported, depending on the system under study and the spectroscopy method. 
Interpreting this body of work has been the subject of considerable debate \cite{Moody2015, Schmidt2016, Kioseoglou2016, Ye2019, Selig2020, Lloyd2021}.
Optical measurements record the excitation energy of the bright states (Fig. \ref{fig:intro1}b)), rendering discernment of the role of dark states difficult.
Critically, optical measurements are also momentum integrated, and can only distinguish between different excitons via the excitation energy and polarization selection rules.

Recently, technological advancements in time- and angle-resolved photoemission spectroscopy (tr-ARPES) have enabled the technique to be applied to small monolayer TMD samples \cite{Madeo2020, Man2021, Wallauer2021a, Karni2021, Schmitt2021},
providing direct momentum-space visualization of exciton wavefunctions as well as previously inaccessible dark states.
In this article, we present
comprehensive tr-ARPES measurements of the exciton dynamics in monolayer WS$_2$ following excitation at 2.4 eV, the nominal B exciton resonance \cite{Li2014a}. 
We measure full 4D ($k_x, k_y, E,t$) photoelectron distributions after both linearly polarized and circularly polarized photoexcitation. 
Resolving exciton binding energy (Fig. \ref{fig:intro1}c)) instead of excitation energy, we observe previously unseen strong mixing between A$_{n>1}$ and B$_{\textss{1s}}$ excitons in the initial photoexcited spectrum. 
With parallel momentum detection across the full Brillouin zone, we provide the first reported momentum-space visualizations of circular dichroism and ultrafast valley depolarization in the monolayer TMDs. 
We also observe that the exciton relaxation is accompanied by significant contraction of the initial exciton distribution in momentum space. 
These measurements report on the time-, energy-, and momentum-dependence of intervalley and intravalley exciton coupling, providing new insights on exciton formation in TMDs and the many-exciton coupled wave function.

\begin{figure*}[t]
	\includegraphics{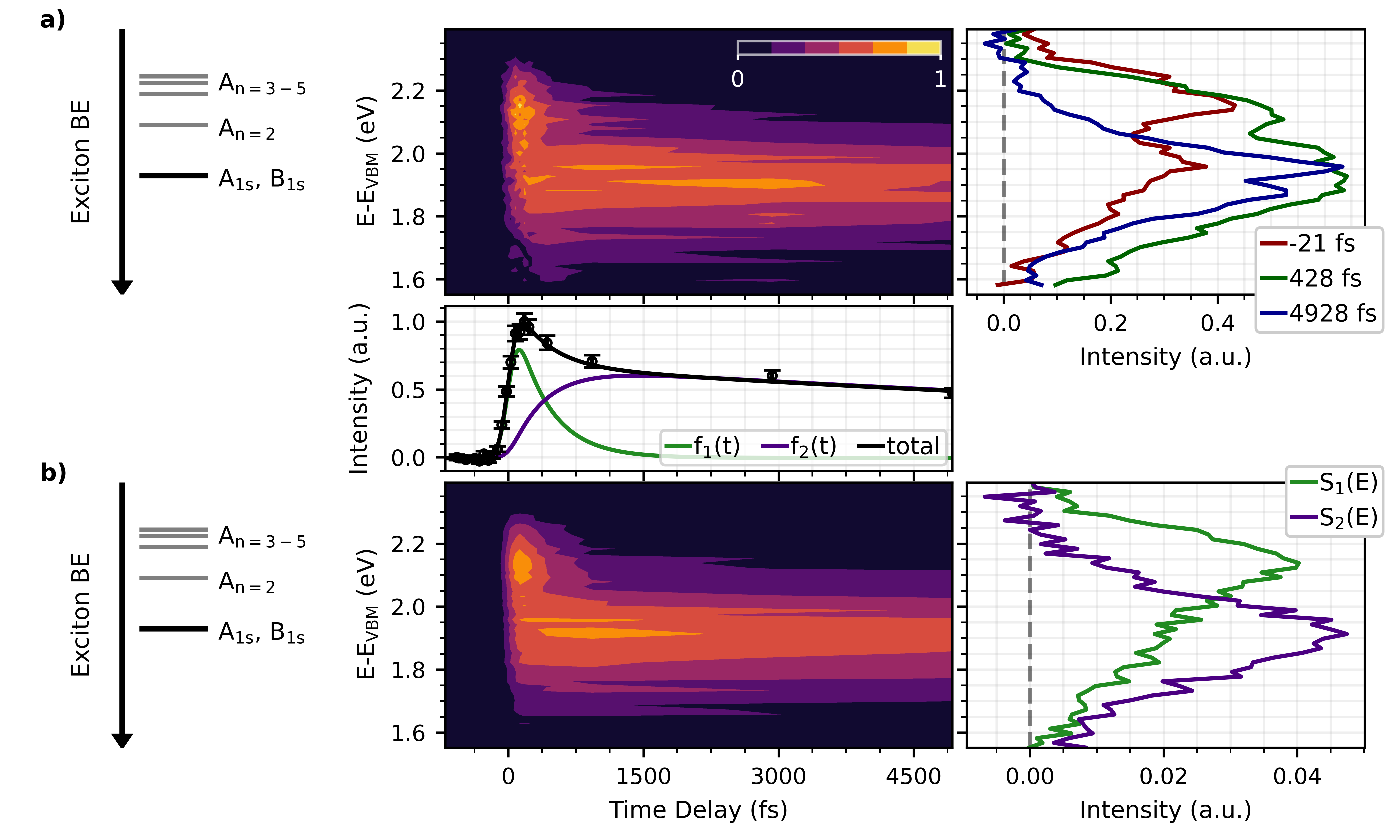}
    \caption{\textbf{Excitons formed by linearly polarized excitation. 
    a)} The time-resolved photoelectron spectrum of the K valley signals following \textit{s}-polarized photoexcitation shows a prompt, short-lived high energy feature consistent with excited A excitons, and a long-lived lower energy feature consistent with A$_{1s}$ and B$_{1s}$ excitons. 
    Energy distribution curves for selected time points are presented at right. $h\nu_{\textss{probe}} = 25.2$ eV.
    \textbf{b)} Two-component global analysis fit of the experimental data with the spectra of the two components (right) and their time-dependence (top).
    The total time dependence of the fit (black line) matches well to the experimental integrated intensities (black points).
    }
    \label{fig:GA2}
\end{figure*}

Our measurement scheme is shown in Fig. \ref{fig:intro1}d). 
Linearly and circularly polarized pump pulses ($h \nu_{\textss{pump}} = 2.4$ eV) and $p$-polarized extreme ultraviolet (XUV) probe pulses ($h\nu_{\textss{probe}} = $ 20$-$30 eV) with variable delay illuminate the sample and photoelectrons are collected by a custom time-of-flight momentum microscope \cite{Chernov2015, Medjanik2017}.
High data rates
are enabled by conducting the experiment at 61 MHz repetition rate with XUV probe pulses produced via cavity-enhanced high-harmonic generation (CE-HHG). 
The laser system and HHG beamline have been previously described in detail in \cite{Li_RSI2016, Corder2018, Corder2018a}. 
The sample is an exfoliated monolayer of WS$_2$ stacked on an exfoliated buffer layer of hexagonal boron nitride on a silicon substrate.
We use the spatial imaging capabilities of the momentum microscope \cite{Schoenhense2021} to isolate the photoelectron signal from the $\sim$10$\times$10 $\upmu$m$^2$ monolayer region of interest of the sample.
The valence band structure of the sample for a cut along the K-$\Gamma$-K axis of WS$_2$ is shown in Fig. \ref{fig:intro1}e)). 
The measured band structure shows that the valence band maximum (VBM) is located at the K$^+$ and K$^-$ valleys at the edges of the WS$_2$ Brillouin zone, as expected for a monolayer sample. 
The energy resolution is broadened to approximately 160 meV due to sample inhomogeneity \cite{Madeo2020}, but the spin-orbit splitting of the valence bands at the K valley is still clearly resolved.
Additional sample characterization and experimental details can be found in the Supplemental Material \cite{SI}.
All measurements are done at room temperature unless stated otherwise.

The 2.4 eV pump pulses produce photoexcited signals at the K$^+$ and K$^-$ valleys (Fig. \ref{fig:intro1}f)). 
In contrast to previous studies on monolayer WSe$_2$/hBN and strongly pumped WS$_2$ on bare silicon \cite{Madeo2020, Wallauer2021a}, the signals we observe at the $\Sigma$ valleys are centered $\sim$100 meV higher than the K valley signal, and are much weaker in intensity than previously reported in WSe$_2$ \cite{Madeo2020}. 
We find that the $\Sigma$/K intensity ratio depends strongly on the probe photon energy, but is always more than 2.5$\times$ smaller than that found in similar measurements of bulk WS$_2$ \cite{SI, Faraday}, where the $\Sigma$ valleys are lower in energy than the K valleys but the photoemission matrix elements are similar.
Thus, we believe there is only minor involvement of excitons with electrons at $\Sigma$ and focus here on the K valley excitons. 
By varying the excitation fluence between 1.3 $\upmu$J/cm$^2$ and 29 $\upmu$J/cm$^2$, we find the tr-ARPES signals to be fluence independent below 5 $\upmu$J/cm$^2$ \cite{SI}. 
Thus, all measurements reported here are conducted at 5 $\upmu$J/cm$^2$ excitation fluence, corresponding to an excitation density of approximately 7 x 10$^{11}$ carriers/cm$^2$ at our pump energy \cite{Li2014a}.
The cross-correlation of the pump and probe pulses yields a Gaussian instrument response function
with 200 $\pm$ 20 fs FWHM. 

Photoexcitation with linearly polarized light populates the K$^+$ and K$^-$ valleys equally and both valleys show the same dynamics.
The time-resolved photoelectron spectrum recorded with \textit{s}-polarized excitation is shown in Fig. \ref{fig:GA2}a). 
No intensity is ever observed in the conduction band at $E_{\textss{VBM}} + h \nu_{\textss{pump}} = 2.4$ eV, indicating the direct formation of bound excitons. 
Exciton signals appear below the conduction band due to the exciton binding energy \cite{Rustagi2018, Christiansen2019, Madeo2020}, as illustrated in Fig. \ref{fig:intro1}c) and the leftmost scales in Fig. \ref{fig:GA2}.
The most prominent feature at early pump-probe delays is the large intensity at energies between 2.05$-$2.3 eV above the VBM in the K valley. This corresponds to exciton binding energies compatible with excited A excitons (A$_{n>1}$) \cite{Chernikov2014, Hill2015}. 
At longer delays, a lower energy feature centered at approximately 1.93 eV grows in and persists beyond the longest pump-probe delays recorded (25 ps). 
This lower energy feature appears at binding energies compatible with those expected for both the A$_{1s}$ and B$_{1s}$ excitons, which are expected to have similar binding energy \cite{Hill2015, Stier2016, Katsch2019}.
Similar results are obtained with $p$-polarized excitation, indicating that excitation of spin-forbidden intravalley excitons
by the out-of-plane component of the electric field has a negligible effect on the observed signals, as expected due to the much smaller transition dipole for these excitations \cite{Echeverry2016, Wang2017a, SI}.

\begin{figure}[t!hb]
	\includegraphics{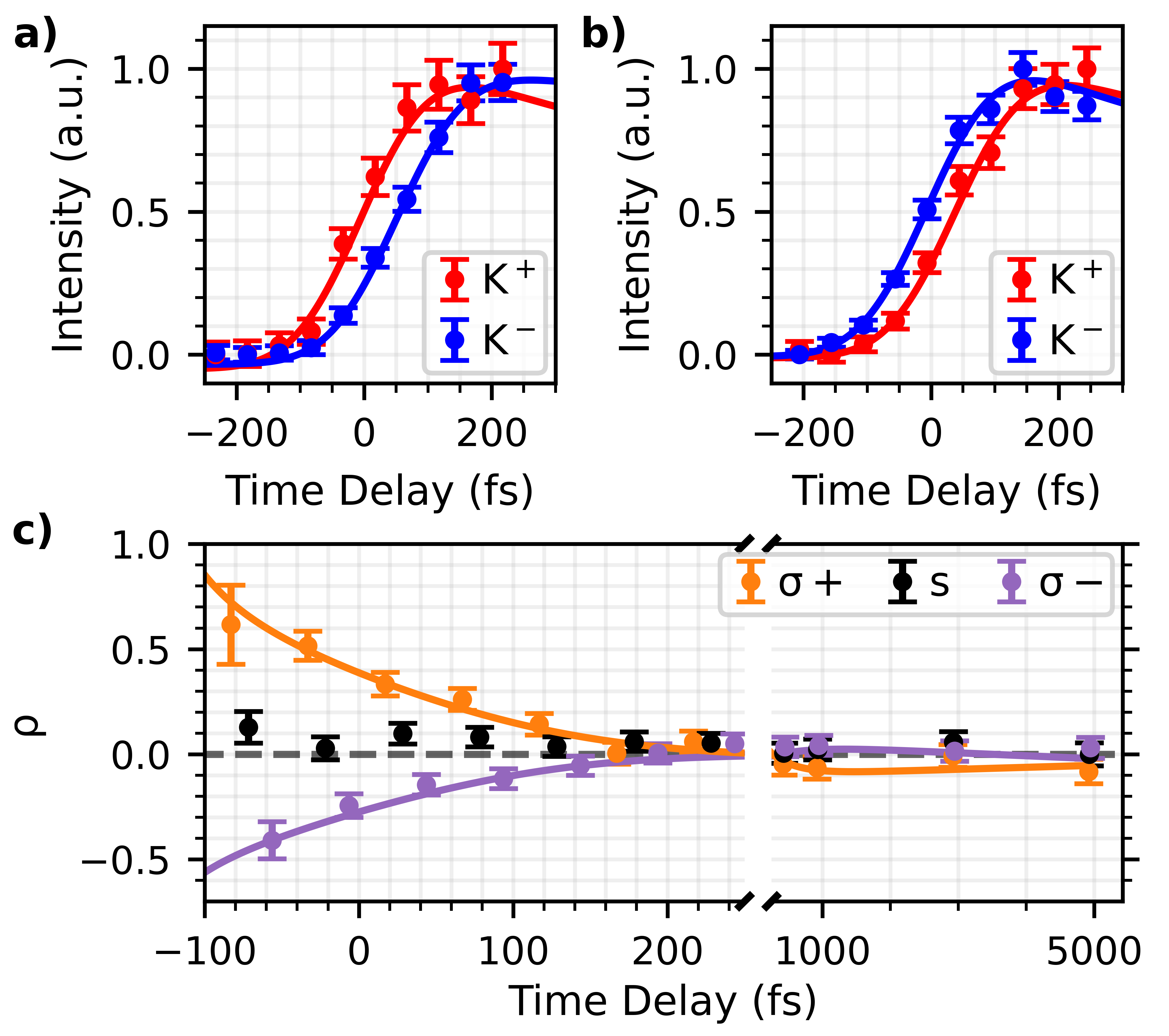}
    \caption{\textbf{Valley asymmetry.}
	Integrated intensities for the K$^+$ and K$^-$ valleys following \textbf{a)} $\sigma^+$ and \textbf{b)} $\sigma^-$ photoexcitation ($h\nu_{\textss{probe}} = 25.2$ eV).
\textbf{c)} The valley asymmetry ($\rho$) for $\sigma^+$, $\sigma^-$, and \textit{s}-polarized photoexcitation.
    Points are experimental data, and lines represent the global analysis fits.
    The observed valley asymmetry decay time scale of $\sim$250 fs is limited by the instrument response.}
    \label{fig:circ3_top}
\end{figure}

\begin{figure*}[b!ht]
	\includegraphics{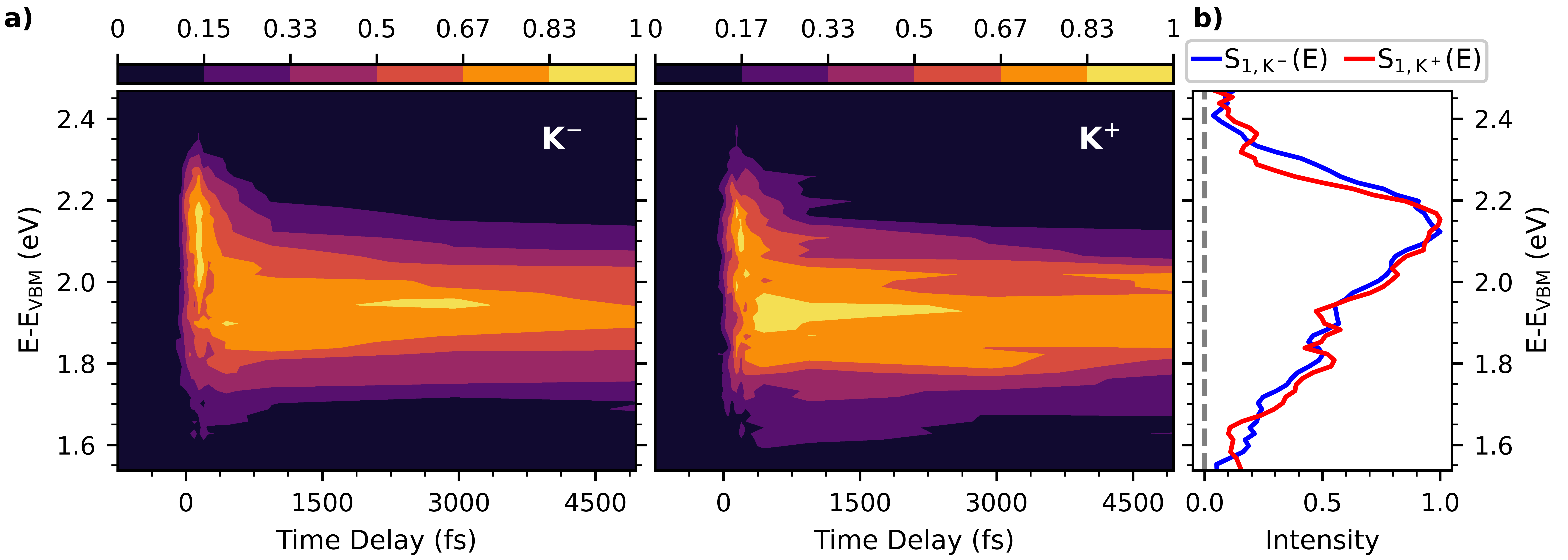}
    \caption{\textbf{Excitons formed by $\sigma^-$ excitation.}
    \textbf{a)} Comparison of the K$^-$ and K$^+$ valleys following $\sigma^-$ excitation shows that the two valleys present nearly identical dynamics, the difference being a $43 \pm 4$ fs delay in the appearance of signal in the unpumped K$^+$ valley. $h\nu_{\textss{probe}} = 25.2$ eV. 
	\textbf{b)} The GA spectral components S$_1$(E) for the K$^-$ and K$^+$ valley signals
show that the population transfer from the pumped valleys to the unpumped valleys does not involve significant changes in the energy distribution.
    }
    \label{fig:circ3_bottom}
\end{figure*}

The spectrum of Fig. \ref{fig:GA2}a) consists of multiple overlapping components. 
To deconvolve the overlapping spectral and temporal components of the experimental data, we have applied global analysis (GA) \cite{Stokkum2004, Bisgaard2009, Jin2012, Beckwith2020, Rebholz2021}, which reduces the signal to a few principal spectral components $S_i(E)$,  each with simple exponential time dynamics $f_i(t)$ convolved with the instrument response, viz. $ I(E,t) = \sum_i^N S_i(E) f_i(t) $. 
We find an excellent fit with only $N= 2$ components as shown in Fig. \ref{fig:GA2}b). 
Component 1 corresponds to the initially excited population and
is peaked at $E-E_{\textss{VBM}} =$ 2.15 eV but also shows a long tail to lower photoelectron energies (larger binding energies) covering the region of the B$_{1s}$ exciton. 
We assign this to an initially excited mixture of A$_{n>1}$ and B$_{1s}$ excitons. 
Despite initial photoexcitation of the B exciton resonance, we clearly observe
strong weighting towards lower binding energies consistent with population of the A$_{n>1}$ excited states.
This is seen both in the GA results \emph{and} in the raw data, with both much more weighted towards the A$_{n>1}$ states than the B exciton than what would be expected from the optical absorption spectrum \cite{Li2014a}. 
This indicates very strong mixing of the B$_{1s}$ states with A$_{n>1}$ states, such that photoexcitation of what is nominally the B exciton resonance promptly populates A$_{n>1}$ exciton states as well. 
Such A/B mixing due to intravalley Coulomb exchange has been discussed before \cite{Guo2018}, although the degree of mixing we observe here is much larger than suggested by this previous work. 

Component 1 decays with a time constant of 378 $\pm$ 40 fs, giving rise to component 2, shown in Fig. \ref{fig:GA2}b). 
Component 2 is centered at the energy of the long-delay photoelectron spectrum and has a GA lifetime longer than 50 ps.  
We assign component 2 to a mixture of relaxed bright and dark 1s excitons with binding energies of approximately 0.35 eV. 
We find adding additional components beyond $N=2$ does not improve the quality of the global fit or offer additional physical insight. 
More details of the GA can be found in the Supplemental Material \cite{SI}.

The dynamics observed under linearly polarized excitation can be due to a mixture of both intervalley and intravalley relaxation mechanisms. To disentangle their relative contributions, we use circularly polarized pump pulses to prepare valley-polarized excitons. 
We excite the sample with both $\sigma^{+}$ and $\sigma^{-}$ polarizations, which preferentially excite K$^+$ and K$^-$ valleys, respectively.
Figs. \ref{fig:circ3_top}a) and \ref{fig:circ3_top}b) show the integrated K$^+$ and K$^-$ valley signals under $\sigma^+$ and $\sigma^-$ polarizations, respectively.
Fig. \ref{fig:circ3_top}c) shows the valley asymmetry, $\rho$(t), defined by:

\begin{equation*}
  \rho(t) = \frac{I_{K^+}(t) - I_{K^-}(t)}{I_{K^+}(t) + I_{K^-}(t)},
\end{equation*}

\noindent where I$_{K^+}$ and I$_{K^-}$ denote the integrated intensity in the K$^+$ and K$^-$ valleys, respectively. 
The valley asymmetry decays in approximately 250 fs, limited by the instrument response. 
We observe similar time scales for the decay of $\rho(t)$ for low-temperature data recorded at 126 K \cite{SI}, suggesting exciton-phonon coupling is not a main driver of the dynamics. 
For comparison, we also show the \textit{s}-polarized data in Fig. \ref{fig:circ3_top}c), which shows no valley asymmetry. 
The K$^+$ and K$^-$ valley signals following \textit{s}-polarized photoexcitation can be found in the Supplemental Material \cite{SI}.

Figs. \ref{fig:circ3_bottom}a) and \ref{fig:circ3_bottom}b) show the time-resolved photoelectron spectra and the 
$S_1(E)$ GA spectral components
for the K$^-$ and K$^+$ valleys after $\sigma^-$ excitation. 
Strikingly, the spectrum in the unpumped K$^+$ valley does not show any appreciable difference to that of the initially pumped K$^-$ valley, except an approximately 50 fs delay between the population of the
two
valleys. 
We quantify this by applying the same GA described above to the K$^+$ and K$^-$ valleys independently in the circularly polarized data. 
For the unpumped valley, we allow for a shift, $\Delta t$, in the onset of the time dynamics $f_i(t) \rightarrow f_i(t - \Delta t)$.
We find the spectral components and exponential rates in the K$^+$ and K$^-$ valleys to be similar to one another and also to those found under $s$-polarized excitation. 
The delayed onset captured by $\Delta t$ was found to be the singular notable difference between the dynamics in the two valleys. 
From the GA fitting, we find  $\Delta t = 43 \pm 4$ fs for $\sigma^{-}$ excitation and $\Delta t = 53 \pm 6$ fs for $\sigma^+$. 
These 50 fs shifts are also apparent in the integrated signals of Figs. \ref{fig:circ3_top}a) and \ref{fig:circ3_top}b).
As a control, we analyzed the $s$-polarized data in the same way and find $\Delta t = 6 \pm 5$ fs \cite{SI}. 
The small 50 fs shift, indicating very rapid valley depolarization, is consistent with the $\sim$250 fs time scale on which $\rho(t)$ becomes zero when the instrument response is considered. 
The integrated GA model results are also shown as the lines in Fig. \ref{fig:circ3_top}. 

Importantly, the prompt valley depolarization we observe in the tr-ARPES signal is not accompanied by energy relaxation. 
This is evident from both the data of Fig. \ref{fig:circ3_bottom}a) 
as well as
the GA analysis in Fig. \ref{fig:circ3_bottom}b), with $S_{1,\textss{K}^+}(E)$ closely resembling $S_{1,\textss{K}^-}(E)$. 
This is consistent with valley depolarization driven by intervalley Coulomb exchange, which couples energetically degenerate bright exciton states, A$^\pm$ $\xleftrightarrow{ }$ A$^\mp$, B$^\pm$ $\xleftrightarrow{ }$ B$^\mp$ \cite{Yu2014, Yu2014a, Qiu2015, Hao2016, Wang2018}, 
but is in contrast to other recently proposed non-degenerate intervalley depolarization mechanisms that couple A$^\pm$ $\xleftrightarrow{ }$ B$^\mp$, B$^\pm$ $\xleftrightarrow{ }$ A$^\mp$ \cite{Schmidt2016, Selig2019, Selig2020, Berghaeuser2018, BernalVillamil2018}.
The observed timescale is also consistent with calculations of intervalley exchange matrix elements. 
For  large $\sim$0.1 \AA$^{-1}$ center-of-mass momentum, valley depolarization via the exchange interaction is expected to be extremely efficient, with eigenstate energy splittings of 10s of meV \cite{Qiu2015} and corresponding valley depolarization predicted in several 10s of fs \cite{Yu2014}. 

\begin{figure}[b!ht]
	\includegraphics{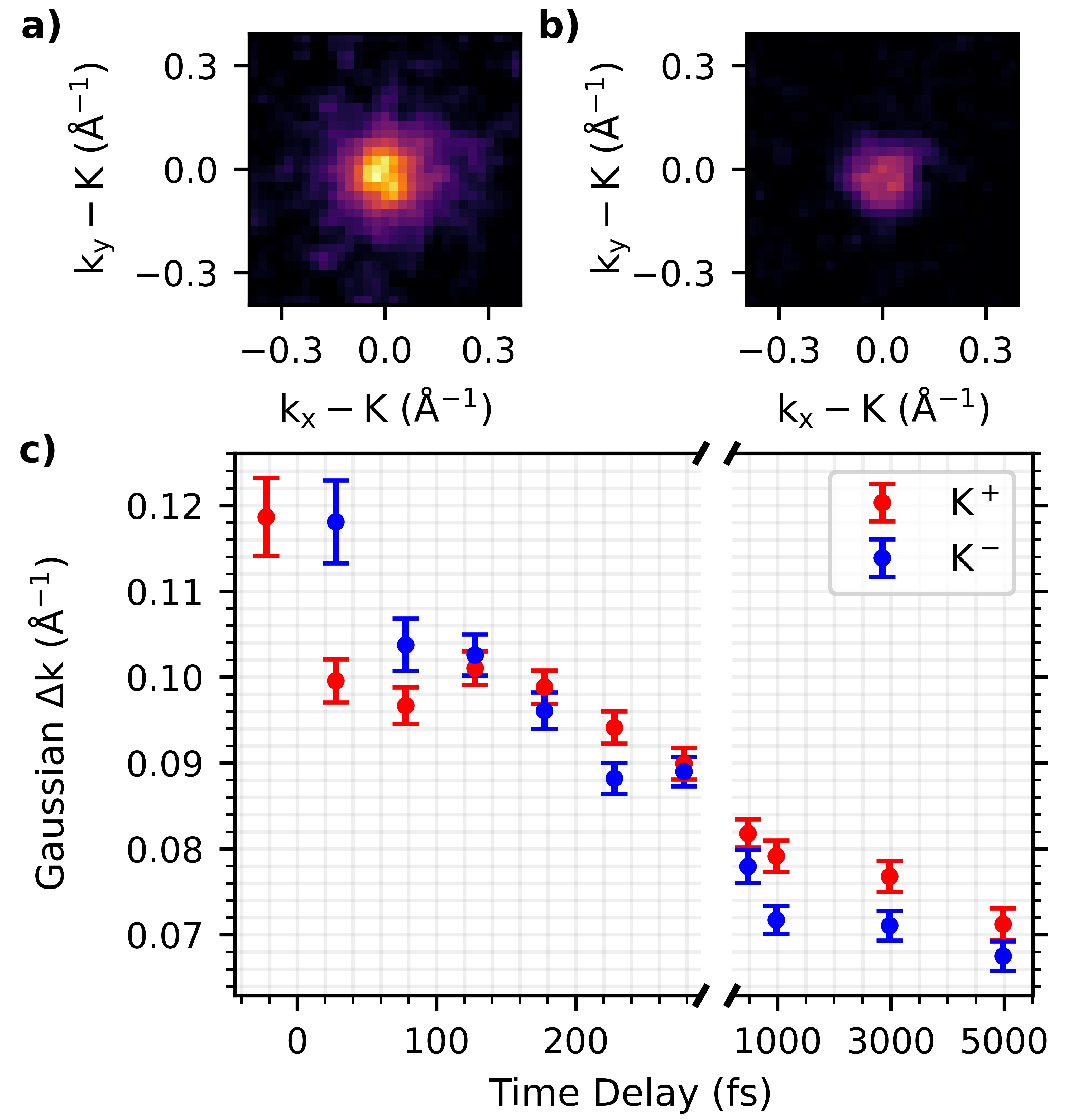}
    \caption{\textbf{Time dependence of the 
    photoelectron
    momentum distributions.}
    Representative images of a K$^+$ valley at
    \textbf{a)} 78 fs delay and
    \textbf{b)} 5 ps delay after $\sigma^+$ polarized photoexcitation, collected with $h\nu_{\textss{probe}} = 30$ eV.
    \textbf{c)} Standard deviation ($\Delta k$) of Gaussian distribution fits to the momentum distributions in the K$^+$ and K$^-$ valleys at each delay time. 
    Significant contraction of the momentum extent of the excitons in each valley is observed.
    }
    \label{fig:analysis4}
\end{figure}

In Fig. \ref{fig:analysis4}, we additionally examine the momentum distributions of the photoelectrons. The data shown are recorded after $\sigma^+$ excitation with 30 eV probe energy.
A representative image of the initial momentum distribution of the K$^+$ valley signal is shown in Fig. \ref{fig:analysis4}a).
At 5 ps, the distribution has relaxed to the narrower one in Fig. \ref{fig:analysis4}b). 
We quantify the extent of the photoelectron momentum distributions in the K$^+$ and K$^-$ valleys as a function of time by fitting the energy-integrated K valley signal with a Gaussian, 
$\exp[-(1/2)|\mathbf{k}-K|^2/(\Delta k)^2]$,
and report the standard deviation, $\Delta k$, in Fig. \ref{fig:analysis4}. 
We observe that the initial photoelectron momentum distribution encompasses nearly twice the extent of the relaxed photoelectron population at approximately 5 ps delay time.
The final distribution width of $\Delta k \sim$ 0.07 \r{A}$^{-1}$ is commensurate with the recent experimental measurement of relaxed exciton states of WSe$_2$ at 90 K \cite{Man2021}.

Remarkably, no large differences are observed in the momentum distributions in the K$^+$ and K$^-$ valleys. For example, the initial K$^+$ valley distribution with $\Delta k = 0.12$\; \AA$^{-1}$ \;arrives at the K$^-$ valley 50 fs later with the same width. 
The Coulomb exchange interaction conserves the total exciton momentum $\mathbf{Q} = \mathbf{k}_e - \mathbf{k}_h$. 
While we do not measure $\mathbf{Q}$ directly in this experiment, we conjecture that the width of the distribution in $\mathbf{Q}$ is correlated with the width of our photoelectron distributions. 
Thus, the conservation of the photoelectron momentum distribution after intervalley coupling suggests conservation of the exciton momentum, consistent with the intervalley exchange coupling mechanism of valley depolarization. 

While energy conservation and momentum conservation during valley depolarization are both consistent with intervalley Coulomb exchange coupling, the similarity of the energy and momentum distributions between the pumped and unpumped valleys suggests that rate of transfer does not appear to depend strongly on the exciton binding energy or exciton momentum. 
The strength of the exchange interaction is expected to scale as $|\mathbf{Q}|$ and the square of the electron-hole wavefunction overlap \cite{Yu2014, Qiu2015, Chen2018}. 
This would indicate faster transfer for excitons with larger momentum or tighter electron-hole binding. 
However, within our experimental resolution, we do not observe such $\mathbf{Q}$- or E-dependence in the population transfer.

In this work, we have used time-of-flight momentum microscopy combined with ultrashort XUV pulses at 61 MHz repetition rate to image the exciton dynamics in monolayer WS$_2$. 
Our measurements record the dynamics in the natural momentum-space basis in which theory and calculations are formulated,
and shed new light on the ultrafast intervalley and intravalley coupling dynamics in monolayer TMDs.
While these dynamics have been the subject of extensive optical spectroscopy, to our knowledge these are the first reported momentum-space measurements of valley depolarization in the monolayer TMDs.
Future work with higher resolution
can address the energy- and momentum-dependence of exciton coupling in further detail and also study these phenomena in 2D heterostructures. \\

This material is based upon work supported by the U.S. Department of Energy, Office of Science, Office of Basic Energy Sciences under award number DE-SC0022004 and the Air Force Office of Scientific Research under FA9550-20-1-0259.
R.K.K. acknowledges support from the U.S. National Science Foundation under Grant No. CHE-1935885.
X.D. acknowledges support from the U.S. National Science Foundation under Grant No. DMR-1808491.
M.G.W. acknowledges support from the U.S. Department of Energy (DOE), Office of Science, Office of Basic Energy Sciences, Chemical Sciences, Geosciences, and Biosciences (CSGB) Division, and the Catalysis Science Program under DOE Contract No. DE-SC0012704. 
Z.H.W. acknowledges support from the U.S. National Science Foundation Graduate Research Fellowship Program.

\FloatBarrier

\bibliographystyle{apsrev4-2}
\bibliography{WS2_ML_paper}

\begin{thebibliography}{89}%
\makeatletter
\providecommand \@ifxundefined [1]{%
 \@ifx{#1\undefined}
}%
\providecommand \@ifnum [1]{%
 \ifnum #1\expandafter \@firstoftwo
 \else \expandafter \@secondoftwo
 \fi
}%
\providecommand \@ifx [1]{%
 \ifx #1\expandafter \@firstoftwo
 \else \expandafter \@secondoftwo
 \fi
}%
\providecommand \natexlab [1]{#1}%
\providecommand \enquote  [1]{``#1''}%
\providecommand \bibnamefont  [1]{#1}%
\providecommand \bibfnamefont [1]{#1}%
\providecommand \citenamefont [1]{#1}%
\providecommand \href@noop [0]{\@secondoftwo}%
\providecommand \href [0]{\begingroup \@sanitize@url \@href}%
\providecommand \@href[1]{\@@startlink{#1}\@@href}%
\providecommand \@@href[1]{\endgroup#1\@@endlink}%
\providecommand \@sanitize@url [0]{\catcode `\\12\catcode `\$12\catcode
  `\&12\catcode `\#12\catcode `\^12\catcode `\_12\catcode `\%12\relax}%
\providecommand \@@startlink[1]{}%
\providecommand \@@endlink[0]{}%
\providecommand \url  [0]{\begingroup\@sanitize@url \@url }%
\providecommand \@url [1]{\endgroup\@href {#1}{\urlprefix }}%
\providecommand \urlprefix  [0]{URL }%
\providecommand \Eprint [0]{\href }%
\providecommand \doibase [0]{https://doi.org/}%
\providecommand \selectlanguage [0]{\@gobble}%
\providecommand \bibinfo  [0]{\@secondoftwo}%
\providecommand \bibfield  [0]{\@secondoftwo}%
\providecommand \translation [1]{[#1]}%
\providecommand \BibitemOpen [0]{}%
\providecommand \bibitemStop [0]{}%
\providecommand \bibitemNoStop [0]{.\EOS\space}%
\providecommand \EOS [0]{\spacefactor3000\relax}%
\providecommand \BibitemShut  [1]{\csname bibitem#1\endcsname}%
\let\auto@bib@innerbib\@empty
\bibitem [{\citenamefont {Xiao}\ \emph {et~al.}(2007)\citenamefont {Xiao},
  \citenamefont {Yao},\ and\ \citenamefont {Niu}}]{Xiao2007}%
  \BibitemOpen
  \bibfield  {author} {\bibinfo {author} {\bibfnamefont {D.}~\bibnamefont
  {Xiao}}, \bibinfo {author} {\bibfnamefont {W.}~\bibnamefont {Yao}},\ and\
  \bibinfo {author} {\bibfnamefont {Q.}~\bibnamefont {Niu}},\ }\href
  {https://doi.org/10.1103/physrevlett.99.236809} {\bibfield  {journal}
  {\bibinfo  {journal} {Physical Review Letters}\ }\textbf {\bibinfo {volume}
  {99}},\ \bibinfo {pages} {236809} (\bibinfo {year} {2007})}\BibitemShut
  {NoStop}%
\bibitem [{\citenamefont {Yao}\ \emph {et~al.}(2008)\citenamefont {Yao},
  \citenamefont {Xiao},\ and\ \citenamefont {Niu}}]{Yao2008}%
  \BibitemOpen
  \bibfield  {author} {\bibinfo {author} {\bibfnamefont {W.}~\bibnamefont
  {Yao}}, \bibinfo {author} {\bibfnamefont {D.}~\bibnamefont {Xiao}},\ and\
  \bibinfo {author} {\bibfnamefont {Q.}~\bibnamefont {Niu}},\ }\href
  {https://doi.org/10.1103/physrevb.77.235406} {\bibfield  {journal} {\bibinfo
  {journal} {Physical Review B}\ }\textbf {\bibinfo {volume} {77}},\ \bibinfo
  {pages} {235406} (\bibinfo {year} {2008})}\BibitemShut {NoStop}%
\bibitem [{\citenamefont {Cao}\ \emph {et~al.}(2012)\citenamefont {Cao},
  \citenamefont {Wang}, \citenamefont {Han}, \citenamefont {Ye}, \citenamefont
  {Zhu}, \citenamefont {Shi}, \citenamefont {Niu}, \citenamefont {Tan},
  \citenamefont {Wang}, \citenamefont {Liu},\ and\ \citenamefont
  {Feng}}]{Cao2012}%
  \BibitemOpen
  \bibfield  {author} {\bibinfo {author} {\bibfnamefont {T.}~\bibnamefont
  {Cao}}, \bibinfo {author} {\bibfnamefont {G.}~\bibnamefont {Wang}}, \bibinfo
  {author} {\bibfnamefont {W.}~\bibnamefont {Han}}, \bibinfo {author}
  {\bibfnamefont {H.}~\bibnamefont {Ye}}, \bibinfo {author} {\bibfnamefont
  {C.}~\bibnamefont {Zhu}}, \bibinfo {author} {\bibfnamefont {J.}~\bibnamefont
  {Shi}}, \bibinfo {author} {\bibfnamefont {Q.}~\bibnamefont {Niu}}, \bibinfo
  {author} {\bibfnamefont {P.}~\bibnamefont {Tan}}, \bibinfo {author}
  {\bibfnamefont {E.}~\bibnamefont {Wang}}, \bibinfo {author} {\bibfnamefont
  {B.}~\bibnamefont {Liu}},\ and\ \bibinfo {author} {\bibfnamefont
  {J.}~\bibnamefont {Feng}},\ }\bibfield  {journal} {\bibinfo  {journal}
  {Nature Communications}\ }\textbf {\bibinfo {volume} {3}},\ \href
  {https://doi.org/10.1038/ncomms1882} {10.1038/ncomms1882} (\bibinfo {year}
  {2012})\BibitemShut {NoStop}%
\bibitem [{\citenamefont {Xiao}\ \emph {et~al.}(2012)\citenamefont {Xiao},
  \citenamefont {Liu}, \citenamefont {Feng}, \citenamefont {Xu},\ and\
  \citenamefont {Yao}}]{Xiao2012}%
  \BibitemOpen
  \bibfield  {author} {\bibinfo {author} {\bibfnamefont {D.}~\bibnamefont
  {Xiao}}, \bibinfo {author} {\bibfnamefont {G.-B.}\ \bibnamefont {Liu}},
  \bibinfo {author} {\bibfnamefont {W.}~\bibnamefont {Feng}}, \bibinfo {author}
  {\bibfnamefont {X.}~\bibnamefont {Xu}},\ and\ \bibinfo {author}
  {\bibfnamefont {W.}~\bibnamefont {Yao}},\ }\href
  {https://doi.org/10.1103/physrevlett.108.196802} {\bibfield  {journal}
  {\bibinfo  {journal} {Physical Review Letters}\ }\textbf {\bibinfo {volume}
  {108}},\ \bibinfo {pages} {196802} (\bibinfo {year} {2012})}\BibitemShut
  {NoStop}%
\bibitem [{\citenamefont {Mak}\ \emph {et~al.}(2012)\citenamefont {Mak},
  \citenamefont {He}, \citenamefont {Shan},\ and\ \citenamefont
  {Heinz}}]{Mak2012}%
  \BibitemOpen
  \bibfield  {author} {\bibinfo {author} {\bibfnamefont {K.~F.}\ \bibnamefont
  {Mak}}, \bibinfo {author} {\bibfnamefont {K.}~\bibnamefont {He}}, \bibinfo
  {author} {\bibfnamefont {J.}~\bibnamefont {Shan}},\ and\ \bibinfo {author}
  {\bibfnamefont {T.~F.}\ \bibnamefont {Heinz}},\ }\href
  {https://doi.org/10.1038/nnano.2012.96} {\bibfield  {journal} {\bibinfo
  {journal} {Nature Nanotechnology}\ }\textbf {\bibinfo {volume} {7}},\
  \bibinfo {pages} {494} (\bibinfo {year} {2012})}\BibitemShut {NoStop}%
\bibitem [{\citenamefont {Zeng}\ \emph {et~al.}(2012)\citenamefont {Zeng},
  \citenamefont {Dai}, \citenamefont {Yao}, \citenamefont {Xiao},\ and\
  \citenamefont {Cui}}]{Zeng2012}%
  \BibitemOpen
  \bibfield  {author} {\bibinfo {author} {\bibfnamefont {H.}~\bibnamefont
  {Zeng}}, \bibinfo {author} {\bibfnamefont {J.}~\bibnamefont {Dai}}, \bibinfo
  {author} {\bibfnamefont {W.}~\bibnamefont {Yao}}, \bibinfo {author}
  {\bibfnamefont {D.}~\bibnamefont {Xiao}},\ and\ \bibinfo {author}
  {\bibfnamefont {X.}~\bibnamefont {Cui}},\ }\href
  {https://doi.org/10.1038/nnano.2012.95} {\bibfield  {journal} {\bibinfo
  {journal} {Nat. Nanotechnol.}\ }\textbf {\bibinfo {volume} {7}},\ \bibinfo
  {pages} {490} (\bibinfo {year} {2012})}\BibitemShut {NoStop}%
\bibitem [{\citenamefont {Kioseoglou}\ \emph {et~al.}(2012)\citenamefont
  {Kioseoglou}, \citenamefont {Hanbicki}, \citenamefont {Currie}, \citenamefont
  {Friedman}, \citenamefont {Gunlycke},\ and\ \citenamefont
  {Jonker}}]{Kioseoglou2012}%
  \BibitemOpen
  \bibfield  {author} {\bibinfo {author} {\bibfnamefont {G.}~\bibnamefont
  {Kioseoglou}}, \bibinfo {author} {\bibfnamefont {A.~T.}\ \bibnamefont
  {Hanbicki}}, \bibinfo {author} {\bibfnamefont {M.}~\bibnamefont {Currie}},
  \bibinfo {author} {\bibfnamefont {A.~L.}\ \bibnamefont {Friedman}}, \bibinfo
  {author} {\bibfnamefont {D.}~\bibnamefont {Gunlycke}},\ and\ \bibinfo
  {author} {\bibfnamefont {B.~T.}\ \bibnamefont {Jonker}},\ }\href
  {https://doi.org/10.1063/1.4768299} {\bibfield  {journal} {\bibinfo
  {journal} {Applied Physics Letters}\ }\textbf {\bibinfo {volume} {101}},\
  \bibinfo {pages} {221907} (\bibinfo {year} {2012})}\BibitemShut {NoStop}%
\bibitem [{\citenamefont {Wang}\ \emph
  {et~al.}(2018{\natexlab{a}})\citenamefont {Wang}, \citenamefont {Chernikov},
  \citenamefont {Glazov}, \citenamefont {Heinz}, \citenamefont {Marie},
  \citenamefont {Amand},\ and\ \citenamefont {Urbaszek}}]{Wang2018}%
  \BibitemOpen
  \bibfield  {author} {\bibinfo {author} {\bibfnamefont {G.}~\bibnamefont
  {Wang}}, \bibinfo {author} {\bibfnamefont {A.}~\bibnamefont {Chernikov}},
  \bibinfo {author} {\bibfnamefont {M.~M.}\ \bibnamefont {Glazov}}, \bibinfo
  {author} {\bibfnamefont {T.~F.}\ \bibnamefont {Heinz}}, \bibinfo {author}
  {\bibfnamefont {X.}~\bibnamefont {Marie}}, \bibinfo {author} {\bibfnamefont
  {T.}~\bibnamefont {Amand}},\ and\ \bibinfo {author} {\bibfnamefont
  {B.}~\bibnamefont {Urbaszek}},\ }\href
  {https://doi.org/10.1103/revmodphys.90.021001} {\bibfield  {journal}
  {\bibinfo  {journal} {Reviews of Modern Physics}\ }\textbf {\bibinfo {volume}
  {90}},\ \bibinfo {pages} {021001} (\bibinfo {year}
  {2018}{\natexlab{a}})}\BibitemShut {NoStop}%
\bibitem [{\citenamefont {Sallen}\ \emph {et~al.}(2012)\citenamefont {Sallen},
  \citenamefont {Bouet}, \citenamefont {Marie}, \citenamefont {Wang},
  \citenamefont {Zhu}, \citenamefont {Han}, \citenamefont {Lu}, \citenamefont
  {Tan}, \citenamefont {Amand}, \citenamefont {Liu},\ and\ \citenamefont
  {Urbaszek}}]{Sallen2012}%
  \BibitemOpen
  \bibfield  {author} {\bibinfo {author} {\bibfnamefont {G.}~\bibnamefont
  {Sallen}}, \bibinfo {author} {\bibfnamefont {L.}~\bibnamefont {Bouet}},
  \bibinfo {author} {\bibfnamefont {X.}~\bibnamefont {Marie}}, \bibinfo
  {author} {\bibfnamefont {G.}~\bibnamefont {Wang}}, \bibinfo {author}
  {\bibfnamefont {C.~R.}\ \bibnamefont {Zhu}}, \bibinfo {author} {\bibfnamefont
  {W.~P.}\ \bibnamefont {Han}}, \bibinfo {author} {\bibfnamefont
  {Y.}~\bibnamefont {Lu}}, \bibinfo {author} {\bibfnamefont {P.~H.}\
  \bibnamefont {Tan}}, \bibinfo {author} {\bibfnamefont {T.}~\bibnamefont
  {Amand}}, \bibinfo {author} {\bibfnamefont {B.~L.}\ \bibnamefont {Liu}},\
  and\ \bibinfo {author} {\bibfnamefont {B.}~\bibnamefont {Urbaszek}},\ }\href
  {https://doi.org/10.1103/physrevb.86.081301} {\bibfield  {journal} {\bibinfo
  {journal} {Physical Review B}\ }\textbf {\bibinfo {volume} {86}},\ \bibinfo
  {pages} {081301} (\bibinfo {year} {2012})}\BibitemShut {NoStop}%
\bibitem [{\citenamefont {Liu}\ \emph {et~al.}(2013)\citenamefont {Liu},
  \citenamefont {Shan}, \citenamefont {Yao}, \citenamefont {Yao},\ and\
  \citenamefont {Xiao}}]{Liu2013}%
  \BibitemOpen
  \bibfield  {author} {\bibinfo {author} {\bibfnamefont {G.-B.}\ \bibnamefont
  {Liu}}, \bibinfo {author} {\bibfnamefont {W.-Y.}\ \bibnamefont {Shan}},
  \bibinfo {author} {\bibfnamefont {Y.}~\bibnamefont {Yao}}, \bibinfo {author}
  {\bibfnamefont {W.}~\bibnamefont {Yao}},\ and\ \bibinfo {author}
  {\bibfnamefont {D.}~\bibnamefont {Xiao}},\ }\href
  {https://doi.org/10.1103/physrevb.88.085433} {\bibfield  {journal} {\bibinfo
  {journal} {Physical Review B}\ }\textbf {\bibinfo {volume} {88}},\ \bibinfo
  {pages} {085433} (\bibinfo {year} {2013})}\BibitemShut {NoStop}%
\bibitem [{\citenamefont {Jones}\ \emph {et~al.}(2013)\citenamefont {Jones},
  \citenamefont {Yu}, \citenamefont {Ghimire}, \citenamefont {Wu},
  \citenamefont {Aivazian}, \citenamefont {Ross}, \citenamefont {Zhao},
  \citenamefont {Yan}, \citenamefont {Mandrus}, \citenamefont {Xiao},
  \citenamefont {Yao},\ and\ \citenamefont {Xu}}]{Jones2013}%
  \BibitemOpen
  \bibfield  {author} {\bibinfo {author} {\bibfnamefont {A.~M.}\ \bibnamefont
  {Jones}}, \bibinfo {author} {\bibfnamefont {H.}~\bibnamefont {Yu}}, \bibinfo
  {author} {\bibfnamefont {N.~J.}\ \bibnamefont {Ghimire}}, \bibinfo {author}
  {\bibfnamefont {S.}~\bibnamefont {Wu}}, \bibinfo {author} {\bibfnamefont
  {G.}~\bibnamefont {Aivazian}}, \bibinfo {author} {\bibfnamefont {J.~S.}\
  \bibnamefont {Ross}}, \bibinfo {author} {\bibfnamefont {B.}~\bibnamefont
  {Zhao}}, \bibinfo {author} {\bibfnamefont {J.}~\bibnamefont {Yan}}, \bibinfo
  {author} {\bibfnamefont {D.~G.}\ \bibnamefont {Mandrus}}, \bibinfo {author}
  {\bibfnamefont {D.}~\bibnamefont {Xiao}}, \bibinfo {author} {\bibfnamefont
  {W.}~\bibnamefont {Yao}},\ and\ \bibinfo {author} {\bibfnamefont
  {X.}~\bibnamefont {Xu}},\ }\href {https://doi.org/10.1038/nnano.2013.151}
  {\bibfield  {journal} {\bibinfo  {journal} {Nature Nanotech}\ }\textbf
  {\bibinfo {volume} {8}},\ \bibinfo {pages} {634} (\bibinfo {year}
  {2013})}\BibitemShut {NoStop}%
\bibitem [{\citenamefont {Qiu}\ \emph {et~al.}(2013)\citenamefont {Qiu},
  \citenamefont {da~Jornada},\ and\ \citenamefont {Louie}}]{Qiu2013}%
  \BibitemOpen
  \bibfield  {author} {\bibinfo {author} {\bibfnamefont {D.~Y.}\ \bibnamefont
  {Qiu}}, \bibinfo {author} {\bibfnamefont {F.~H.}\ \bibnamefont
  {da~Jornada}},\ and\ \bibinfo {author} {\bibfnamefont {S.~G.}\ \bibnamefont
  {Louie}},\ }\href {https://doi.org/10.1103/physrevlett.111.216805} {\bibfield
   {journal} {\bibinfo  {journal} {Physical Review Letters}\ }\textbf {\bibinfo
  {volume} {111}},\ \bibinfo {pages} {216805} (\bibinfo {year}
  {2013})}\BibitemShut {NoStop}%
\bibitem [{\citenamefont {Xu}\ \emph {et~al.}(2014)\citenamefont {Xu},
  \citenamefont {Yao}, \citenamefont {Xiao},\ and\ \citenamefont
  {Heinz}}]{Xu2014}%
  \BibitemOpen
  \bibfield  {author} {\bibinfo {author} {\bibfnamefont {X.}~\bibnamefont
  {Xu}}, \bibinfo {author} {\bibfnamefont {W.}~\bibnamefont {Yao}}, \bibinfo
  {author} {\bibfnamefont {D.}~\bibnamefont {Xiao}},\ and\ \bibinfo {author}
  {\bibfnamefont {T.~F.}\ \bibnamefont {Heinz}},\ }\href
  {https://doi.org/10.1038/nphys2942} {\bibfield  {journal} {\bibinfo
  {journal} {Nat. Phys.}\ }\textbf {\bibinfo {volume} {10}},\ \bibinfo {pages}
  {343} (\bibinfo {year} {2014})}\BibitemShut {NoStop}%
\bibitem [{\citenamefont {Glazov}\ \emph {et~al.}(2015)\citenamefont {Glazov},
  \citenamefont {Ivchenko}, \citenamefont {Wang}, \citenamefont {Amand},
  \citenamefont {Marie}, \citenamefont {Urbaszek},\ and\ \citenamefont
  {Liu}}]{Glazov2015}%
  \BibitemOpen
  \bibfield  {author} {\bibinfo {author} {\bibfnamefont {M.~M.}\ \bibnamefont
  {Glazov}}, \bibinfo {author} {\bibfnamefont {E.~L.}\ \bibnamefont
  {Ivchenko}}, \bibinfo {author} {\bibfnamefont {G.}~\bibnamefont {Wang}},
  \bibinfo {author} {\bibfnamefont {T.}~\bibnamefont {Amand}}, \bibinfo
  {author} {\bibfnamefont {X.}~\bibnamefont {Marie}}, \bibinfo {author}
  {\bibfnamefont {B.}~\bibnamefont {Urbaszek}},\ and\ \bibinfo {author}
  {\bibfnamefont {B.~L.}\ \bibnamefont {Liu}},\ }\href
  {https://doi.org/10.1002/pssb.201552211} {\bibfield  {journal} {\bibinfo
  {journal} {physica status solidi (b)}\ }\textbf {\bibinfo {volume} {252}},\
  \bibinfo {pages} {2349} (\bibinfo {year} {2015})}\BibitemShut {NoStop}%
\bibitem [{\citenamefont {Koperski}\ \emph {et~al.}(2017)\citenamefont
  {Koperski}, \citenamefont {Molas}, \citenamefont {Arora}, \citenamefont
  {Nogajewski}, \citenamefont {Slobodeniuk}, \citenamefont {Faugeras},\ and\
  \citenamefont {Potemski}}]{Koperski2017}%
  \BibitemOpen
  \bibfield  {author} {\bibinfo {author} {\bibfnamefont {M.}~\bibnamefont
  {Koperski}}, \bibinfo {author} {\bibfnamefont {M.~R.}\ \bibnamefont {Molas}},
  \bibinfo {author} {\bibfnamefont {A.}~\bibnamefont {Arora}}, \bibinfo
  {author} {\bibfnamefont {K.}~\bibnamefont {Nogajewski}}, \bibinfo {author}
  {\bibfnamefont {A.~O.}\ \bibnamefont {Slobodeniuk}}, \bibinfo {author}
  {\bibfnamefont {C.}~\bibnamefont {Faugeras}},\ and\ \bibinfo {author}
  {\bibfnamefont {M.}~\bibnamefont {Potemski}},\ }\href
  {https://doi.org/10.1515/nanoph-2016-0165} {\bibfield  {journal} {\bibinfo
  {journal} {Nanophotonics}\ }\textbf {\bibinfo {volume} {6}},\ \bibinfo
  {pages} {1289} (\bibinfo {year} {2017})}\BibitemShut {NoStop}%
\bibitem [{\citenamefont {Li}\ \emph {et~al.}(2019)\citenamefont {Li},
  \citenamefont {Fong}, \citenamefont {Zhu}, \citenamefont {Li}, \citenamefont
  {Wang}, \citenamefont {Yang}, \citenamefont {Wang},\ and\ \citenamefont
  {Zhang}}]{Li2019b}%
  \BibitemOpen
  \bibfield  {author} {\bibinfo {author} {\bibfnamefont {H.-K.}\ \bibnamefont
  {Li}}, \bibinfo {author} {\bibfnamefont {K.~Y.}\ \bibnamefont {Fong}},
  \bibinfo {author} {\bibfnamefont {H.}~\bibnamefont {Zhu}}, \bibinfo {author}
  {\bibfnamefont {Q.}~\bibnamefont {Li}}, \bibinfo {author} {\bibfnamefont
  {S.}~\bibnamefont {Wang}}, \bibinfo {author} {\bibfnamefont {S.}~\bibnamefont
  {Yang}}, \bibinfo {author} {\bibfnamefont {Y.}~\bibnamefont {Wang}},\ and\
  \bibinfo {author} {\bibfnamefont {X.}~\bibnamefont {Zhang}},\ }\href
  {https://doi.org/10.1038/s41566-019-0428-0} {\bibfield  {journal} {\bibinfo
  {journal} {Nat. Photonics}\ }\textbf {\bibinfo {volume} {13}},\ \bibinfo
  {pages} {397} (\bibinfo {year} {2019})}\BibitemShut {NoStop}%
\bibitem [{\citenamefont {Jariwala}\ \emph {et~al.}(2014)\citenamefont
  {Jariwala}, \citenamefont {Sangwan}, \citenamefont {Lauhon}, \citenamefont
  {Marks},\ and\ \citenamefont {Hersam}}]{Jariwala2014}%
  \BibitemOpen
  \bibfield  {author} {\bibinfo {author} {\bibfnamefont {D.}~\bibnamefont
  {Jariwala}}, \bibinfo {author} {\bibfnamefont {V.~K.}\ \bibnamefont
  {Sangwan}}, \bibinfo {author} {\bibfnamefont {L.~J.}\ \bibnamefont {Lauhon}},
  \bibinfo {author} {\bibfnamefont {T.~J.}\ \bibnamefont {Marks}},\ and\
  \bibinfo {author} {\bibfnamefont {M.~C.}\ \bibnamefont {Hersam}},\ }\href
  {https://doi.org/10.1021/nn500064s} {\bibfield  {journal} {\bibinfo
  {journal} {{ACS} Nano}\ }\textbf {\bibinfo {volume} {8}},\ \bibinfo {pages}
  {1102} (\bibinfo {year} {2014})}\BibitemShut {NoStop}%
\bibitem [{\citenamefont {Schaibley}\ \emph {et~al.}(2016)\citenamefont
  {Schaibley}, \citenamefont {Yu}, \citenamefont {Clark}, \citenamefont
  {Rivera}, \citenamefont {Ross}, \citenamefont {Seyler}, \citenamefont {Yao},\
  and\ \citenamefont {Xu}}]{Schaibley2016}%
  \BibitemOpen
  \bibfield  {author} {\bibinfo {author} {\bibfnamefont {J.~R.}\ \bibnamefont
  {Schaibley}}, \bibinfo {author} {\bibfnamefont {H.}~\bibnamefont {Yu}},
  \bibinfo {author} {\bibfnamefont {G.}~\bibnamefont {Clark}}, \bibinfo
  {author} {\bibfnamefont {P.}~\bibnamefont {Rivera}}, \bibinfo {author}
  {\bibfnamefont {J.~S.}\ \bibnamefont {Ross}}, \bibinfo {author}
  {\bibfnamefont {K.~L.}\ \bibnamefont {Seyler}}, \bibinfo {author}
  {\bibfnamefont {W.}~\bibnamefont {Yao}},\ and\ \bibinfo {author}
  {\bibfnamefont {X.}~\bibnamefont {Xu}},\ }\bibfield  {journal} {\bibinfo
  {journal} {Nature Reviews Materials}\ }\textbf {\bibinfo {volume} {1}},\
  \href {https://doi.org/10.1038/natrevmats.2016.55}
  {10.1038/natrevmats.2016.55} (\bibinfo {year} {2016})\BibitemShut {NoStop}%
\bibitem [{\citenamefont {Mak}\ and\ \citenamefont {Shan}(2016)}]{Mak2016}%
  \BibitemOpen
  \bibfield  {author} {\bibinfo {author} {\bibfnamefont {K.~F.}\ \bibnamefont
  {Mak}}\ and\ \bibinfo {author} {\bibfnamefont {J.}~\bibnamefont {Shan}},\
  }\href {https://doi.org/10.1038/nphoton.2015.282} {\bibfield  {journal}
  {\bibinfo  {journal} {Nat. Photonics}\ }\textbf {\bibinfo {volume} {10}},\
  \bibinfo {pages} {216} (\bibinfo {year} {2016})}\BibitemShut {NoStop}%
\bibitem [{\citenamefont {Xiao}\ \emph {et~al.}(2017)\citenamefont {Xiao},
  \citenamefont {Zhao}, \citenamefont {Wang},\ and\ \citenamefont
  {Zhang}}]{Xiao2017}%
  \BibitemOpen
  \bibfield  {author} {\bibinfo {author} {\bibfnamefont {J.}~\bibnamefont
  {Xiao}}, \bibinfo {author} {\bibfnamefont {M.}~\bibnamefont {Zhao}}, \bibinfo
  {author} {\bibfnamefont {Y.}~\bibnamefont {Wang}},\ and\ \bibinfo {author}
  {\bibfnamefont {X.}~\bibnamefont {Zhang}},\ }\href
  {https://doi.org/10.1515/nanoph-2016-0160} {\bibfield  {journal} {\bibinfo
  {journal} {Nanophotonics}\ }\textbf {\bibinfo {volume} {6}},\ \bibinfo
  {pages} {1309} (\bibinfo {year} {2017})}\BibitemShut {NoStop}%
\bibitem [{\citenamefont {Mueller}\ and\ \citenamefont
  {Malic}(2018)}]{Mueller2018}%
  \BibitemOpen
  \bibfield  {author} {\bibinfo {author} {\bibfnamefont {T.}~\bibnamefont
  {Mueller}}\ and\ \bibinfo {author} {\bibfnamefont {E.}~\bibnamefont
  {Malic}},\ }\href {https://doi.org/10.1038/s41699-018-0074-2} {\bibfield
  {journal} {\bibinfo  {journal} {NPJ 2D Mater. Appl.}\ }\textbf {\bibinfo
  {volume} {2}},\ \bibinfo {pages} {29} (\bibinfo {year} {2018})}\BibitemShut
  {NoStop}%
\bibitem [{\citenamefont {Vitale}\ \emph {et~al.}(2018)\citenamefont {Vitale},
  \citenamefont {Nezich}, \citenamefont {Varghese}, \citenamefont {Kim},
  \citenamefont {Gedik}, \citenamefont {Jarillo-Herrero}, \citenamefont
  {Xiao},\ and\ \citenamefont {Rothschild}}]{Vitale2018}%
  \BibitemOpen
  \bibfield  {author} {\bibinfo {author} {\bibfnamefont {S.~A.}\ \bibnamefont
  {Vitale}}, \bibinfo {author} {\bibfnamefont {D.}~\bibnamefont {Nezich}},
  \bibinfo {author} {\bibfnamefont {J.~O.}\ \bibnamefont {Varghese}}, \bibinfo
  {author} {\bibfnamefont {P.}~\bibnamefont {Kim}}, \bibinfo {author}
  {\bibfnamefont {N.}~\bibnamefont {Gedik}}, \bibinfo {author} {\bibfnamefont
  {P.}~\bibnamefont {Jarillo-Herrero}}, \bibinfo {author} {\bibfnamefont
  {D.}~\bibnamefont {Xiao}},\ and\ \bibinfo {author} {\bibfnamefont
  {M.}~\bibnamefont {Rothschild}},\ }\href
  {https://doi.org/10.1002/smll.201801483} {\bibfield  {journal} {\bibinfo
  {journal} {Small}\ }\textbf {\bibinfo {volume} {14}},\ \bibinfo {pages}
  {1801483} (\bibinfo {year} {2018})}\BibitemShut {NoStop}%
\bibitem [{\citenamefont {Chernikov}\ \emph {et~al.}(2014)\citenamefont
  {Chernikov}, \citenamefont {Berkelbach}, \citenamefont {Hill}, \citenamefont
  {Rigosi}, \citenamefont {Li}, \citenamefont {Aslan}, \citenamefont
  {Reichman}, \citenamefont {Hybertsen},\ and\ \citenamefont
  {Heinz}}]{Chernikov2014}%
  \BibitemOpen
  \bibfield  {author} {\bibinfo {author} {\bibfnamefont {A.}~\bibnamefont
  {Chernikov}}, \bibinfo {author} {\bibfnamefont {T.~C.}\ \bibnamefont
  {Berkelbach}}, \bibinfo {author} {\bibfnamefont {H.~M.}\ \bibnamefont
  {Hill}}, \bibinfo {author} {\bibfnamefont {A.}~\bibnamefont {Rigosi}},
  \bibinfo {author} {\bibfnamefont {Y.}~\bibnamefont {Li}}, \bibinfo {author}
  {\bibfnamefont {O.~B.}\ \bibnamefont {Aslan}}, \bibinfo {author}
  {\bibfnamefont {D.~R.}\ \bibnamefont {Reichman}}, \bibinfo {author}
  {\bibfnamefont {M.~S.}\ \bibnamefont {Hybertsen}},\ and\ \bibinfo {author}
  {\bibfnamefont {T.~F.}\ \bibnamefont {Heinz}},\ }\href
  {https://doi.org/10.1103/physrevlett.113.076802} {\bibfield  {journal}
  {\bibinfo  {journal} {Physical Review Letters}\ }\textbf {\bibinfo {volume}
  {113}},\ \bibinfo {pages} {076802} (\bibinfo {year} {2014})}\BibitemShut
  {NoStop}%
\bibitem [{\citenamefont {Hill}\ \emph {et~al.}(2015)\citenamefont {Hill},
  \citenamefont {Rigosi}, \citenamefont {Roquelet}, \citenamefont {Chernikov},
  \citenamefont {Berkelbach}, \citenamefont {Reichman}, \citenamefont
  {Hybertsen}, \citenamefont {Brus},\ and\ \citenamefont {Heinz}}]{Hill2015}%
  \BibitemOpen
  \bibfield  {author} {\bibinfo {author} {\bibfnamefont {H.~M.}\ \bibnamefont
  {Hill}}, \bibinfo {author} {\bibfnamefont {A.~F.}\ \bibnamefont {Rigosi}},
  \bibinfo {author} {\bibfnamefont {C.}~\bibnamefont {Roquelet}}, \bibinfo
  {author} {\bibfnamefont {A.}~\bibnamefont {Chernikov}}, \bibinfo {author}
  {\bibfnamefont {T.~C.}\ \bibnamefont {Berkelbach}}, \bibinfo {author}
  {\bibfnamefont {D.~R.}\ \bibnamefont {Reichman}}, \bibinfo {author}
  {\bibfnamefont {M.~S.}\ \bibnamefont {Hybertsen}}, \bibinfo {author}
  {\bibfnamefont {L.~E.}\ \bibnamefont {Brus}},\ and\ \bibinfo {author}
  {\bibfnamefont {T.~F.}\ \bibnamefont {Heinz}},\ }\href
  {https://doi.org/10.1021/nl504868p} {\bibfield  {journal} {\bibinfo
  {journal} {Nano Letters}\ }\textbf {\bibinfo {volume} {15}},\ \bibinfo
  {pages} {2992} (\bibinfo {year} {2015})}\BibitemShut {NoStop}%
\bibitem [{\citenamefont {Pikus}\ and\ \citenamefont {Bir}(1971)}]{Pikus1971}%
  \BibitemOpen
  \bibfield  {author} {\bibinfo {author} {\bibfnamefont {G.~E.}\ \bibnamefont
  {Pikus}}\ and\ \bibinfo {author} {\bibfnamefont {G.~L.}\ \bibnamefont
  {Bir}},\ }\href@noop {} {\bibfield  {journal} {\bibinfo  {journal} {Zh. Eksp.
  Teor. Fiz.}\ }\textbf {\bibinfo {volume} {60}},\ \bibinfo {pages} {195}
  (\bibinfo {year} {1971})}\BibitemShut {NoStop}%
\bibitem [{\citenamefont {Yu}\ \emph {et~al.}(2014)\citenamefont {Yu},
  \citenamefont {Liu}, \citenamefont {Gong}, \citenamefont {Xu},\ and\
  \citenamefont {Yao}}]{Yu2014a}%
  \BibitemOpen
  \bibfield  {author} {\bibinfo {author} {\bibfnamefont {H.}~\bibnamefont
  {Yu}}, \bibinfo {author} {\bibfnamefont {G.-B.}\ \bibnamefont {Liu}},
  \bibinfo {author} {\bibfnamefont {P.}~\bibnamefont {Gong}}, \bibinfo {author}
  {\bibfnamefont {X.}~\bibnamefont {Xu}},\ and\ \bibinfo {author}
  {\bibfnamefont {W.}~\bibnamefont {Yao}},\ }\bibfield  {journal} {\bibinfo
  {journal} {Nature Communications}\ }\textbf {\bibinfo {volume} {5}},\ \href
  {https://doi.org/10.1038/ncomms4876} {10.1038/ncomms4876} (\bibinfo {year}
  {2014})\BibitemShut {NoStop}%
\bibitem [{\citenamefont {Glazov}\ \emph {et~al.}(2014)\citenamefont {Glazov},
  \citenamefont {Amand}, \citenamefont {Marie}, \citenamefont {Lagarde},
  \citenamefont {Bouet},\ and\ \citenamefont {Urbaszek}}]{Glazov2014}%
  \BibitemOpen
  \bibfield  {author} {\bibinfo {author} {\bibfnamefont {M.~M.}\ \bibnamefont
  {Glazov}}, \bibinfo {author} {\bibfnamefont {T.}~\bibnamefont {Amand}},
  \bibinfo {author} {\bibfnamefont {X.}~\bibnamefont {Marie}}, \bibinfo
  {author} {\bibfnamefont {D.}~\bibnamefont {Lagarde}}, \bibinfo {author}
  {\bibfnamefont {L.}~\bibnamefont {Bouet}},\ and\ \bibinfo {author}
  {\bibfnamefont {B.}~\bibnamefont {Urbaszek}},\ }\href
  {https://doi.org/10.1103/physrevb.89.201302} {\bibfield  {journal} {\bibinfo
  {journal} {Physical Review B}\ }\textbf {\bibinfo {volume} {89}},\ \bibinfo
  {pages} {201302} (\bibinfo {year} {2014})}\BibitemShut {NoStop}%
\bibitem [{\citenamefont {Qiu}\ \emph {et~al.}(2015)\citenamefont {Qiu},
  \citenamefont {Cao},\ and\ \citenamefont {Louie}}]{Qiu2015}%
  \BibitemOpen
  \bibfield  {author} {\bibinfo {author} {\bibfnamefont {D.~Y.}\ \bibnamefont
  {Qiu}}, \bibinfo {author} {\bibfnamefont {T.}~\bibnamefont {Cao}},\ and\
  \bibinfo {author} {\bibfnamefont {S.~G.}\ \bibnamefont {Louie}},\ }\href
  {https://doi.org/10.1103/physrevlett.115.176801} {\bibfield  {journal}
  {\bibinfo  {journal} {Physical Review Letters}\ }\textbf {\bibinfo {volume}
  {115}},\ \bibinfo {pages} {176801} (\bibinfo {year} {2015})}\BibitemShut
  {NoStop}%
\bibitem [{\citenamefont {Zhu}\ \emph {et~al.}(2014{\natexlab{a}})\citenamefont
  {Zhu}, \citenamefont {Zhang}, \citenamefont {Glazov}, \citenamefont
  {Urbaszek}, \citenamefont {Amand}, \citenamefont {Ji}, \citenamefont {Liu},\
  and\ \citenamefont {Marie}}]{Zhu2014a}%
  \BibitemOpen
  \bibfield  {author} {\bibinfo {author} {\bibfnamefont {C.~R.}\ \bibnamefont
  {Zhu}}, \bibinfo {author} {\bibfnamefont {K.}~\bibnamefont {Zhang}}, \bibinfo
  {author} {\bibfnamefont {M.}~\bibnamefont {Glazov}}, \bibinfo {author}
  {\bibfnamefont {B.}~\bibnamefont {Urbaszek}}, \bibinfo {author}
  {\bibfnamefont {T.}~\bibnamefont {Amand}}, \bibinfo {author} {\bibfnamefont
  {Z.~W.}\ \bibnamefont {Ji}}, \bibinfo {author} {\bibfnamefont {B.~L.}\
  \bibnamefont {Liu}},\ and\ \bibinfo {author} {\bibfnamefont {X.}~\bibnamefont
  {Marie}},\ }\href {https://doi.org/10.1103/physrevb.90.161302} {\bibfield
  {journal} {\bibinfo  {journal} {Physical Review B}\ }\textbf {\bibinfo
  {volume} {90}},\ \bibinfo {pages} {161302} (\bibinfo {year}
  {2014}{\natexlab{a}})}\BibitemShut {NoStop}%
\bibitem [{\citenamefont {Hao}\ \emph {et~al.}(2016)\citenamefont {Hao},
  \citenamefont {Moody}, \citenamefont {Wu}, \citenamefont {Dass},
  \citenamefont {Xu}, \citenamefont {Chen}, \citenamefont {Sun}, \citenamefont
  {Li}, \citenamefont {Li}, \citenamefont {MacDonald},\ and\ \citenamefont
  {Li}}]{Hao2016}%
  \BibitemOpen
  \bibfield  {author} {\bibinfo {author} {\bibfnamefont {K.}~\bibnamefont
  {Hao}}, \bibinfo {author} {\bibfnamefont {G.}~\bibnamefont {Moody}}, \bibinfo
  {author} {\bibfnamefont {F.}~\bibnamefont {Wu}}, \bibinfo {author}
  {\bibfnamefont {C.~K.}\ \bibnamefont {Dass}}, \bibinfo {author}
  {\bibfnamefont {L.}~\bibnamefont {Xu}}, \bibinfo {author} {\bibfnamefont
  {C.-H.}\ \bibnamefont {Chen}}, \bibinfo {author} {\bibfnamefont
  {L.}~\bibnamefont {Sun}}, \bibinfo {author} {\bibfnamefont {M.-Y.}\
  \bibnamefont {Li}}, \bibinfo {author} {\bibfnamefont {L.-J.}\ \bibnamefont
  {Li}}, \bibinfo {author} {\bibfnamefont {A.~H.}\ \bibnamefont {MacDonald}},\
  and\ \bibinfo {author} {\bibfnamefont {X.}~\bibnamefont {Li}},\ }\href
  {https://doi.org/10.1038/nphys3674} {\bibfield  {journal} {\bibinfo
  {journal} {Nature Physics}\ }\textbf {\bibinfo {volume} {12}},\ \bibinfo
  {pages} {677} (\bibinfo {year} {2016})}\BibitemShut {NoStop}%
\bibitem [{\citenamefont {Glazov}\ \emph {et~al.}(2017)\citenamefont {Glazov},
  \citenamefont {Golub}, \citenamefont {Wang}, \citenamefont {Marie},
  \citenamefont {Amand},\ and\ \citenamefont {Urbaszek}}]{Glazov2017}%
  \BibitemOpen
  \bibfield  {author} {\bibinfo {author} {\bibfnamefont {M.~M.}\ \bibnamefont
  {Glazov}}, \bibinfo {author} {\bibfnamefont {L.~E.}\ \bibnamefont {Golub}},
  \bibinfo {author} {\bibfnamefont {G.}~\bibnamefont {Wang}}, \bibinfo {author}
  {\bibfnamefont {X.}~\bibnamefont {Marie}}, \bibinfo {author} {\bibfnamefont
  {T.}~\bibnamefont {Amand}},\ and\ \bibinfo {author} {\bibfnamefont
  {B.}~\bibnamefont {Urbaszek}},\ }\href
  {https://doi.org/10.1103/physrevb.95.035311} {\bibfield  {journal} {\bibinfo
  {journal} {Physical Review B}\ }\textbf {\bibinfo {volume} {95}},\ \bibinfo
  {pages} {035311} (\bibinfo {year} {2017})}\BibitemShut {NoStop}%
\bibitem [{\citenamefont {Guo}\ \emph {et~al.}(2018)\citenamefont {Guo},
  \citenamefont {Wu}, \citenamefont {Cao}, \citenamefont {Monahan},
  \citenamefont {Lee}, \citenamefont {Louie},\ and\ \citenamefont
  {Fleming}}]{Guo2018}%
  \BibitemOpen
  \bibfield  {author} {\bibinfo {author} {\bibfnamefont {L.}~\bibnamefont
  {Guo}}, \bibinfo {author} {\bibfnamefont {M.}~\bibnamefont {Wu}}, \bibinfo
  {author} {\bibfnamefont {T.}~\bibnamefont {Cao}}, \bibinfo {author}
  {\bibfnamefont {D.~M.}\ \bibnamefont {Monahan}}, \bibinfo {author}
  {\bibfnamefont {Y.-H.}\ \bibnamefont {Lee}}, \bibinfo {author} {\bibfnamefont
  {S.~G.}\ \bibnamefont {Louie}},\ and\ \bibinfo {author} {\bibfnamefont
  {G.~R.}\ \bibnamefont {Fleming}},\ }\href
  {https://doi.org/10.1038/s41567-018-0362-y} {\bibfield  {journal} {\bibinfo
  {journal} {Nature Physics}\ }\textbf {\bibinfo {volume} {15}},\ \bibinfo
  {pages} {228} (\bibinfo {year} {2018})}\BibitemShut {NoStop}%
\bibitem [{\citenamefont {Yu}\ and\ \citenamefont {Wu}(2014)}]{Yu2014}%
  \BibitemOpen
  \bibfield  {author} {\bibinfo {author} {\bibfnamefont {T.}~\bibnamefont
  {Yu}}\ and\ \bibinfo {author} {\bibfnamefont {M.~W.}\ \bibnamefont {Wu}},\
  }\href {https://doi.org/10.1103/physrevb.89.205303} {\bibfield  {journal}
  {\bibinfo  {journal} {Physical Review B}\ }\textbf {\bibinfo {volume} {89}},\
  \bibinfo {pages} {205303} (\bibinfo {year} {2014})}\BibitemShut {NoStop}%
\bibitem [{\citenamefont {Korm{\'{a}}nyos}\ \emph {et~al.}(2015)\citenamefont
  {Korm{\'{a}}nyos}, \citenamefont {Burkard}, \citenamefont {Gmitra},
  \citenamefont {Fabian}, \citenamefont {Z{\'{o}}lyomi}, \citenamefont
  {Drummond},\ and\ \citenamefont {Falko}}]{Kormanyos2015}%
  \BibitemOpen
  \bibfield  {author} {\bibinfo {author} {\bibfnamefont {A.}~\bibnamefont
  {Korm{\'{a}}nyos}}, \bibinfo {author} {\bibfnamefont {G.}~\bibnamefont
  {Burkard}}, \bibinfo {author} {\bibfnamefont {M.}~\bibnamefont {Gmitra}},
  \bibinfo {author} {\bibfnamefont {J.}~\bibnamefont {Fabian}}, \bibinfo
  {author} {\bibfnamefont {V.}~\bibnamefont {Z{\'{o}}lyomi}}, \bibinfo {author}
  {\bibfnamefont {N.~D.}\ \bibnamefont {Drummond}},\ and\ \bibinfo {author}
  {\bibfnamefont {V.}~\bibnamefont {Falko}},\ }\href
  {https://doi.org/10.1088/2053-1583/2/2/022001} {\bibfield  {journal}
  {\bibinfo  {journal} {2D Materials}\ }\textbf {\bibinfo {volume} {2}},\
  \bibinfo {pages} {022001} (\bibinfo {year} {2015})}\BibitemShut {NoStop}%
\bibitem [{\citenamefont {Moody}\ \emph {et~al.}(2016)\citenamefont {Moody},
  \citenamefont {Schaibley},\ and\ \citenamefont {Xu}}]{Moody2016}%
  \BibitemOpen
  \bibfield  {author} {\bibinfo {author} {\bibfnamefont {G.}~\bibnamefont
  {Moody}}, \bibinfo {author} {\bibfnamefont {J.}~\bibnamefont {Schaibley}},\
  and\ \bibinfo {author} {\bibfnamefont {X.}~\bibnamefont {Xu}},\ }\href
  {https://doi.org/10.1364/josab.33.000c39} {\bibfield  {journal} {\bibinfo
  {journal} {J. Opt. Soc. Am. B}\ }\textbf {\bibinfo {volume} {33}},\ \bibinfo
  {pages} {C39} (\bibinfo {year} {2016})}\BibitemShut {NoStop}%
\bibitem [{\citenamefont {Wang}\ \emph
  {et~al.}(2018{\natexlab{b}})\citenamefont {Wang}, \citenamefont
  {Molina-S{\'{a}}nchez}, \citenamefont {Altmann}, \citenamefont {Sangalli},
  \citenamefont {Fazio}, \citenamefont {Soavi}, \citenamefont {Sassi},
  \citenamefont {Bottegoni}, \citenamefont {Ciccacci}, \citenamefont {Finazzi},
  \citenamefont {Wirtz}, \citenamefont {Ferrari}, \citenamefont {Marini},
  \citenamefont {Cerullo},\ and\ \citenamefont {Conte}}]{Wang2018a}%
  \BibitemOpen
  \bibfield  {author} {\bibinfo {author} {\bibfnamefont {Z.}~\bibnamefont
  {Wang}}, \bibinfo {author} {\bibfnamefont {A.}~\bibnamefont
  {Molina-S{\'{a}}nchez}}, \bibinfo {author} {\bibfnamefont {P.}~\bibnamefont
  {Altmann}}, \bibinfo {author} {\bibfnamefont {D.}~\bibnamefont {Sangalli}},
  \bibinfo {author} {\bibfnamefont {D.~D.}\ \bibnamefont {Fazio}}, \bibinfo
  {author} {\bibfnamefont {G.}~\bibnamefont {Soavi}}, \bibinfo {author}
  {\bibfnamefont {U.}~\bibnamefont {Sassi}}, \bibinfo {author} {\bibfnamefont
  {F.}~\bibnamefont {Bottegoni}}, \bibinfo {author} {\bibfnamefont
  {F.}~\bibnamefont {Ciccacci}}, \bibinfo {author} {\bibfnamefont
  {M.}~\bibnamefont {Finazzi}}, \bibinfo {author} {\bibfnamefont
  {L.}~\bibnamefont {Wirtz}}, \bibinfo {author} {\bibfnamefont {A.~C.}\
  \bibnamefont {Ferrari}}, \bibinfo {author} {\bibfnamefont {A.}~\bibnamefont
  {Marini}}, \bibinfo {author} {\bibfnamefont {G.}~\bibnamefont {Cerullo}},\
  and\ \bibinfo {author} {\bibfnamefont {S.~D.}\ \bibnamefont {Conte}},\ }\href
  {https://doi.org/10.1021/acs.nanolett.8b02774} {\bibfield  {journal}
  {\bibinfo  {journal} {Nano Letters}\ }\textbf {\bibinfo {volume} {18}},\
  \bibinfo {pages} {6882} (\bibinfo {year} {2018}{\natexlab{b}})}\BibitemShut
  {NoStop}%
\bibitem [{\citenamefont {Selig}\ \emph {et~al.}(2018)\citenamefont {Selig},
  \citenamefont {Bergh{\"a}user}, \citenamefont {Richter}, \citenamefont
  {Bratschitsch}, \citenamefont {Knorr},\ and\ \citenamefont
  {Malic}}]{Selig2018}%
  \BibitemOpen
  \bibfield  {author} {\bibinfo {author} {\bibfnamefont {M.}~\bibnamefont
  {Selig}}, \bibinfo {author} {\bibfnamefont {G.}~\bibnamefont
  {Bergh{\"a}user}}, \bibinfo {author} {\bibfnamefont {M.}~\bibnamefont
  {Richter}}, \bibinfo {author} {\bibfnamefont {R.}~\bibnamefont
  {Bratschitsch}}, \bibinfo {author} {\bibfnamefont {A.}~\bibnamefont
  {Knorr}},\ and\ \bibinfo {author} {\bibfnamefont {E.}~\bibnamefont {Malic}},\
  }\href {https://doi.org/10.1088/2053-1583/aabea3} {\bibfield  {journal}
  {\bibinfo  {journal} {2D Materials}\ }\textbf {\bibinfo {volume} {5}},\
  \bibinfo {pages} {035017} (\bibinfo {year} {2018})}\BibitemShut {NoStop}%
\bibitem [{\citenamefont {He}\ \emph {et~al.}(2020)\citenamefont {He},
  \citenamefont {Rivera}, \citenamefont {Tuan}, \citenamefont {Wilson},
  \citenamefont {Yang}, \citenamefont {Taniguchi}, \citenamefont {Watanabe},
  \citenamefont {Yan}, \citenamefont {Mandrus}, \citenamefont {Yu},
  \citenamefont {Dery}, \citenamefont {Yao},\ and\ \citenamefont
  {Xu}}]{He2020}%
  \BibitemOpen
  \bibfield  {author} {\bibinfo {author} {\bibfnamefont {M.}~\bibnamefont
  {He}}, \bibinfo {author} {\bibfnamefont {P.}~\bibnamefont {Rivera}}, \bibinfo
  {author} {\bibfnamefont {D.~V.}\ \bibnamefont {Tuan}}, \bibinfo {author}
  {\bibfnamefont {N.~P.}\ \bibnamefont {Wilson}}, \bibinfo {author}
  {\bibfnamefont {M.}~\bibnamefont {Yang}}, \bibinfo {author} {\bibfnamefont
  {T.}~\bibnamefont {Taniguchi}}, \bibinfo {author} {\bibfnamefont
  {K.}~\bibnamefont {Watanabe}}, \bibinfo {author} {\bibfnamefont
  {J.}~\bibnamefont {Yan}}, \bibinfo {author} {\bibfnamefont {D.~G.}\
  \bibnamefont {Mandrus}}, \bibinfo {author} {\bibfnamefont {H.}~\bibnamefont
  {Yu}}, \bibinfo {author} {\bibfnamefont {H.}~\bibnamefont {Dery}}, \bibinfo
  {author} {\bibfnamefont {W.}~\bibnamefont {Yao}},\ and\ \bibinfo {author}
  {\bibfnamefont {X.}~\bibnamefont {Xu}},\ }\bibfield  {journal} {\bibinfo
  {journal} {Nature Communications}\ }\textbf {\bibinfo {volume} {11}},\ \href
  {https://doi.org/10.1038/s41467-020-14472-0} {10.1038/s41467-020-14472-0}
  (\bibinfo {year} {2020})\BibitemShut {NoStop}%
\bibitem [{\citenamefont {Jiang}\ \emph {et~al.}(2021)\citenamefont {Jiang},
  \citenamefont {Zheng}, \citenamefont {Lan}, \citenamefont {Saidi},
  \citenamefont {Ren},\ and\ \citenamefont {Zhao}}]{Jiang2021}%
  \BibitemOpen
  \bibfield  {author} {\bibinfo {author} {\bibfnamefont {X.}~\bibnamefont
  {Jiang}}, \bibinfo {author} {\bibfnamefont {Q.}~\bibnamefont {Zheng}},
  \bibinfo {author} {\bibfnamefont {Z.}~\bibnamefont {Lan}}, \bibinfo {author}
  {\bibfnamefont {W.~A.}\ \bibnamefont {Saidi}}, \bibinfo {author}
  {\bibfnamefont {X.}~\bibnamefont {Ren}},\ and\ \bibinfo {author}
  {\bibfnamefont {J.}~\bibnamefont {Zhao}},\ }\bibfield  {journal} {\bibinfo
  {journal} {Science Advances}\ }\textbf {\bibinfo {volume} {7}},\ \href
  {https://doi.org/10.1126/sciadv.abf3759} {10.1126/sciadv.abf3759} (\bibinfo
  {year} {2021})\BibitemShut {NoStop}%
\bibitem [{\citenamefont {Zhu}\ \emph {et~al.}(2014{\natexlab{b}})\citenamefont
  {Zhu}, \citenamefont {Zeng}, \citenamefont {Dai}, \citenamefont {Gong},\ and\
  \citenamefont {Cui}}]{Zhu2014}%
  \BibitemOpen
  \bibfield  {author} {\bibinfo {author} {\bibfnamefont {B.}~\bibnamefont
  {Zhu}}, \bibinfo {author} {\bibfnamefont {H.}~\bibnamefont {Zeng}}, \bibinfo
  {author} {\bibfnamefont {J.}~\bibnamefont {Dai}}, \bibinfo {author}
  {\bibfnamefont {Z.}~\bibnamefont {Gong}},\ and\ \bibinfo {author}
  {\bibfnamefont {X.}~\bibnamefont {Cui}},\ }\href
  {https://doi.org/10.1073/pnas.1406960111} {\bibfield  {journal} {\bibinfo
  {journal} {Proceedings of the National Academy of Sciences of the United
  States of America}\ }\textbf {\bibinfo {volume} {111}},\ \bibinfo {pages}
  {11606} (\bibinfo {year} {2014}{\natexlab{b}})}\BibitemShut {NoStop}%
\bibitem [{\citenamefont {Wang}\ \emph {et~al.}(2014)\citenamefont {Wang},
  \citenamefont {Bouet}, \citenamefont {Lagarde}, \citenamefont {Vidal},
  \citenamefont {Balocchi}, \citenamefont {Amand}, \citenamefont {Marie},\ and\
  \citenamefont {Urbaszek}}]{Wang2014}%
  \BibitemOpen
  \bibfield  {author} {\bibinfo {author} {\bibfnamefont {G.}~\bibnamefont
  {Wang}}, \bibinfo {author} {\bibfnamefont {L.}~\bibnamefont {Bouet}},
  \bibinfo {author} {\bibfnamefont {D.}~\bibnamefont {Lagarde}}, \bibinfo
  {author} {\bibfnamefont {M.}~\bibnamefont {Vidal}}, \bibinfo {author}
  {\bibfnamefont {A.}~\bibnamefont {Balocchi}}, \bibinfo {author}
  {\bibfnamefont {T.}~\bibnamefont {Amand}}, \bibinfo {author} {\bibfnamefont
  {X.}~\bibnamefont {Marie}},\ and\ \bibinfo {author} {\bibfnamefont
  {B.}~\bibnamefont {Urbaszek}},\ }\href
  {https://doi.org/10.1103/physrevb.90.075413} {\bibfield  {journal} {\bibinfo
  {journal} {Physical Review B}\ }\textbf {\bibinfo {volume} {90}},\ \bibinfo
  {pages} {075413} (\bibinfo {year} {2014})}\BibitemShut {NoStop}%
\bibitem [{\citenamefont {Lagarde}\ \emph {et~al.}(2014)\citenamefont
  {Lagarde}, \citenamefont {Bouet}, \citenamefont {Marie}, \citenamefont {Zhu},
  \citenamefont {Liu}, \citenamefont {Amand}, \citenamefont {Tan},\ and\
  \citenamefont {Urbaszek}}]{Lagarde2014}%
  \BibitemOpen
  \bibfield  {author} {\bibinfo {author} {\bibfnamefont {D.}~\bibnamefont
  {Lagarde}}, \bibinfo {author} {\bibfnamefont {L.}~\bibnamefont {Bouet}},
  \bibinfo {author} {\bibfnamefont {X.}~\bibnamefont {Marie}}, \bibinfo
  {author} {\bibfnamefont {C.~R.}\ \bibnamefont {Zhu}}, \bibinfo {author}
  {\bibfnamefont {B.~L.}\ \bibnamefont {Liu}}, \bibinfo {author} {\bibfnamefont
  {T.}~\bibnamefont {Amand}}, \bibinfo {author} {\bibfnamefont {P.~H.}\
  \bibnamefont {Tan}},\ and\ \bibinfo {author} {\bibfnamefont {B.}~\bibnamefont
  {Urbaszek}},\ }\href {https://doi.org/10.1103/physrevlett.112.047401}
  {\bibfield  {journal} {\bibinfo  {journal} {Physical Review Letters}\
  }\textbf {\bibinfo {volume} {112}},\ \bibinfo {pages} {047401} (\bibinfo
  {year} {2014})}\BibitemShut {NoStop}%
\bibitem [{\citenamefont {Yan}\ \emph {et~al.}(2015)\citenamefont {Yan},
  \citenamefont {Qiao}, \citenamefont {Tan},\ and\ \citenamefont
  {Zhang}}]{Yan2015}%
  \BibitemOpen
  \bibfield  {author} {\bibinfo {author} {\bibfnamefont {T.}~\bibnamefont
  {Yan}}, \bibinfo {author} {\bibfnamefont {X.}~\bibnamefont {Qiao}}, \bibinfo
  {author} {\bibfnamefont {P.}~\bibnamefont {Tan}},\ and\ \bibinfo {author}
  {\bibfnamefont {X.}~\bibnamefont {Zhang}},\ }\bibfield  {journal} {\bibinfo
  {journal} {Scientific Reports}\ }\textbf {\bibinfo {volume} {5}},\ \href
  {https://doi.org/10.1038/srep15625} {10.1038/srep15625} (\bibinfo {year}
  {2015})\BibitemShut {NoStop}%
\bibitem [{\citenamefont {Mai}\ \emph {et~al.}(2014)\citenamefont {Mai},
  \citenamefont {Semenov}, \citenamefont {Barrette}, \citenamefont {Yu},
  \citenamefont {Jin}, \citenamefont {Cao}, \citenamefont {Kim},\ and\
  \citenamefont {Gundogdu}}]{Mai2014}%
  \BibitemOpen
  \bibfield  {author} {\bibinfo {author} {\bibfnamefont {C.}~\bibnamefont
  {Mai}}, \bibinfo {author} {\bibfnamefont {Y.~G.}\ \bibnamefont {Semenov}},
  \bibinfo {author} {\bibfnamefont {A.}~\bibnamefont {Barrette}}, \bibinfo
  {author} {\bibfnamefont {Y.}~\bibnamefont {Yu}}, \bibinfo {author}
  {\bibfnamefont {Z.}~\bibnamefont {Jin}}, \bibinfo {author} {\bibfnamefont
  {L.}~\bibnamefont {Cao}}, \bibinfo {author} {\bibfnamefont {K.~W.}\
  \bibnamefont {Kim}},\ and\ \bibinfo {author} {\bibfnamefont {K.}~\bibnamefont
  {Gundogdu}},\ }\href {https://doi.org/10.1103/physrevb.90.041414} {\bibfield
  {journal} {\bibinfo  {journal} {Physical Review B}\ }\textbf {\bibinfo
  {volume} {90}},\ \bibinfo {pages} {041414} (\bibinfo {year}
  {2014})}\BibitemShut {NoStop}%
\bibitem [{\citenamefont {Schmidt}\ \emph {et~al.}(2016)\citenamefont
  {Schmidt}, \citenamefont {Bergh\"auser}, \citenamefont {Schneider},
  \citenamefont {Selig}, \citenamefont {Tonndorf}, \citenamefont {Mali{\'{c}}},
  \citenamefont {Knorr}, \citenamefont {de~Vasconcellos},\ and\ \citenamefont
  {Bratschitsch}}]{Schmidt2016}%
  \BibitemOpen
  \bibfield  {author} {\bibinfo {author} {\bibfnamefont {R.}~\bibnamefont
  {Schmidt}}, \bibinfo {author} {\bibfnamefont {G.}~\bibnamefont
  {Bergh\"auser}}, \bibinfo {author} {\bibfnamefont {R.}~\bibnamefont
  {Schneider}}, \bibinfo {author} {\bibfnamefont {M.}~\bibnamefont {Selig}},
  \bibinfo {author} {\bibfnamefont {P.}~\bibnamefont {Tonndorf}}, \bibinfo
  {author} {\bibfnamefont {E.}~\bibnamefont {Mali{\'{c}}}}, \bibinfo {author}
  {\bibfnamefont {A.}~\bibnamefont {Knorr}}, \bibinfo {author} {\bibfnamefont
  {S.~M.}\ \bibnamefont {de~Vasconcellos}},\ and\ \bibinfo {author}
  {\bibfnamefont {R.}~\bibnamefont {Bratschitsch}},\ }\href
  {https://doi.org/10.1021/acs.nanolett.5b04733} {\bibfield  {journal}
  {\bibinfo  {journal} {Nano Lett.}\ }\textbf {\bibinfo {volume} {16}},\
  \bibinfo {pages} {2945} (\bibinfo {year} {2016})}\BibitemShut {NoStop}%
\bibitem [{\citenamefont {Conte}\ \emph {et~al.}(2015)\citenamefont {Conte},
  \citenamefont {Bottegoni}, \citenamefont {Pogna}, \citenamefont {Fazio},
  \citenamefont {Ambrogio}, \citenamefont {Bargigia}, \citenamefont {Andrea},
  \citenamefont {Lombardo}, \citenamefont {Bruna}, \citenamefont {Ciccacci},
  \citenamefont {Ferrari}, \citenamefont {Cerullo},\ and\ \citenamefont
  {Finazzi}}]{Conte2015}%
  \BibitemOpen
  \bibfield  {author} {\bibinfo {author} {\bibfnamefont {S.~D.}\ \bibnamefont
  {Conte}}, \bibinfo {author} {\bibfnamefont {F.}~\bibnamefont {Bottegoni}},
  \bibinfo {author} {\bibfnamefont {E.~A.~A.}\ \bibnamefont {Pogna}}, \bibinfo
  {author} {\bibfnamefont {D.~D.}\ \bibnamefont {Fazio}}, \bibinfo {author}
  {\bibfnamefont {S.}~\bibnamefont {Ambrogio}}, \bibinfo {author}
  {\bibfnamefont {I.}~\bibnamefont {Bargigia}}, \bibinfo {author}
  {\bibfnamefont {C.~D.}\ \bibnamefont {Andrea}}, \bibinfo {author}
  {\bibfnamefont {A.}~\bibnamefont {Lombardo}}, \bibinfo {author}
  {\bibfnamefont {M.}~\bibnamefont {Bruna}}, \bibinfo {author} {\bibfnamefont
  {F.}~\bibnamefont {Ciccacci}}, \bibinfo {author} {\bibfnamefont {A.~C.}\
  \bibnamefont {Ferrari}}, \bibinfo {author} {\bibfnamefont {G.}~\bibnamefont
  {Cerullo}},\ and\ \bibinfo {author} {\bibfnamefont {M.}~\bibnamefont
  {Finazzi}},\ }\href {https://doi.org/10.1103/physrevb.92.235425} {\bibfield
  {journal} {\bibinfo  {journal} {Physical Review B}\ }\textbf {\bibinfo
  {volume} {92}},\ \bibinfo {pages} {235425} (\bibinfo {year}
  {2015})}\BibitemShut {NoStop}%
\bibitem [{\citenamefont {Hsu}\ \emph {et~al.}(2015)\citenamefont {Hsu},
  \citenamefont {Chen}, \citenamefont {Chen}, \citenamefont {Liu},
  \citenamefont {Hou}, \citenamefont {Li},\ and\ \citenamefont
  {Chang}}]{Hsu2015}%
  \BibitemOpen
  \bibfield  {author} {\bibinfo {author} {\bibfnamefont {W.-T.}\ \bibnamefont
  {Hsu}}, \bibinfo {author} {\bibfnamefont {Y.-L.}\ \bibnamefont {Chen}},
  \bibinfo {author} {\bibfnamefont {C.-H.}\ \bibnamefont {Chen}}, \bibinfo
  {author} {\bibfnamefont {P.-S.}\ \bibnamefont {Liu}}, \bibinfo {author}
  {\bibfnamefont {T.-H.}\ \bibnamefont {Hou}}, \bibinfo {author} {\bibfnamefont
  {L.-J.}\ \bibnamefont {Li}},\ and\ \bibinfo {author} {\bibfnamefont {W.-H.}\
  \bibnamefont {Chang}},\ }\bibfield  {journal} {\bibinfo  {journal} {Nature
  Communications}\ }\textbf {\bibinfo {volume} {6}},\ \href
  {https://doi.org/10.1038/ncomms9963} {10.1038/ncomms9963} (\bibinfo {year}
  {2015})\BibitemShut {NoStop}%
\bibitem [{\citenamefont {Plechinger}\ \emph {et~al.}(2016)\citenamefont
  {Plechinger}, \citenamefont {Nagler}, \citenamefont {Arora}, \citenamefont
  {Schmidt}, \citenamefont {Chernikov}, \citenamefont {del {\'{A}}guila},
  \citenamefont {Christianen}, \citenamefont {Bratschitsch}, \citenamefont
  {Sch\"uller},\ and\ \citenamefont {Korn}}]{Plechinger2016}%
  \BibitemOpen
  \bibfield  {author} {\bibinfo {author} {\bibfnamefont {G.}~\bibnamefont
  {Plechinger}}, \bibinfo {author} {\bibfnamefont {P.}~\bibnamefont {Nagler}},
  \bibinfo {author} {\bibfnamefont {A.}~\bibnamefont {Arora}}, \bibinfo
  {author} {\bibfnamefont {R.}~\bibnamefont {Schmidt}}, \bibinfo {author}
  {\bibfnamefont {A.}~\bibnamefont {Chernikov}}, \bibinfo {author}
  {\bibfnamefont {A.~G.}\ \bibnamefont {del {\'{A}}guila}}, \bibinfo {author}
  {\bibfnamefont {P.~C.}\ \bibnamefont {Christianen}}, \bibinfo {author}
  {\bibfnamefont {R.}~\bibnamefont {Bratschitsch}}, \bibinfo {author}
  {\bibfnamefont {C.}~\bibnamefont {Sch\"uller}},\ and\ \bibinfo {author}
  {\bibfnamefont {T.}~\bibnamefont {Korn}},\ }\href
  {https://doi.org/10.1038/ncomms12715} {\bibfield  {journal} {\bibinfo
  {journal} {Nat. Commun.}\ }\textbf {\bibinfo {volume} {7}},\ \bibinfo {pages}
  {12715} (\bibinfo {year} {2016})}\BibitemShut {NoStop}%
\bibitem [{\citenamefont {Plechinger}\ \emph {et~al.}(2017)\citenamefont
  {Plechinger}, \citenamefont {Korn},\ and\ \citenamefont
  {Lupton}}]{Plechinger2017}%
  \BibitemOpen
  \bibfield  {author} {\bibinfo {author} {\bibfnamefont {G.}~\bibnamefont
  {Plechinger}}, \bibinfo {author} {\bibfnamefont {T.}~\bibnamefont {Korn}},\
  and\ \bibinfo {author} {\bibfnamefont {J.~M.}\ \bibnamefont {Lupton}},\
  }\href {https://doi.org/10.1021/acs.jpcc.7b01468} {\bibfield  {journal}
  {\bibinfo  {journal} {The Journal of Physical Chemistry C}\ }\textbf
  {\bibinfo {volume} {121}},\ \bibinfo {pages} {6409} (\bibinfo {year}
  {2017})}\BibitemShut {NoStop}%
\bibitem [{\citenamefont {McCormick}\ \emph {et~al.}(2017)\citenamefont
  {McCormick}, \citenamefont {Newburger}, \citenamefont {Luo}, \citenamefont
  {McCreary}, \citenamefont {Singh}, \citenamefont {Martin}, \citenamefont
  {Cichewicz}, \citenamefont {Jonker},\ and\ \citenamefont
  {Kawakami}}]{McCormick2017}%
  \BibitemOpen
  \bibfield  {author} {\bibinfo {author} {\bibfnamefont {E.~J.}\ \bibnamefont
  {McCormick}}, \bibinfo {author} {\bibfnamefont {M.~J.}\ \bibnamefont
  {Newburger}}, \bibinfo {author} {\bibfnamefont {Y.~K.}\ \bibnamefont {Luo}},
  \bibinfo {author} {\bibfnamefont {K.~M.}\ \bibnamefont {McCreary}}, \bibinfo
  {author} {\bibfnamefont {S.}~\bibnamefont {Singh}}, \bibinfo {author}
  {\bibfnamefont {I.~B.}\ \bibnamefont {Martin}}, \bibinfo {author}
  {\bibfnamefont {E.~J.}\ \bibnamefont {Cichewicz}}, \bibinfo {author}
  {\bibfnamefont {B.~T.}\ \bibnamefont {Jonker}},\ and\ \bibinfo {author}
  {\bibfnamefont {R.~K.}\ \bibnamefont {Kawakami}},\ }\href
  {https://doi.org/10.1088/2053-1583/aa98ae} {\bibfield  {journal} {\bibinfo
  {journal} {2D Materials}\ }\textbf {\bibinfo {volume} {5}},\ \bibinfo {pages}
  {011010} (\bibinfo {year} {2017})}\BibitemShut {NoStop}%
\bibitem [{\citenamefont {Schwemmer}\ \emph {et~al.}(2017)\citenamefont
  {Schwemmer}, \citenamefont {Nagler}, \citenamefont {Hanninger}, \citenamefont
  {Sch{\"u}ller},\ and\ \citenamefont {Korn}}]{Schwemmer2017}%
  \BibitemOpen
  \bibfield  {author} {\bibinfo {author} {\bibfnamefont {M.}~\bibnamefont
  {Schwemmer}}, \bibinfo {author} {\bibfnamefont {P.}~\bibnamefont {Nagler}},
  \bibinfo {author} {\bibfnamefont {A.}~\bibnamefont {Hanninger}}, \bibinfo
  {author} {\bibfnamefont {C.}~\bibnamefont {Sch{\"u}ller}},\ and\ \bibinfo
  {author} {\bibfnamefont {T.}~\bibnamefont {Korn}},\ }\href
  {https://doi.org/10.1063/1.4987000} {\bibfield  {journal} {\bibinfo
  {journal} {Applied Physics Letters}\ }\textbf {\bibinfo {volume} {111}},\
  \bibinfo {pages} {082404} (\bibinfo {year} {2017})}\BibitemShut {NoStop}%
\bibitem [{\citenamefont {Dey}\ \emph {et~al.}(2017)\citenamefont {Dey},
  \citenamefont {Yang}, \citenamefont {Robert}, \citenamefont {Wang},
  \citenamefont {Urbaszek}, \citenamefont {Marie},\ and\ \citenamefont
  {Crooker}}]{Dey2017}%
  \BibitemOpen
  \bibfield  {author} {\bibinfo {author} {\bibfnamefont {P.}~\bibnamefont
  {Dey}}, \bibinfo {author} {\bibfnamefont {L.}~\bibnamefont {Yang}}, \bibinfo
  {author} {\bibfnamefont {C.}~\bibnamefont {Robert}}, \bibinfo {author}
  {\bibfnamefont {G.}~\bibnamefont {Wang}}, \bibinfo {author} {\bibfnamefont
  {B.}~\bibnamefont {Urbaszek}}, \bibinfo {author} {\bibfnamefont
  {X.}~\bibnamefont {Marie}},\ and\ \bibinfo {author} {\bibfnamefont
  {S.}~\bibnamefont {Crooker}},\ }\href
  {https://doi.org/10.1103/physrevlett.119.137401} {\bibfield  {journal}
  {\bibinfo  {journal} {Physical Review Letters}\ }\textbf {\bibinfo {volume}
  {119}},\ \bibinfo {pages} {137401} (\bibinfo {year} {2017})}\BibitemShut
  {NoStop}%
\bibitem [{\citenamefont {Smallwood}\ and\ \citenamefont
  {Cundiff}(2018)}]{Smallwood2018}%
  \BibitemOpen
  \bibfield  {author} {\bibinfo {author} {\bibfnamefont {C.~L.}\ \bibnamefont
  {Smallwood}}\ and\ \bibinfo {author} {\bibfnamefont {S.~T.}\ \bibnamefont
  {Cundiff}},\ }\href {https://doi.org/10.1002/lpor.201800171} {\bibfield
  {journal} {\bibinfo  {journal} {Laser {\&} Photonics Reviews}\ }\textbf
  {\bibinfo {volume} {12}},\ \bibinfo {pages} {1800171} (\bibinfo {year}
  {2018})}\BibitemShut {NoStop}%
\bibitem [{\citenamefont {Lloyd}\ \emph {et~al.}(2021)\citenamefont {Lloyd},
  \citenamefont {Wood}, \citenamefont {Mujid}, \citenamefont {Sohoni},
  \citenamefont {Ji}, \citenamefont {Ting}, \citenamefont {Higgins},
  \citenamefont {Park},\ and\ \citenamefont {Engel}}]{Lloyd2021}%
  \BibitemOpen
  \bibfield  {author} {\bibinfo {author} {\bibfnamefont {L.~T.}\ \bibnamefont
  {Lloyd}}, \bibinfo {author} {\bibfnamefont {R.~E.}\ \bibnamefont {Wood}},
  \bibinfo {author} {\bibfnamefont {F.}~\bibnamefont {Mujid}}, \bibinfo
  {author} {\bibfnamefont {S.}~\bibnamefont {Sohoni}}, \bibinfo {author}
  {\bibfnamefont {K.~L.}\ \bibnamefont {Ji}}, \bibinfo {author} {\bibfnamefont
  {P.-C.}\ \bibnamefont {Ting}}, \bibinfo {author} {\bibfnamefont {J.~S.}\
  \bibnamefont {Higgins}}, \bibinfo {author} {\bibfnamefont {J.}~\bibnamefont
  {Park}},\ and\ \bibinfo {author} {\bibfnamefont {G.~S.}\ \bibnamefont
  {Engel}},\ }\href {https://doi.org/10.1021/acsnano.1c02381} {\bibfield
  {journal} {\bibinfo  {journal} {ACS Nano}\ }\textbf {\bibinfo {volume}
  {15}},\ \bibinfo {pages} {10253} (\bibinfo {year} {2021})}\BibitemShut
  {NoStop}%
\bibitem [{\citenamefont {Purz}\ \emph {et~al.}(2021)\citenamefont {Purz},
  \citenamefont {Martin}, \citenamefont {Rivera}, \citenamefont {Holtzmann},
  \citenamefont {Xu},\ and\ \citenamefont {Cundiff}}]{Purz2021}%
  \BibitemOpen
  \bibfield  {author} {\bibinfo {author} {\bibfnamefont {T.~L.}\ \bibnamefont
  {Purz}}, \bibinfo {author} {\bibfnamefont {E.~W.}\ \bibnamefont {Martin}},
  \bibinfo {author} {\bibfnamefont {P.}~\bibnamefont {Rivera}}, \bibinfo
  {author} {\bibfnamefont {W.~G.}\ \bibnamefont {Holtzmann}}, \bibinfo {author}
  {\bibfnamefont {X.}~\bibnamefont {Xu}},\ and\ \bibinfo {author}
  {\bibfnamefont {S.~T.}\ \bibnamefont {Cundiff}},\ }\href
  {https://doi.org/10.1103/physrevb.104.l241302} {\bibfield  {journal}
  {\bibinfo  {journal} {Physical Review B}\ }\textbf {\bibinfo {volume}
  {104}},\ \bibinfo {pages} {l241302} (\bibinfo {year} {2021})}\BibitemShut
  {NoStop}%
\bibitem [{\citenamefont {Mahmood}\ \emph {et~al.}(2017)\citenamefont
  {Mahmood}, \citenamefont {Alpichshev}, \citenamefont {Lee}, \citenamefont
  {Kong},\ and\ \citenamefont {Gedik}}]{Mahmood2017}%
  \BibitemOpen
  \bibfield  {author} {\bibinfo {author} {\bibfnamefont {F.}~\bibnamefont
  {Mahmood}}, \bibinfo {author} {\bibfnamefont {Z.}~\bibnamefont {Alpichshev}},
  \bibinfo {author} {\bibfnamefont {Y.-H.}\ \bibnamefont {Lee}}, \bibinfo
  {author} {\bibfnamefont {J.}~\bibnamefont {Kong}},\ and\ \bibinfo {author}
  {\bibfnamefont {N.}~\bibnamefont {Gedik}},\ }\href
  {https://doi.org/10.1021/acs.nanolett.7b03953} {\bibfield  {journal}
  {\bibinfo  {journal} {Nano Letters}\ }\textbf {\bibinfo {volume} {18}},\
  \bibinfo {pages} {223} (\bibinfo {year} {2017})}\BibitemShut {NoStop}%
\bibitem [{\citenamefont {Moody}\ \emph {et~al.}(2015)\citenamefont {Moody},
  \citenamefont {Dass}, \citenamefont {Hao}, \citenamefont {Chen},
  \citenamefont {Li}, \citenamefont {Singh}, \citenamefont {Tran},
  \citenamefont {Clark}, \citenamefont {Xu}, \citenamefont {Bergh{\"a}user},
  \citenamefont {Malic}, \citenamefont {Knorr},\ and\ \citenamefont
  {Li}}]{Moody2015}%
  \BibitemOpen
  \bibfield  {author} {\bibinfo {author} {\bibfnamefont {G.}~\bibnamefont
  {Moody}}, \bibinfo {author} {\bibfnamefont {C.~K.}\ \bibnamefont {Dass}},
  \bibinfo {author} {\bibfnamefont {K.}~\bibnamefont {Hao}}, \bibinfo {author}
  {\bibfnamefont {C.-H.}\ \bibnamefont {Chen}}, \bibinfo {author}
  {\bibfnamefont {L.-J.}\ \bibnamefont {Li}}, \bibinfo {author} {\bibfnamefont
  {A.}~\bibnamefont {Singh}}, \bibinfo {author} {\bibfnamefont
  {K.}~\bibnamefont {Tran}}, \bibinfo {author} {\bibfnamefont {G.}~\bibnamefont
  {Clark}}, \bibinfo {author} {\bibfnamefont {X.}~\bibnamefont {Xu}}, \bibinfo
  {author} {\bibfnamefont {G.}~\bibnamefont {Bergh{\"a}user}}, \bibinfo
  {author} {\bibfnamefont {E.}~\bibnamefont {Malic}}, \bibinfo {author}
  {\bibfnamefont {A.}~\bibnamefont {Knorr}},\ and\ \bibinfo {author}
  {\bibfnamefont {X.}~\bibnamefont {Li}},\ }\bibfield  {journal} {\bibinfo
  {journal} {Nature Communications}\ }\textbf {\bibinfo {volume} {6}},\ \href
  {https://doi.org/10.1038/ncomms9315} {10.1038/ncomms9315} (\bibinfo {year}
  {2015})\BibitemShut {NoStop}%
\bibitem [{\citenamefont {Kioseoglou}\ \emph {et~al.}(2016)\citenamefont
  {Kioseoglou}, \citenamefont {Hanbicki}, \citenamefont {Currie}, \citenamefont
  {Friedman},\ and\ \citenamefont {Jonker}}]{Kioseoglou2016}%
  \BibitemOpen
  \bibfield  {author} {\bibinfo {author} {\bibfnamefont {G.}~\bibnamefont
  {Kioseoglou}}, \bibinfo {author} {\bibfnamefont {A.~T.}\ \bibnamefont
  {Hanbicki}}, \bibinfo {author} {\bibfnamefont {M.}~\bibnamefont {Currie}},
  \bibinfo {author} {\bibfnamefont {A.~L.}\ \bibnamefont {Friedman}},\ and\
  \bibinfo {author} {\bibfnamefont {B.~T.}\ \bibnamefont {Jonker}},\ }\bibfield
   {journal} {\bibinfo  {journal} {Scientific Reports}\ }\textbf {\bibinfo
  {volume} {6}},\ \href {https://doi.org/10.1038/srep25041} {10.1038/srep25041}
  (\bibinfo {year} {2016})\BibitemShut {NoStop}%
\bibitem [{\citenamefont {Ye}\ \emph {et~al.}(2019)\citenamefont {Ye},
  \citenamefont {Li}, \citenamefont {Yan}, \citenamefont {Zhai},\ and\
  \citenamefont {Zhang}}]{Ye2019}%
  \BibitemOpen
  \bibfield  {author} {\bibinfo {author} {\bibfnamefont {J.}~\bibnamefont
  {Ye}}, \bibinfo {author} {\bibfnamefont {Y.}~\bibnamefont {Li}}, \bibinfo
  {author} {\bibfnamefont {T.}~\bibnamefont {Yan}}, \bibinfo {author}
  {\bibfnamefont {G.}~\bibnamefont {Zhai}},\ and\ \bibinfo {author}
  {\bibfnamefont {X.}~\bibnamefont {Zhang}},\ }\href
  {https://doi.org/10.1021/acs.jpclett.9b01068} {\bibfield  {journal} {\bibinfo
   {journal} {The Journal of Physical Chemistry Letters}\ }\textbf {\bibinfo
  {volume} {10}},\ \bibinfo {pages} {2963} (\bibinfo {year}
  {2019})}\BibitemShut {NoStop}%
\bibitem [{\citenamefont {Selig}\ \emph {et~al.}(2020)\citenamefont {Selig},
  \citenamefont {Katsch}, \citenamefont {Brem}, \citenamefont {Mkrtchian},
  \citenamefont {Malic},\ and\ \citenamefont {Knorr}}]{Selig2020}%
  \BibitemOpen
  \bibfield  {author} {\bibinfo {author} {\bibfnamefont {M.}~\bibnamefont
  {Selig}}, \bibinfo {author} {\bibfnamefont {F.}~\bibnamefont {Katsch}},
  \bibinfo {author} {\bibfnamefont {S.}~\bibnamefont {Brem}}, \bibinfo {author}
  {\bibfnamefont {G.~F.}\ \bibnamefont {Mkrtchian}}, \bibinfo {author}
  {\bibfnamefont {E.}~\bibnamefont {Malic}},\ and\ \bibinfo {author}
  {\bibfnamefont {A.}~\bibnamefont {Knorr}},\ }\href
  {https://doi.org/10.1103/physrevresearch.2.023322} {\bibfield  {journal}
  {\bibinfo  {journal} {Physical Review Research}\ }\textbf {\bibinfo {volume}
  {2}},\ \bibinfo {pages} {023322} (\bibinfo {year} {2020})}\BibitemShut
  {NoStop}%
\bibitem [{\citenamefont {Li}\ \emph {et~al.}(2014)\citenamefont {Li},
  \citenamefont {Chernikov}, \citenamefont {Zhang}, \citenamefont {Rigosi},
  \citenamefont {Hill}, \citenamefont {van~der Zande}, \citenamefont {Chenet},
  \citenamefont {Shih}, \citenamefont {Hone},\ and\ \citenamefont
  {Heinz}}]{Li2014a}%
  \BibitemOpen
  \bibfield  {author} {\bibinfo {author} {\bibfnamefont {Y.}~\bibnamefont
  {Li}}, \bibinfo {author} {\bibfnamefont {A.}~\bibnamefont {Chernikov}},
  \bibinfo {author} {\bibfnamefont {X.}~\bibnamefont {Zhang}}, \bibinfo
  {author} {\bibfnamefont {A.}~\bibnamefont {Rigosi}}, \bibinfo {author}
  {\bibfnamefont {H.~M.}\ \bibnamefont {Hill}}, \bibinfo {author}
  {\bibfnamefont {A.~M.}\ \bibnamefont {van~der Zande}}, \bibinfo {author}
  {\bibfnamefont {D.~A.}\ \bibnamefont {Chenet}}, \bibinfo {author}
  {\bibfnamefont {E.-M.}\ \bibnamefont {Shih}}, \bibinfo {author}
  {\bibfnamefont {J.}~\bibnamefont {Hone}},\ and\ \bibinfo {author}
  {\bibfnamefont {T.~F.}\ \bibnamefont {Heinz}},\ }\href
  {https://doi.org/10.1103/physrevb.90.205422} {\bibfield  {journal} {\bibinfo
  {journal} {Phys. Rev. B}\ }\textbf {\bibinfo {volume} {90}},\ \bibinfo
  {pages} {205422} (\bibinfo {year} {2014})}\BibitemShut {NoStop}%
\bibitem [{\citenamefont {Mad{\'{e}}o}\ \emph {et~al.}(2020)\citenamefont
  {Mad{\'{e}}o}, \citenamefont {Man}, \citenamefont {Sahoo}, \citenamefont
  {Campbell}, \citenamefont {Pareek}, \citenamefont {Wong}, \citenamefont
  {Al-Mahboob}, \citenamefont {Chan}, \citenamefont {Karmakar}, \citenamefont
  {Mariserla}, \citenamefont {Li}, \citenamefont {Heinz}, \citenamefont {Cao},\
  and\ \citenamefont {Dani}}]{Madeo2020}%
  \BibitemOpen
  \bibfield  {author} {\bibinfo {author} {\bibfnamefont {J.}~\bibnamefont
  {Mad{\'{e}}o}}, \bibinfo {author} {\bibfnamefont {M.~K.~L.}\ \bibnamefont
  {Man}}, \bibinfo {author} {\bibfnamefont {C.}~\bibnamefont {Sahoo}}, \bibinfo
  {author} {\bibfnamefont {M.}~\bibnamefont {Campbell}}, \bibinfo {author}
  {\bibfnamefont {V.}~\bibnamefont {Pareek}}, \bibinfo {author} {\bibfnamefont
  {E.~L.}\ \bibnamefont {Wong}}, \bibinfo {author} {\bibfnamefont
  {A.}~\bibnamefont {Al-Mahboob}}, \bibinfo {author} {\bibfnamefont {N.~S.}\
  \bibnamefont {Chan}}, \bibinfo {author} {\bibfnamefont {A.}~\bibnamefont
  {Karmakar}}, \bibinfo {author} {\bibfnamefont {B.~M.~K.}\ \bibnamefont
  {Mariserla}}, \bibinfo {author} {\bibfnamefont {X.}~\bibnamefont {Li}},
  \bibinfo {author} {\bibfnamefont {T.~F.}\ \bibnamefont {Heinz}}, \bibinfo
  {author} {\bibfnamefont {T.}~\bibnamefont {Cao}},\ and\ \bibinfo {author}
  {\bibfnamefont {K.~M.}\ \bibnamefont {Dani}},\ }\href
  {https://doi.org/10.1126/science.aba1029} {\bibfield  {journal} {\bibinfo
  {journal} {Science}\ }\textbf {\bibinfo {volume} {370}},\ \bibinfo {pages}
  {1199} (\bibinfo {year} {2020})}\BibitemShut {NoStop}%
\bibitem [{\citenamefont {Man}\ \emph {et~al.}(2021)\citenamefont {Man},
  \citenamefont {Mad{\'{e}}o}, \citenamefont {Sahoo}, \citenamefont {Xie},
  \citenamefont {Campbell}, \citenamefont {Pareek}, \citenamefont {Karmakar},
  \citenamefont {Wong}, \citenamefont {Al-Mahboob}, \citenamefont {Chan},
  \citenamefont {Bacon}, \citenamefont {Zhu}, \citenamefont {Abdelrasoul},
  \citenamefont {Li}, \citenamefont {Heinz}, \citenamefont {da~Jornada},
  \citenamefont {Cao},\ and\ \citenamefont {Dani}}]{Man2021}%
  \BibitemOpen
  \bibfield  {author} {\bibinfo {author} {\bibfnamefont {M.~K.~L.}\
  \bibnamefont {Man}}, \bibinfo {author} {\bibfnamefont {J.}~\bibnamefont
  {Mad{\'{e}}o}}, \bibinfo {author} {\bibfnamefont {C.}~\bibnamefont {Sahoo}},
  \bibinfo {author} {\bibfnamefont {K.}~\bibnamefont {Xie}}, \bibinfo {author}
  {\bibfnamefont {M.}~\bibnamefont {Campbell}}, \bibinfo {author}
  {\bibfnamefont {V.}~\bibnamefont {Pareek}}, \bibinfo {author} {\bibfnamefont
  {A.}~\bibnamefont {Karmakar}}, \bibinfo {author} {\bibfnamefont {E.~L.}\
  \bibnamefont {Wong}}, \bibinfo {author} {\bibfnamefont {A.}~\bibnamefont
  {Al-Mahboob}}, \bibinfo {author} {\bibfnamefont {N.~S.}\ \bibnamefont
  {Chan}}, \bibinfo {author} {\bibfnamefont {D.~R.}\ \bibnamefont {Bacon}},
  \bibinfo {author} {\bibfnamefont {X.}~\bibnamefont {Zhu}}, \bibinfo {author}
  {\bibfnamefont {M.~M.~M.}\ \bibnamefont {Abdelrasoul}}, \bibinfo {author}
  {\bibfnamefont {X.}~\bibnamefont {Li}}, \bibinfo {author} {\bibfnamefont
  {T.~F.}\ \bibnamefont {Heinz}}, \bibinfo {author} {\bibfnamefont {F.~H.}\
  \bibnamefont {da~Jornada}}, \bibinfo {author} {\bibfnamefont
  {T.}~\bibnamefont {Cao}},\ and\ \bibinfo {author} {\bibfnamefont {K.~M.}\
  \bibnamefont {Dani}},\ }\href {https://doi.org/10.1126/sciadv.abg0192}
  {\bibfield  {journal} {\bibinfo  {journal} {Science Advances}\ }\textbf
  {\bibinfo {volume} {7}},\ \bibinfo {pages} {eabg0192} (\bibinfo {year}
  {2021})}\BibitemShut {NoStop}%
\bibitem [{\citenamefont {Wallauer}\ \emph {et~al.}(2021)\citenamefont
  {Wallauer}, \citenamefont {Perea-Causin}, \citenamefont {M\"unster},
  \citenamefont {Zajusch}, \citenamefont {Brem}, \citenamefont {G\"udde},
  \citenamefont {Tanimura}, \citenamefont {Lin}, \citenamefont {Huber},
  \citenamefont {Malic},\ and\ \citenamefont {H\"ofer}}]{Wallauer2021a}%
  \BibitemOpen
  \bibfield  {author} {\bibinfo {author} {\bibfnamefont {R.}~\bibnamefont
  {Wallauer}}, \bibinfo {author} {\bibfnamefont {R.}~\bibnamefont
  {Perea-Causin}}, \bibinfo {author} {\bibfnamefont {L.}~\bibnamefont
  {M\"unster}}, \bibinfo {author} {\bibfnamefont {S.}~\bibnamefont {Zajusch}},
  \bibinfo {author} {\bibfnamefont {S.}~\bibnamefont {Brem}}, \bibinfo {author}
  {\bibfnamefont {J.}~\bibnamefont {G\"udde}}, \bibinfo {author} {\bibfnamefont
  {K.}~\bibnamefont {Tanimura}}, \bibinfo {author} {\bibfnamefont {K.-Q.}\
  \bibnamefont {Lin}}, \bibinfo {author} {\bibfnamefont {R.}~\bibnamefont
  {Huber}}, \bibinfo {author} {\bibfnamefont {E.}~\bibnamefont {Malic}},\ and\
  \bibinfo {author} {\bibfnamefont {U.}~\bibnamefont {H\"ofer}},\ }\href
  {https://doi.org/10.1021/acs.nanolett.1c01839} {\bibfield  {journal}
  {\bibinfo  {journal} {Nano Letters}\ }\textbf {\bibinfo {volume} {21}},\
  \bibinfo {pages} {5867} (\bibinfo {year} {2021})}\BibitemShut {NoStop}%
\bibitem [{\citenamefont {Karni}\ \emph {et~al.}(2021)\citenamefont {Karni},
  \citenamefont {Barr{\'e}}, \citenamefont {Pareek}, \citenamefont {Georgaras},
  \citenamefont {Man}, \citenamefont {Sahoo}, \citenamefont {Bacon},
  \citenamefont {Zhu}, \citenamefont {Ribeiro}, \citenamefont {O'Beirne},
  \citenamefont {Hu}, \citenamefont {Al-Mahboob}, \citenamefont {Abdelrasoul},
  \citenamefont {Chan}, \citenamefont {Karmakar}, \citenamefont {Winchester},
  \citenamefont {Kim}, \citenamefont {Watanabe}, \citenamefont {Taniguchi},
  \citenamefont {Barmak}, \citenamefont {Mad{\'e}o}, \citenamefont
  {da~Jornada}, \citenamefont {Heinz},\ and\ \citenamefont {Dani}}]{Karni2021}%
  \BibitemOpen
  \bibfield  {author} {\bibinfo {author} {\bibfnamefont {O.}~\bibnamefont
  {Karni}}, \bibinfo {author} {\bibfnamefont {E.}~\bibnamefont {Barr{\'e}}},
  \bibinfo {author} {\bibfnamefont {V.}~\bibnamefont {Pareek}}, \bibinfo
  {author} {\bibfnamefont {J.~D.}\ \bibnamefont {Georgaras}}, \bibinfo {author}
  {\bibfnamefont {M.~K.~L.}\ \bibnamefont {Man}}, \bibinfo {author}
  {\bibfnamefont {C.}~\bibnamefont {Sahoo}}, \bibinfo {author} {\bibfnamefont
  {D.~R.}\ \bibnamefont {Bacon}}, \bibinfo {author} {\bibfnamefont
  {X.}~\bibnamefont {Zhu}}, \bibinfo {author} {\bibfnamefont {H.~B.}\
  \bibnamefont {Ribeiro}}, \bibinfo {author} {\bibfnamefont {A.~L.}\
  \bibnamefont {O'Beirne}}, \bibinfo {author} {\bibfnamefont {J.}~\bibnamefont
  {Hu}}, \bibinfo {author} {\bibfnamefont {A.}~\bibnamefont {Al-Mahboob}},
  \bibinfo {author} {\bibfnamefont {M.~M.~M.}\ \bibnamefont {Abdelrasoul}},
  \bibinfo {author} {\bibfnamefont {N.~S.}\ \bibnamefont {Chan}}, \bibinfo
  {author} {\bibfnamefont {A.}~\bibnamefont {Karmakar}}, \bibinfo {author}
  {\bibfnamefont {A.~J.}\ \bibnamefont {Winchester}}, \bibinfo {author}
  {\bibfnamefont {B.}~\bibnamefont {Kim}}, \bibinfo {author} {\bibfnamefont
  {K.}~\bibnamefont {Watanabe}}, \bibinfo {author} {\bibfnamefont
  {T.}~\bibnamefont {Taniguchi}}, \bibinfo {author} {\bibfnamefont
  {K.}~\bibnamefont {Barmak}}, \bibinfo {author} {\bibfnamefont
  {J.}~\bibnamefont {Mad{\'e}o}}, \bibinfo {author} {\bibfnamefont {F.~H.}\
  \bibnamefont {da~Jornada}}, \bibinfo {author} {\bibfnamefont {T.~F.}\
  \bibnamefont {Heinz}},\ and\ \bibinfo {author} {\bibfnamefont {K.~M.}\
  \bibnamefont {Dani}},\ }\href@noop {} {\bibfield  {journal} {\bibinfo
  {journal} {arXiv}\ } (\bibinfo {year} {2021})},\ \Eprint
  {https://arxiv.org/abs/2108.01933} {arXiv:2108.01933 [physics.optics]}
  \BibitemShut {NoStop}%
\bibitem [{\citenamefont {Schmitt}\ \emph {et~al.}(2021)\citenamefont
  {Schmitt}, \citenamefont {Bange}, \citenamefont {Bennecke}, \citenamefont
  {AlMutairi}, \citenamefont {Watanabe}, \citenamefont {Taniguchi},
  \citenamefont {Steil}, \citenamefont {Luke}, \citenamefont {Weitz},
  \citenamefont {Steil}, \citenamefont {Jansen}, \citenamefont {Hofmann},
  \citenamefont {Reutzel},\ and\ \citenamefont {Mathias}}]{Schmitt2021}%
  \BibitemOpen
  \bibfield  {author} {\bibinfo {author} {\bibfnamefont {D.}~\bibnamefont
  {Schmitt}}, \bibinfo {author} {\bibfnamefont {J.~P.}\ \bibnamefont {Bange}},
  \bibinfo {author} {\bibfnamefont {W.}~\bibnamefont {Bennecke}}, \bibinfo
  {author} {\bibfnamefont {A.}~\bibnamefont {AlMutairi}}, \bibinfo {author}
  {\bibfnamefont {K.}~\bibnamefont {Watanabe}}, \bibinfo {author}
  {\bibfnamefont {T.}~\bibnamefont {Taniguchi}}, \bibinfo {author}
  {\bibfnamefont {D.}~\bibnamefont {Steil}}, \bibinfo {author} {\bibfnamefont
  {D.~R.}\ \bibnamefont {Luke}}, \bibinfo {author} {\bibfnamefont {R.~T.}\
  \bibnamefont {Weitz}}, \bibinfo {author} {\bibfnamefont {S.}~\bibnamefont
  {Steil}}, \bibinfo {author} {\bibfnamefont {G.~S.~M.}\ \bibnamefont
  {Jansen}}, \bibinfo {author} {\bibfnamefont {S.}~\bibnamefont {Hofmann}},
  \bibinfo {author} {\bibfnamefont {M.}~\bibnamefont {Reutzel}},\ and\ \bibinfo
  {author} {\bibfnamefont {S.}~\bibnamefont {Mathias}},\ }\href@noop {}
  {\bibfield  {journal} {\bibinfo  {journal} {arXiv}\ } (\bibinfo {year}
  {2021})},\ \Eprint {https://arxiv.org/abs/2112.05011} {arXiv:2112.05011
  [cond-mat.mes-hall]} \BibitemShut {NoStop}%
\bibitem [{\citenamefont {Chernov}\ \emph {et~al.}(2015)\citenamefont
  {Chernov}, \citenamefont {Medjanik}, \citenamefont {Tusche}, \citenamefont
  {Kutnyakhov}, \citenamefont {Nepijko}, \citenamefont {Oelsner}, \citenamefont
  {Braun}, \citenamefont {Min{\'{a}}r}, \citenamefont {Borek}, \citenamefont
  {Ebert}, \citenamefont {Elmers}, \citenamefont {Kirschner},\ and\
  \citenamefont {Sch{\"o}nhense}}]{Chernov2015}%
  \BibitemOpen
  \bibfield  {author} {\bibinfo {author} {\bibfnamefont {S.}~\bibnamefont
  {Chernov}}, \bibinfo {author} {\bibfnamefont {K.}~\bibnamefont {Medjanik}},
  \bibinfo {author} {\bibfnamefont {C.}~\bibnamefont {Tusche}}, \bibinfo
  {author} {\bibfnamefont {D.}~\bibnamefont {Kutnyakhov}}, \bibinfo {author}
  {\bibfnamefont {S.}~\bibnamefont {Nepijko}}, \bibinfo {author} {\bibfnamefont
  {A.}~\bibnamefont {Oelsner}}, \bibinfo {author} {\bibfnamefont
  {J.}~\bibnamefont {Braun}}, \bibinfo {author} {\bibfnamefont
  {J.}~\bibnamefont {Min{\'{a}}r}}, \bibinfo {author} {\bibfnamefont
  {S.}~\bibnamefont {Borek}}, \bibinfo {author} {\bibfnamefont
  {H.}~\bibnamefont {Ebert}}, \bibinfo {author} {\bibfnamefont
  {H.}~\bibnamefont {Elmers}}, \bibinfo {author} {\bibfnamefont
  {J.}~\bibnamefont {Kirschner}},\ and\ \bibinfo {author} {\bibfnamefont
  {G.}~\bibnamefont {Sch{\"o}nhense}},\ }\href
  {https://doi.org/10.1016/j.ultramic.2015.07.008} {\bibfield  {journal}
  {\bibinfo  {journal} {Ultramicroscopy}\ }\textbf {\bibinfo {volume} {159}},\
  \bibinfo {pages} {453} (\bibinfo {year} {2015})}\BibitemShut {NoStop}%
\bibitem [{\citenamefont {Medjanik}\ \emph {et~al.}(2017)\citenamefont
  {Medjanik}, \citenamefont {Fedchenko}, \citenamefont {Chernov}, \citenamefont
  {Kutnyakhov}, \citenamefont {Ellguth}, \citenamefont {Oelsner}, \citenamefont
  {Sch\"onhense}, \citenamefont {Peixoto}, \citenamefont {Lutz}, \citenamefont
  {Min}, \citenamefont {Reinert}, \citenamefont {D\"aster}, \citenamefont
  {Acremann}, \citenamefont {Viefhaus}, \citenamefont {Wurth}, \citenamefont
  {Elmers},\ and\ \citenamefont {Sch\"onhense}}]{Medjanik2017}%
  \BibitemOpen
  \bibfield  {author} {\bibinfo {author} {\bibfnamefont {K.}~\bibnamefont
  {Medjanik}}, \bibinfo {author} {\bibfnamefont {O.}~\bibnamefont {Fedchenko}},
  \bibinfo {author} {\bibfnamefont {S.}~\bibnamefont {Chernov}}, \bibinfo
  {author} {\bibfnamefont {D.}~\bibnamefont {Kutnyakhov}}, \bibinfo {author}
  {\bibfnamefont {M.}~\bibnamefont {Ellguth}}, \bibinfo {author} {\bibfnamefont
  {A.}~\bibnamefont {Oelsner}}, \bibinfo {author} {\bibfnamefont
  {B.}~\bibnamefont {Sch\"onhense}}, \bibinfo {author} {\bibfnamefont
  {T.~R.~F.}\ \bibnamefont {Peixoto}}, \bibinfo {author} {\bibfnamefont
  {P.}~\bibnamefont {Lutz}}, \bibinfo {author} {\bibfnamefont {C.-H.}\
  \bibnamefont {Min}}, \bibinfo {author} {\bibfnamefont {F.}~\bibnamefont
  {Reinert}}, \bibinfo {author} {\bibfnamefont {S.}~\bibnamefont {D\"aster}},
  \bibinfo {author} {\bibfnamefont {Y.}~\bibnamefont {Acremann}}, \bibinfo
  {author} {\bibfnamefont {J.}~\bibnamefont {Viefhaus}}, \bibinfo {author}
  {\bibfnamefont {W.}~\bibnamefont {Wurth}}, \bibinfo {author} {\bibfnamefont
  {H.~J.}\ \bibnamefont {Elmers}},\ and\ \bibinfo {author} {\bibfnamefont
  {G.}~\bibnamefont {Sch\"onhense}},\ }\href {https://doi.org/10.1038/nmat4875}
  {\bibfield  {journal} {\bibinfo  {journal} {Nat. Mater.}\ }\textbf {\bibinfo
  {volume} {16}},\ \bibinfo {pages} {615} (\bibinfo {year} {2017})}\BibitemShut
  {NoStop}%
\bibitem [{\citenamefont {Li}\ \emph {et~al.}(2016)\citenamefont {Li},
  \citenamefont {Reber}, \citenamefont {Corder}, \citenamefont {Chen},
  \citenamefont {Zhao},\ and\ \citenamefont {Allison}}]{Li_RSI2016}%
  \BibitemOpen
  \bibfield  {author} {\bibinfo {author} {\bibfnamefont {X.}~\bibnamefont
  {Li}}, \bibinfo {author} {\bibfnamefont {M.~A.~R.}\ \bibnamefont {Reber}},
  \bibinfo {author} {\bibfnamefont {C.}~\bibnamefont {Corder}}, \bibinfo
  {author} {\bibfnamefont {Y.}~\bibnamefont {Chen}}, \bibinfo {author}
  {\bibfnamefont {P.}~\bibnamefont {Zhao}},\ and\ \bibinfo {author}
  {\bibfnamefont {T.~K.}\ \bibnamefont {Allison}},\ }\href
  {https://doi.org/http://dx.doi.org/10.1063/1.4962867} {\bibfield  {journal}
  {\bibinfo  {journal} {Review of Scientific Instruments}\ }\textbf {\bibinfo
  {volume} {87}},\ \bibinfo {eid} {093114} (\bibinfo {year}
  {2016})}\BibitemShut {NoStop}%
\bibitem [{\citenamefont {Corder}\ \emph
  {et~al.}(2018{\natexlab{a}})\citenamefont {Corder}, \citenamefont {Zhao},
  \citenamefont {Bakalis}, \citenamefont {Li}, \citenamefont {Kershis},
  \citenamefont {Muraca}, \citenamefont {White},\ and\ \citenamefont
  {Allison}}]{Corder2018}%
  \BibitemOpen
  \bibfield  {author} {\bibinfo {author} {\bibfnamefont {C.}~\bibnamefont
  {Corder}}, \bibinfo {author} {\bibfnamefont {P.}~\bibnamefont {Zhao}},
  \bibinfo {author} {\bibfnamefont {J.}~\bibnamefont {Bakalis}}, \bibinfo
  {author} {\bibfnamefont {X.}~\bibnamefont {Li}}, \bibinfo {author}
  {\bibfnamefont {M.~D.}\ \bibnamefont {Kershis}}, \bibinfo {author}
  {\bibfnamefont {A.~R.}\ \bibnamefont {Muraca}}, \bibinfo {author}
  {\bibfnamefont {M.~G.}\ \bibnamefont {White}},\ and\ \bibinfo {author}
  {\bibfnamefont {T.~K.}\ \bibnamefont {Allison}},\ }\href
  {https://doi.org/10.1063/1.5045578} {\bibfield  {journal} {\bibinfo
  {journal} {Struct. Dyn.}\ }\textbf {\bibinfo {volume} {5}},\ \bibinfo {pages}
  {054301} (\bibinfo {year} {2018}{\natexlab{a}})}\BibitemShut {NoStop}%
\bibitem [{\citenamefont {Corder}\ \emph
  {et~al.}(2018{\natexlab{b}})\citenamefont {Corder}, \citenamefont {Zhao},
  \citenamefont {Li}, \citenamefont {Kershis}, \citenamefont {White},\ and\
  \citenamefont {Allison}}]{Corder2018a}%
  \BibitemOpen
  \bibfield  {author} {\bibinfo {author} {\bibfnamefont {C.}~\bibnamefont
  {Corder}}, \bibinfo {author} {\bibfnamefont {P.}~\bibnamefont {Zhao}},
  \bibinfo {author} {\bibfnamefont {X.}~\bibnamefont {Li}}, \bibinfo {author}
  {\bibfnamefont {M.~D.}\ \bibnamefont {Kershis}}, \bibinfo {author}
  {\bibfnamefont {M.~G.}\ \bibnamefont {White}},\ and\ \bibinfo {author}
  {\bibfnamefont {T.~K.}\ \bibnamefont {Allison}},\ }in\ \href
  {https://doi.org/10.1117/12.2295232} {\emph {\bibinfo {booktitle} {{Laser
  Applications in Microelectronic and Optoelectronic Manufacturing} ({LAMOM})
  {XXIII}}}},\ \bibinfo {editor} {edited by\ \bibinfo {editor} {\bibfnamefont
  {B.}~\bibnamefont {Neuenschwander}}, \bibinfo {editor} {\bibfnamefont
  {G.}~\bibnamefont {Ra{\v{c}}iukaitis}}, \bibinfo {editor} {\bibfnamefont
  {T.}~\bibnamefont {Makimura}},\ and\ \bibinfo {editor} {\bibfnamefont
  {C.~P.}\ \bibnamefont {Grigoropoulos}}}\ (\bibinfo  {publisher} {{SPIE}},\
  \bibinfo {year} {2018})\BibitemShut {NoStop}%
\bibitem [{\citenamefont {Schönhense}\ \emph {et~al.}(2021)\citenamefont
  {Schönhense}, \citenamefont {Medjanik}, \citenamefont {Fedchenko},
  \citenamefont {Zymakov{\'{a}}}, \citenamefont {Chernov}, \citenamefont
  {Kutnyakhov}, \citenamefont {Vasilyev}, \citenamefont {Babenkov},
  \citenamefont {Elmers}, \citenamefont {Baumgärtel}, \citenamefont
  {Goslawski}, \citenamefont {Öhrwall}, \citenamefont {Grunske}, \citenamefont
  {Kauerhof}, \citenamefont {von Volkmann}, \citenamefont {Kallmayer},
  \citenamefont {Ellguth},\ and\ \citenamefont {Oelsner}}]{Schoenhense2021}%
  \BibitemOpen
  \bibfield  {author} {\bibinfo {author} {\bibfnamefont {G.}~\bibnamefont
  {Schönhense}}, \bibinfo {author} {\bibfnamefont {K.}~\bibnamefont
  {Medjanik}}, \bibinfo {author} {\bibfnamefont {O.}~\bibnamefont {Fedchenko}},
  \bibinfo {author} {\bibfnamefont {A.}~\bibnamefont {Zymakov{\'{a}}}},
  \bibinfo {author} {\bibfnamefont {S.}~\bibnamefont {Chernov}}, \bibinfo
  {author} {\bibfnamefont {D.}~\bibnamefont {Kutnyakhov}}, \bibinfo {author}
  {\bibfnamefont {D.}~\bibnamefont {Vasilyev}}, \bibinfo {author}
  {\bibfnamefont {S.}~\bibnamefont {Babenkov}}, \bibinfo {author}
  {\bibfnamefont {H.~J.}\ \bibnamefont {Elmers}}, \bibinfo {author}
  {\bibfnamefont {P.}~\bibnamefont {Baumgärtel}}, \bibinfo {author}
  {\bibfnamefont {P.}~\bibnamefont {Goslawski}}, \bibinfo {author}
  {\bibfnamefont {G.}~\bibnamefont {Öhrwall}}, \bibinfo {author}
  {\bibfnamefont {T.}~\bibnamefont {Grunske}}, \bibinfo {author} {\bibfnamefont
  {T.}~\bibnamefont {Kauerhof}}, \bibinfo {author} {\bibfnamefont
  {K.}~\bibnamefont {von Volkmann}}, \bibinfo {author} {\bibfnamefont
  {M.}~\bibnamefont {Kallmayer}}, \bibinfo {author} {\bibfnamefont
  {M.}~\bibnamefont {Ellguth}},\ and\ \bibinfo {author} {\bibfnamefont
  {A.}~\bibnamefont {Oelsner}},\ }\href
  {https://doi.org/10.1107/s1600577521010511} {\bibfield  {journal} {\bibinfo
  {journal} {Journal of Synchrotron Radiation}\ }\textbf {\bibinfo {volume}
  {28}},\ \bibinfo {pages} {1891} (\bibinfo {year} {2021})}\BibitemShut
  {NoStop}%
\bibitem [{SI()}]{SI}%
  \BibitemOpen
  \href@noop {} {\bibinfo  {journal} {{S}ee {S}upplemental {M}aterial at
  [http://...] for experimental details, additional GA spectra, $\Sigma$ valley
  dynamics, and temperature-dependent valley asymmetry}\ }\BibitemShut
  {NoStop}%
\bibitem [{\citenamefont {Kunin}\ \emph {et~al.}()\citenamefont {Kunin},
  \citenamefont {Chernov}, \citenamefont {Bakalis}, \citenamefont {Corder},
  \citenamefont {Zhao}, \citenamefont {White}, \citenamefont {Sch\"onhense},\
  and\ \citenamefont {Allison}}]{Faraday}%
  \BibitemOpen
\bibfield  {journal} {  }\bibfield  {author} {\bibinfo {author} {\bibfnamefont
  {A.}~\bibnamefont {Kunin}}, \bibinfo {author} {\bibfnamefont
  {S.}~\bibnamefont {Chernov}}, \bibinfo {author} {\bibfnamefont
  {J.}~\bibnamefont {Bakalis}}, \bibinfo {author} {\bibfnamefont
  {C.}~\bibnamefont {Corder}}, \bibinfo {author} {\bibfnamefont
  {P.}~\bibnamefont {Zhao}}, \bibinfo {author} {\bibfnamefont {M.~G.}\
  \bibnamefont {White}}, \bibinfo {author} {\bibfnamefont {G.}~\bibnamefont
  {Sch\"onhense}},\ and\ \bibinfo {author} {\bibfnamefont {T.~K.}\ \bibnamefont
  {Allison}},\ }\href@noop {} {\bibinfo  {journal} {In Preparation}\
  }\BibitemShut {NoStop}%
\bibitem [{\citenamefont {Rustagi}\ and\ \citenamefont
  {Kemper}(2018)}]{Rustagi2018}%
  \BibitemOpen
\bibfield  {journal} {  }\bibfield  {author} {\bibinfo {author} {\bibfnamefont
  {A.}~\bibnamefont {Rustagi}}\ and\ \bibinfo {author} {\bibfnamefont {A.~F.}\
  \bibnamefont {Kemper}},\ }\href {https://doi.org/10.1103/physrevb.97.235310}
  {\bibfield  {journal} {\bibinfo  {journal} {Phys. Rev. B}\ }\textbf {\bibinfo
  {volume} {97}},\ \bibinfo {pages} {235310} (\bibinfo {year}
  {2018})}\BibitemShut {NoStop}%
\bibitem [{\citenamefont {Christiansen}\ \emph {et~al.}(2019)\citenamefont
  {Christiansen}, \citenamefont {Selig}, \citenamefont {Malic}, \citenamefont
  {Ernstorfer},\ and\ \citenamefont {Knorr}}]{Christiansen2019}%
  \BibitemOpen
  \bibfield  {author} {\bibinfo {author} {\bibfnamefont {D.}~\bibnamefont
  {Christiansen}}, \bibinfo {author} {\bibfnamefont {M.}~\bibnamefont {Selig}},
  \bibinfo {author} {\bibfnamefont {E.}~\bibnamefont {Malic}}, \bibinfo
  {author} {\bibfnamefont {R.}~\bibnamefont {Ernstorfer}},\ and\ \bibinfo
  {author} {\bibfnamefont {A.}~\bibnamefont {Knorr}},\ }\href
  {https://doi.org/10.1103/physrevb.100.205401} {\bibfield  {journal} {\bibinfo
   {journal} {Physical Review B}\ }\textbf {\bibinfo {volume} {100}},\ \bibinfo
  {pages} {205401} (\bibinfo {year} {2019})}\BibitemShut {NoStop}%
\bibitem [{\citenamefont {Stier}\ \emph {et~al.}(2016)\citenamefont {Stier},
  \citenamefont {McCreary}, \citenamefont {Jonker}, \citenamefont {Kono},\ and\
  \citenamefont {Crooker}}]{Stier2016}%
  \BibitemOpen
  \bibfield  {author} {\bibinfo {author} {\bibfnamefont {A.~V.}\ \bibnamefont
  {Stier}}, \bibinfo {author} {\bibfnamefont {K.~M.}\ \bibnamefont {McCreary}},
  \bibinfo {author} {\bibfnamefont {B.~T.}\ \bibnamefont {Jonker}}, \bibinfo
  {author} {\bibfnamefont {J.}~\bibnamefont {Kono}},\ and\ \bibinfo {author}
  {\bibfnamefont {S.~A.}\ \bibnamefont {Crooker}},\ }\bibfield  {journal}
  {\bibinfo  {journal} {Nature Communications}\ }\textbf {\bibinfo {volume}
  {7}},\ \href {https://doi.org/10.1038/ncomms10643} {10.1038/ncomms10643}
  (\bibinfo {year} {2016})\BibitemShut {NoStop}%
\bibitem [{\citenamefont {Katsch}\ \emph {et~al.}(2019)\citenamefont {Katsch},
  \citenamefont {Selig},\ and\ \citenamefont {Knorr}}]{Katsch2019}%
  \BibitemOpen
  \bibfield  {author} {\bibinfo {author} {\bibfnamefont {F.}~\bibnamefont
  {Katsch}}, \bibinfo {author} {\bibfnamefont {M.}~\bibnamefont {Selig}},\ and\
  \bibinfo {author} {\bibfnamefont {A.}~\bibnamefont {Knorr}},\ }\href
  {https://doi.org/10.1088/2053-1583/ab5407} {\bibfield  {journal} {\bibinfo
  {journal} {2D Materials}\ }\textbf {\bibinfo {volume} {7}},\ \bibinfo {pages}
  {015021} (\bibinfo {year} {2019})}\BibitemShut {NoStop}%
\bibitem [{\citenamefont {Echeverry}\ \emph {et~al.}(2016)\citenamefont
  {Echeverry}, \citenamefont {Urbaszek}, \citenamefont {Amand}, \citenamefont
  {Marie},\ and\ \citenamefont {Gerber}}]{Echeverry2016}%
  \BibitemOpen
  \bibfield  {author} {\bibinfo {author} {\bibfnamefont {J.~P.}\ \bibnamefont
  {Echeverry}}, \bibinfo {author} {\bibfnamefont {B.}~\bibnamefont {Urbaszek}},
  \bibinfo {author} {\bibfnamefont {T.}~\bibnamefont {Amand}}, \bibinfo
  {author} {\bibfnamefont {X.}~\bibnamefont {Marie}},\ and\ \bibinfo {author}
  {\bibfnamefont {I.~C.}\ \bibnamefont {Gerber}},\ }\href
  {https://doi.org/10.1103/physrevb.93.121107} {\bibfield  {journal} {\bibinfo
  {journal} {Physical Review B}\ }\textbf {\bibinfo {volume} {93}},\ \bibinfo
  {pages} {121107} (\bibinfo {year} {2016})}\BibitemShut {NoStop}%
\bibitem [{\citenamefont {Wang}\ \emph {et~al.}(2017)\citenamefont {Wang},
  \citenamefont {Robert}, \citenamefont {Glazov}, \citenamefont {Cadiz},
  \citenamefont {Courtade}, \citenamefont {Amand}, \citenamefont {Lagarde},
  \citenamefont {Taniguchi}, \citenamefont {Watanabe}, \citenamefont
  {Urbaszek},\ and\ \citenamefont {Marie}}]{Wang2017a}%
  \BibitemOpen
  \bibfield  {author} {\bibinfo {author} {\bibfnamefont {G.}~\bibnamefont
  {Wang}}, \bibinfo {author} {\bibfnamefont {C.}~\bibnamefont {Robert}},
  \bibinfo {author} {\bibfnamefont {M.}~\bibnamefont {Glazov}}, \bibinfo
  {author} {\bibfnamefont {F.}~\bibnamefont {Cadiz}}, \bibinfo {author}
  {\bibfnamefont {E.}~\bibnamefont {Courtade}}, \bibinfo {author}
  {\bibfnamefont {T.}~\bibnamefont {Amand}}, \bibinfo {author} {\bibfnamefont
  {D.}~\bibnamefont {Lagarde}}, \bibinfo {author} {\bibfnamefont
  {T.}~\bibnamefont {Taniguchi}}, \bibinfo {author} {\bibfnamefont
  {K.}~\bibnamefont {Watanabe}}, \bibinfo {author} {\bibfnamefont
  {B.}~\bibnamefont {Urbaszek}},\ and\ \bibinfo {author} {\bibfnamefont
  {X.}~\bibnamefont {Marie}},\ }\href
  {https://doi.org/10.1103/physrevlett.119.047401} {\bibfield  {journal}
  {\bibinfo  {journal} {Physical Review Letters}\ }\textbf {\bibinfo {volume}
  {119}},\ \bibinfo {pages} {047401} (\bibinfo {year} {2017})}\BibitemShut
  {NoStop}%
\bibitem [{\citenamefont {van Stokkum}\ \emph {et~al.}(2004)\citenamefont {van
  Stokkum}, \citenamefont {Larsen},\ and\ \citenamefont {van
  Grondelle}}]{Stokkum2004}%
  \BibitemOpen
  \bibfield  {author} {\bibinfo {author} {\bibfnamefont {I.~H.}\ \bibnamefont
  {van Stokkum}}, \bibinfo {author} {\bibfnamefont {D.~S.}\ \bibnamefont
  {Larsen}},\ and\ \bibinfo {author} {\bibfnamefont {R.}~\bibnamefont {van
  Grondelle}},\ }\href {https://doi.org/10.1016/j.bbabio.2004.04.011}
  {\bibfield  {journal} {\bibinfo  {journal} {Biochimica et Biophysica Acta
  ({BBA}) - Bioenergetics}\ }\textbf {\bibinfo {volume} {1657}},\ \bibinfo
  {pages} {82} (\bibinfo {year} {2004})}\BibitemShut {NoStop}%
\bibitem [{\citenamefont {Bisgaard}\ \emph {et~al.}(2009)\citenamefont
  {Bisgaard}, \citenamefont {Clarkin}, \citenamefont {Wu}, \citenamefont {Lee},
  \citenamefont {Ge{\ss}ner}, \citenamefont {Hayden},\ and\ \citenamefont
  {Stolow}}]{Bisgaard2009}%
  \BibitemOpen
  \bibfield  {author} {\bibinfo {author} {\bibfnamefont {C.~Z.}\ \bibnamefont
  {Bisgaard}}, \bibinfo {author} {\bibfnamefont {O.~J.}\ \bibnamefont
  {Clarkin}}, \bibinfo {author} {\bibfnamefont {G.}~\bibnamefont {Wu}},
  \bibinfo {author} {\bibfnamefont {A.~M.~D.}\ \bibnamefont {Lee}}, \bibinfo
  {author} {\bibfnamefont {O.}~\bibnamefont {Ge{\ss}ner}}, \bibinfo {author}
  {\bibfnamefont {C.~C.}\ \bibnamefont {Hayden}},\ and\ \bibinfo {author}
  {\bibfnamefont {A.}~\bibnamefont {Stolow}},\ }\href
  {https://doi.org/10.1126/science.1169183} {\bibfield  {journal} {\bibinfo
  {journal} {Science}\ }\textbf {\bibinfo {volume} {323}},\ \bibinfo {pages}
  {1464} (\bibinfo {year} {2009})}\BibitemShut {NoStop}%
\bibitem [{\citenamefont {Jin}\ \emph {et~al.}(2012)\citenamefont {Jin},
  \citenamefont {Rupp}, \citenamefont {Chevalier}, \citenamefont {Wolf},
  \citenamefont {Rojas}, \citenamefont {Lefkidis}, \citenamefont {Kr{\"u}ger},
  \citenamefont {Diller},\ and\ \citenamefont {H{\"u}bner}}]{Jin2012}%
  \BibitemOpen
  \bibfield  {author} {\bibinfo {author} {\bibfnamefont {W.}~\bibnamefont
  {Jin}}, \bibinfo {author} {\bibfnamefont {F.}~\bibnamefont {Rupp}}, \bibinfo
  {author} {\bibfnamefont {K.}~\bibnamefont {Chevalier}}, \bibinfo {author}
  {\bibfnamefont {M.~M.~N.}\ \bibnamefont {Wolf}}, \bibinfo {author}
  {\bibfnamefont {M.~C.}\ \bibnamefont {Rojas}}, \bibinfo {author}
  {\bibfnamefont {G.}~\bibnamefont {Lefkidis}}, \bibinfo {author}
  {\bibfnamefont {H.-J.}\ \bibnamefont {Kr{\"u}ger}}, \bibinfo {author}
  {\bibfnamefont {R.}~\bibnamefont {Diller}},\ and\ \bibinfo {author}
  {\bibfnamefont {W.}~\bibnamefont {H{\"u}bner}},\ }\href
  {https://doi.org/10.1103/physrevlett.109.267209} {\bibfield  {journal}
  {\bibinfo  {journal} {Physical Review Letters}\ }\textbf {\bibinfo {volume}
  {109}},\ \bibinfo {pages} {267209} (\bibinfo {year} {2012})}\BibitemShut
  {NoStop}%
\bibitem [{\citenamefont {Beckwith}\ \emph {et~al.}(2020)\citenamefont
  {Beckwith}, \citenamefont {Rumble},\ and\ \citenamefont
  {Vauthey}}]{Beckwith2020}%
  \BibitemOpen
  \bibfield  {author} {\bibinfo {author} {\bibfnamefont {J.~S.}\ \bibnamefont
  {Beckwith}}, \bibinfo {author} {\bibfnamefont {C.~A.}\ \bibnamefont
  {Rumble}},\ and\ \bibinfo {author} {\bibfnamefont {E.}~\bibnamefont
  {Vauthey}},\ }\href {https://doi.org/10.1080/0144235x.2020.1757942}
  {\bibfield  {journal} {\bibinfo  {journal} {International Reviews in Physical
  Chemistry}\ }\textbf {\bibinfo {volume} {39}},\ \bibinfo {pages} {135}
  (\bibinfo {year} {2020})}\BibitemShut {NoStop}%
\bibitem [{\citenamefont {Rebholz}\ \emph {et~al.}(2021)\citenamefont
  {Rebholz}, \citenamefont {Ding}, \citenamefont {Despr{\'{e}}}, \citenamefont
  {Aufleger}, \citenamefont {Hartmann}, \citenamefont {Meyer}, \citenamefont
  {Stoo{\ss}}, \citenamefont {Magunia}, \citenamefont {Wachs}, \citenamefont
  {Birk}, \citenamefont {Mi}, \citenamefont {Borisova}, \citenamefont
  {da~Costa~Castanheira}, \citenamefont {Rupprecht}, \citenamefont {Schmid},
  \citenamefont {Schnorr}, \citenamefont {Schr{\"o}ter}, \citenamefont
  {Moshammer}, \citenamefont {Loh}, \citenamefont {Attar}, \citenamefont
  {Leone}, \citenamefont {Gaumnitz}, \citenamefont {W{\"o}rner}, \citenamefont
  {Roling}, \citenamefont {Butz}, \citenamefont {Zacharias}, \citenamefont
  {D{\"u}sterer}, \citenamefont {Treusch}, \citenamefont {Brenner},
  \citenamefont {Vester}, \citenamefont {Kuleff}, \citenamefont {Ott},\ and\
  \citenamefont {Pfeifer}}]{Rebholz2021}%
  \BibitemOpen
  \bibfield  {author} {\bibinfo {author} {\bibfnamefont {M.}~\bibnamefont
  {Rebholz}}, \bibinfo {author} {\bibfnamefont {T.}~\bibnamefont {Ding}},
  \bibinfo {author} {\bibfnamefont {V.}~\bibnamefont {Despr{\'{e}}}}, \bibinfo
  {author} {\bibfnamefont {L.}~\bibnamefont {Aufleger}}, \bibinfo {author}
  {\bibfnamefont {M.}~\bibnamefont {Hartmann}}, \bibinfo {author}
  {\bibfnamefont {K.}~\bibnamefont {Meyer}}, \bibinfo {author} {\bibfnamefont
  {V.}~\bibnamefont {Stoo{\ss}}}, \bibinfo {author} {\bibfnamefont
  {A.}~\bibnamefont {Magunia}}, \bibinfo {author} {\bibfnamefont
  {D.}~\bibnamefont {Wachs}}, \bibinfo {author} {\bibfnamefont
  {P.}~\bibnamefont {Birk}}, \bibinfo {author} {\bibfnamefont {Y.}~\bibnamefont
  {Mi}}, \bibinfo {author} {\bibfnamefont {G.~D.}\ \bibnamefont {Borisova}},
  \bibinfo {author} {\bibfnamefont {C.}~\bibnamefont {da~Costa~Castanheira}},
  \bibinfo {author} {\bibfnamefont {P.}~\bibnamefont {Rupprecht}}, \bibinfo
  {author} {\bibfnamefont {G.}~\bibnamefont {Schmid}}, \bibinfo {author}
  {\bibfnamefont {K.}~\bibnamefont {Schnorr}}, \bibinfo {author} {\bibfnamefont
  {C.~D.}\ \bibnamefont {Schr{\"o}ter}}, \bibinfo {author} {\bibfnamefont
  {R.}~\bibnamefont {Moshammer}}, \bibinfo {author} {\bibfnamefont {Z.-H.}\
  \bibnamefont {Loh}}, \bibinfo {author} {\bibfnamefont {A.~R.}\ \bibnamefont
  {Attar}}, \bibinfo {author} {\bibfnamefont {S.~R.}\ \bibnamefont {Leone}},
  \bibinfo {author} {\bibfnamefont {T.}~\bibnamefont {Gaumnitz}}, \bibinfo
  {author} {\bibfnamefont {H.~J.}\ \bibnamefont {W{\"o}rner}}, \bibinfo
  {author} {\bibfnamefont {S.}~\bibnamefont {Roling}}, \bibinfo {author}
  {\bibfnamefont {M.}~\bibnamefont {Butz}}, \bibinfo {author} {\bibfnamefont
  {H.}~\bibnamefont {Zacharias}}, \bibinfo {author} {\bibfnamefont
  {S.}~\bibnamefont {D{\"u}sterer}}, \bibinfo {author} {\bibfnamefont
  {R.}~\bibnamefont {Treusch}}, \bibinfo {author} {\bibfnamefont
  {G.}~\bibnamefont {Brenner}}, \bibinfo {author} {\bibfnamefont
  {J.}~\bibnamefont {Vester}}, \bibinfo {author} {\bibfnamefont {A.~I.}\
  \bibnamefont {Kuleff}}, \bibinfo {author} {\bibfnamefont {C.}~\bibnamefont
  {Ott}},\ and\ \bibinfo {author} {\bibfnamefont {T.}~\bibnamefont {Pfeifer}},\
  }\href {https://doi.org/10.1103/physrevx.11.031001} {\bibfield  {journal}
  {\bibinfo  {journal} {Physical Review X}\ }\textbf {\bibinfo {volume} {11}},\
  \bibinfo {pages} {031001} (\bibinfo {year} {2021})}\BibitemShut {NoStop}%
\bibitem [{\citenamefont {Selig}\ \emph {et~al.}(2019)\citenamefont {Selig},
  \citenamefont {Katsch}, \citenamefont {Schmidt}, \citenamefont
  {de~Vasconcellos}, \citenamefont {Bratschitsch}, \citenamefont {Malic},\ and\
  \citenamefont {Knorr}}]{Selig2019}%
  \BibitemOpen
  \bibfield  {author} {\bibinfo {author} {\bibfnamefont {M.}~\bibnamefont
  {Selig}}, \bibinfo {author} {\bibfnamefont {F.}~\bibnamefont {Katsch}},
  \bibinfo {author} {\bibfnamefont {R.}~\bibnamefont {Schmidt}}, \bibinfo
  {author} {\bibfnamefont {S.~M.}\ \bibnamefont {de~Vasconcellos}}, \bibinfo
  {author} {\bibfnamefont {R.}~\bibnamefont {Bratschitsch}}, \bibinfo {author}
  {\bibfnamefont {E.}~\bibnamefont {Malic}},\ and\ \bibinfo {author}
  {\bibfnamefont {A.}~\bibnamefont {Knorr}},\ }\href
  {https://doi.org/10.1103/physrevresearch.1.022007} {\bibfield  {journal}
  {\bibinfo  {journal} {Physical Review Research}\ }\textbf {\bibinfo {volume}
  {1}},\ \bibinfo {pages} {022007} (\bibinfo {year} {2019})}\BibitemShut
  {NoStop}%
\bibitem [{\citenamefont {Bergh\"auser}\ \emph {et~al.}(2018)\citenamefont
  {Bergh\"auser}, \citenamefont {Bernal-Villamil}, \citenamefont {Schmidt},
  \citenamefont {Schneider}, \citenamefont {Niehues}, \citenamefont {Erhart},
  \citenamefont {de~Vasconcellos}, \citenamefont {Bratschitsch}, \citenamefont
  {Knorr},\ and\ \citenamefont {Malic}}]{Berghaeuser2018}%
  \BibitemOpen
  \bibfield  {author} {\bibinfo {author} {\bibfnamefont {G.}~\bibnamefont
  {Bergh\"auser}}, \bibinfo {author} {\bibfnamefont {I.}~\bibnamefont
  {Bernal-Villamil}}, \bibinfo {author} {\bibfnamefont {R.}~\bibnamefont
  {Schmidt}}, \bibinfo {author} {\bibfnamefont {R.}~\bibnamefont {Schneider}},
  \bibinfo {author} {\bibfnamefont {I.}~\bibnamefont {Niehues}}, \bibinfo
  {author} {\bibfnamefont {P.}~\bibnamefont {Erhart}}, \bibinfo {author}
  {\bibfnamefont {S.~M.}\ \bibnamefont {de~Vasconcellos}}, \bibinfo {author}
  {\bibfnamefont {R.}~\bibnamefont {Bratschitsch}}, \bibinfo {author}
  {\bibfnamefont {A.}~\bibnamefont {Knorr}},\ and\ \bibinfo {author}
  {\bibfnamefont {E.}~\bibnamefont {Malic}},\ }\bibfield  {journal} {\bibinfo
  {journal} {Nature Communications}\ }\textbf {\bibinfo {volume} {9}},\ \href
  {https://doi.org/10.1038/s41467-018-03354-1} {10.1038/s41467-018-03354-1}
  (\bibinfo {year} {2018})\BibitemShut {NoStop}%
\bibitem [{\citenamefont {Bernal-Villamil}\ \emph {et~al.}(2018)\citenamefont
  {Bernal-Villamil}, \citenamefont {Bergh\"auser}, \citenamefont {Selig},
  \citenamefont {Niehues}, \citenamefont {Schmidt}, \citenamefont {Schneider},
  \citenamefont {Tonndorf}, \citenamefont {Erhart}, \citenamefont
  {de~Vasconcellos}, \citenamefont {Bratschitsch}, \citenamefont {Knorr},\ and\
  \citenamefont {Malic}}]{BernalVillamil2018}%
  \BibitemOpen
  \bibfield  {author} {\bibinfo {author} {\bibfnamefont {I.}~\bibnamefont
  {Bernal-Villamil}}, \bibinfo {author} {\bibfnamefont {G.}~\bibnamefont
  {Bergh\"auser}}, \bibinfo {author} {\bibfnamefont {M.}~\bibnamefont {Selig}},
  \bibinfo {author} {\bibfnamefont {I.}~\bibnamefont {Niehues}}, \bibinfo
  {author} {\bibfnamefont {R.}~\bibnamefont {Schmidt}}, \bibinfo {author}
  {\bibfnamefont {R.}~\bibnamefont {Schneider}}, \bibinfo {author}
  {\bibfnamefont {P.}~\bibnamefont {Tonndorf}}, \bibinfo {author}
  {\bibfnamefont {P.}~\bibnamefont {Erhart}}, \bibinfo {author} {\bibfnamefont
  {S.~M.}\ \bibnamefont {de~Vasconcellos}}, \bibinfo {author} {\bibfnamefont
  {R.}~\bibnamefont {Bratschitsch}}, \bibinfo {author} {\bibfnamefont
  {A.}~\bibnamefont {Knorr}},\ and\ \bibinfo {author} {\bibfnamefont
  {E.}~\bibnamefont {Malic}},\ }\href
  {https://doi.org/10.1088/2053-1583/aaaa8b} {\bibfield  {journal} {\bibinfo
  {journal} {2D Mater.}\ }\textbf {\bibinfo {volume} {5}},\ \bibinfo {pages}
  {025011} (\bibinfo {year} {2018})}\BibitemShut {NoStop}%
\bibitem [{\citenamefont {Chen}\ \emph {et~al.}(2018)\citenamefont {Chen},
  \citenamefont {Goldstein}, \citenamefont {Tong}, \citenamefont {Taniguchi},
  \citenamefont {Watanabe},\ and\ \citenamefont {Yan}}]{Chen2018}%
  \BibitemOpen
  \bibfield  {author} {\bibinfo {author} {\bibfnamefont {S.-Y.}\ \bibnamefont
  {Chen}}, \bibinfo {author} {\bibfnamefont {T.}~\bibnamefont {Goldstein}},
  \bibinfo {author} {\bibfnamefont {J.}~\bibnamefont {Tong}}, \bibinfo {author}
  {\bibfnamefont {T.}~\bibnamefont {Taniguchi}}, \bibinfo {author}
  {\bibfnamefont {K.}~\bibnamefont {Watanabe}},\ and\ \bibinfo {author}
  {\bibfnamefont {J.}~\bibnamefont {Yan}},\ }\href
  {https://doi.org/10.1103/physrevlett.120.046402} {\bibfield  {journal}
  {\bibinfo  {journal} {Physical Review Letters}\ }\textbf {\bibinfo {volume}
  {120}},\ \bibinfo {pages} {046402} (\bibinfo {year} {2018})}\BibitemShut
  {NoStop}%
\end{thebibliography}%


\begin{thebibliography}{20}%
\makeatletter
\providecommand \@ifxundefined [1]{%
 \@ifx{#1\undefined}
}%
\providecommand \@ifnum [1]{%
 \ifnum #1\expandafter \@firstoftwo
 \else \expandafter \@secondoftwo
 \fi
}%
\providecommand \@ifx [1]{%
 \ifx #1\expandafter \@firstoftwo
 \else \expandafter \@secondoftwo
 \fi
}%
\providecommand \natexlab [1]{#1}%
\providecommand \enquote  [1]{``#1''}%
\providecommand \bibnamefont  [1]{#1}%
\providecommand \bibfnamefont [1]{#1}%
\providecommand \citenamefont [1]{#1}%
\providecommand \href@noop [0]{\@secondoftwo}%
\providecommand \href [0]{\begingroup \@sanitize@url \@href}%
\providecommand \@href[1]{\@@startlink{#1}\@@href}%
\providecommand \@@href[1]{\endgroup#1\@@endlink}%
\providecommand \@sanitize@url [0]{\catcode `\\12\catcode `\$12\catcode
  `\&12\catcode `\#12\catcode `\^12\catcode `\_12\catcode `\%12\relax}%
\providecommand \@@startlink[1]{}%
\providecommand \@@endlink[0]{}%
\providecommand \url  [0]{\begingroup\@sanitize@url \@url }%
\providecommand \@url [1]{\endgroup\@href {#1}{\urlprefix }}%
\providecommand \urlprefix  [0]{URL }%
\providecommand \Eprint [0]{\href }%
\providecommand \doibase [0]{https://doi.org/}%
\providecommand \selectlanguage [0]{\@gobble}%
\providecommand \bibinfo  [0]{\@secondoftwo}%
\providecommand \bibfield  [0]{\@secondoftwo}%
\providecommand \translation [1]{[#1]}%
\providecommand \BibitemOpen [0]{}%
\providecommand \bibitemStop [0]{}%
\providecommand \bibitemNoStop [0]{.\EOS\space}%
\providecommand \EOS [0]{\spacefactor3000\relax}%
\providecommand \BibitemShut  [1]{\csname bibitem#1\endcsname}%
\let\auto@bib@innerbib\@empty
\bibitem [{\citenamefont {Berkdemir}\ \emph {et~al.}(2013)\citenamefont
  {Berkdemir}, \citenamefont {Guti{\'{e}}rrez}, \citenamefont
  {Botello-M{\'{e}}ndez}, \citenamefont {Perea-L{\'{o}}pez}, \citenamefont
  {El{\'{\i}}as}, \citenamefont {Chia}, \citenamefont {Wang}, \citenamefont
  {Crespi}, \citenamefont {L{\'{o}}pez-Ur{\'{\i}}as}, \citenamefont {Charlier},
  \citenamefont {Terrones},\ and\ \citenamefont {Terrones}}]{Berkdemir2013}%
  \BibitemOpen
  \bibfield  {author} {\bibinfo {author} {\bibfnamefont {A.}~\bibnamefont
  {Berkdemir}}, \bibinfo {author} {\bibfnamefont {H.~R.}\ \bibnamefont
  {Guti{\'{e}}rrez}}, \bibinfo {author} {\bibfnamefont {A.~R.}\ \bibnamefont
  {Botello-M{\'{e}}ndez}}, \bibinfo {author} {\bibfnamefont {N.}~\bibnamefont
  {Perea-L{\'{o}}pez}}, \bibinfo {author} {\bibfnamefont {A.~L.}\ \bibnamefont
  {El{\'{\i}}as}}, \bibinfo {author} {\bibfnamefont {C.-I.}\ \bibnamefont
  {Chia}}, \bibinfo {author} {\bibfnamefont {B.}~\bibnamefont {Wang}}, \bibinfo
  {author} {\bibfnamefont {V.~H.}\ \bibnamefont {Crespi}}, \bibinfo {author}
  {\bibfnamefont {F.}~\bibnamefont {L{\'{o}}pez-Ur{\'{\i}}as}}, \bibinfo
  {author} {\bibfnamefont {J.-C.}\ \bibnamefont {Charlier}}, \bibinfo {author}
  {\bibfnamefont {H.}~\bibnamefont {Terrones}},\ and\ \bibinfo {author}
  {\bibfnamefont {M.}~\bibnamefont {Terrones}},\ }\bibfield  {journal}
  {\bibinfo  {journal} {Scientific Reports}\ }\textbf {\bibinfo {volume} {3}},\
  \href {https://doi.org/10.1038/srep01755} {10.1038/srep01755} (\bibinfo
  {year} {2013})\BibitemShut {NoStop}%
\bibitem [{\citenamefont {Saito}\ \emph {et~al.}(2016)\citenamefont {Saito},
  \citenamefont {Tatsumi}, \citenamefont {Huang}, \citenamefont {Ling},\ and\
  \citenamefont {Dresselhaus}}]{Saito2016}%
  \BibitemOpen
  \bibfield  {author} {\bibinfo {author} {\bibfnamefont {R.}~\bibnamefont
  {Saito}}, \bibinfo {author} {\bibfnamefont {Y.}~\bibnamefont {Tatsumi}},
  \bibinfo {author} {\bibfnamefont {S.}~\bibnamefont {Huang}}, \bibinfo
  {author} {\bibfnamefont {X.}~\bibnamefont {Ling}},\ and\ \bibinfo {author}
  {\bibfnamefont {M.~S.}\ \bibnamefont {Dresselhaus}},\ }\href
  {https://doi.org/10.1088/0953-8984/28/35/353002} {\bibfield  {journal}
  {\bibinfo  {journal} {Journal of Physics: Condensed Matter}\ }\textbf
  {\bibinfo {volume} {28}},\ \bibinfo {pages} {353002} (\bibinfo {year}
  {2016})}\BibitemShut {NoStop}%
\bibitem [{\citenamefont {Man}\ \emph {et~al.}(2016)\citenamefont {Man},
  \citenamefont {Deckoff-Jones}, \citenamefont {Winchester}, \citenamefont
  {Shi}, \citenamefont {Gupta}, \citenamefont {Mohite}, \citenamefont {Kar},
  \citenamefont {Kioupakis}, \citenamefont {Talapatra},\ and\ \citenamefont
  {Dani}}]{Man2016}%
  \BibitemOpen
  \bibfield  {author} {\bibinfo {author} {\bibfnamefont {M.~K.~L.}\
  \bibnamefont {Man}}, \bibinfo {author} {\bibfnamefont {S.}~\bibnamefont
  {Deckoff-Jones}}, \bibinfo {author} {\bibfnamefont {A.}~\bibnamefont
  {Winchester}}, \bibinfo {author} {\bibfnamefont {G.}~\bibnamefont {Shi}},
  \bibinfo {author} {\bibfnamefont {G.}~\bibnamefont {Gupta}}, \bibinfo
  {author} {\bibfnamefont {A.~D.}\ \bibnamefont {Mohite}}, \bibinfo {author}
  {\bibfnamefont {S.}~\bibnamefont {Kar}}, \bibinfo {author} {\bibfnamefont
  {E.}~\bibnamefont {Kioupakis}}, \bibinfo {author} {\bibfnamefont
  {S.}~\bibnamefont {Talapatra}},\ and\ \bibinfo {author} {\bibfnamefont
  {K.~M.}\ \bibnamefont {Dani}},\ }\bibfield  {journal} {\bibinfo  {journal}
  {Scientific Reports}\ }\textbf {\bibinfo {volume} {6}},\ \href
  {https://doi.org/10.1038/srep20890} {10.1038/srep20890} (\bibinfo {year}
  {2016})\BibitemShut {NoStop}%
\bibitem [{\citenamefont {Li}\ \emph {et~al.}(2016)\citenamefont {Li},
  \citenamefont {Reber}, \citenamefont {Corder}, \citenamefont {Chen},
  \citenamefont {Zhao},\ and\ \citenamefont {Allison}}]{Li_RSI2016}%
  \BibitemOpen
  \bibfield  {author} {\bibinfo {author} {\bibfnamefont {X.}~\bibnamefont
  {Li}}, \bibinfo {author} {\bibfnamefont {M.~A.~R.}\ \bibnamefont {Reber}},
  \bibinfo {author} {\bibfnamefont {C.}~\bibnamefont {Corder}}, \bibinfo
  {author} {\bibfnamefont {Y.}~\bibnamefont {Chen}}, \bibinfo {author}
  {\bibfnamefont {P.}~\bibnamefont {Zhao}},\ and\ \bibinfo {author}
  {\bibfnamefont {T.~K.}\ \bibnamefont {Allison}},\ }\href
  {https://doi.org/http://dx.doi.org/10.1063/1.4962867} {\bibfield  {journal}
  {\bibinfo  {journal} {Review of Scientific Instruments}\ }\textbf {\bibinfo
  {volume} {87}},\ \bibinfo {eid} {093114} (\bibinfo {year}
  {2016})}\BibitemShut {NoStop}%
\bibitem [{\citenamefont {Corder}\ \emph
  {et~al.}(2018{\natexlab{a}})\citenamefont {Corder}, \citenamefont {Zhao},
  \citenamefont {Li}, \citenamefont {Kershis}, \citenamefont {White},\ and\
  \citenamefont {Allison}}]{Corder2018a}%
  \BibitemOpen
  \bibfield  {author} {\bibinfo {author} {\bibfnamefont {C.}~\bibnamefont
  {Corder}}, \bibinfo {author} {\bibfnamefont {P.}~\bibnamefont {Zhao}},
  \bibinfo {author} {\bibfnamefont {X.}~\bibnamefont {Li}}, \bibinfo {author}
  {\bibfnamefont {M.~D.}\ \bibnamefont {Kershis}}, \bibinfo {author}
  {\bibfnamefont {M.~G.}\ \bibnamefont {White}},\ and\ \bibinfo {author}
  {\bibfnamefont {T.~K.}\ \bibnamefont {Allison}},\ }in\ \href
  {https://doi.org/10.1117/12.2295232} {\emph {\bibinfo {booktitle} {{Laser
  Applications in Microelectronic and Optoelectronic Manufacturing} ({LAMOM})
  {XXIII}}}},\ \bibinfo {editor} {edited by\ \bibinfo {editor} {\bibfnamefont
  {B.}~\bibnamefont {Neuenschwander}}, \bibinfo {editor} {\bibfnamefont
  {G.}~\bibnamefont {Ra{\v{c}}iukaitis}}, \bibinfo {editor} {\bibfnamefont
  {T.}~\bibnamefont {Makimura}},\ and\ \bibinfo {editor} {\bibfnamefont
  {C.~P.}\ \bibnamefont {Grigoropoulos}}}\ (\bibinfo  {publisher} {{SPIE}},\
  \bibinfo {year} {2018})\BibitemShut {NoStop}%
\bibitem [{\citenamefont {Corder}\ \emph
  {et~al.}(2018{\natexlab{b}})\citenamefont {Corder}, \citenamefont {Zhao},
  \citenamefont {Bakalis}, \citenamefont {Li}, \citenamefont {Kershis},
  \citenamefont {Muraca}, \citenamefont {White},\ and\ \citenamefont
  {Allison}}]{Corder2018}%
  \BibitemOpen
  \bibfield  {author} {\bibinfo {author} {\bibfnamefont {C.}~\bibnamefont
  {Corder}}, \bibinfo {author} {\bibfnamefont {P.}~\bibnamefont {Zhao}},
  \bibinfo {author} {\bibfnamefont {J.}~\bibnamefont {Bakalis}}, \bibinfo
  {author} {\bibfnamefont {X.}~\bibnamefont {Li}}, \bibinfo {author}
  {\bibfnamefont {M.~D.}\ \bibnamefont {Kershis}}, \bibinfo {author}
  {\bibfnamefont {A.~R.}\ \bibnamefont {Muraca}}, \bibinfo {author}
  {\bibfnamefont {M.~G.}\ \bibnamefont {White}},\ and\ \bibinfo {author}
  {\bibfnamefont {T.~K.}\ \bibnamefont {Allison}},\ }\href
  {https://doi.org/10.1063/1.5045578} {\bibfield  {journal} {\bibinfo
  {journal} {Struct. Dyn.}\ }\textbf {\bibinfo {volume} {5}},\ \bibinfo {pages}
  {054301} (\bibinfo {year} {2018}{\natexlab{b}})}\BibitemShut {NoStop}%
\bibitem [{\citenamefont {Chernov}\ \emph {et~al.}(2015)\citenamefont
  {Chernov}, \citenamefont {Medjanik}, \citenamefont {Tusche}, \citenamefont
  {Kutnyakhov}, \citenamefont {Nepijko}, \citenamefont {Oelsner}, \citenamefont
  {Braun}, \citenamefont {Min{\'{a}}r}, \citenamefont {Borek}, \citenamefont
  {Ebert}, \citenamefont {Elmers}, \citenamefont {Kirschner},\ and\
  \citenamefont {Sch{\"o}nhense}}]{Chernov2015}%
  \BibitemOpen
  \bibfield  {author} {\bibinfo {author} {\bibfnamefont {S.}~\bibnamefont
  {Chernov}}, \bibinfo {author} {\bibfnamefont {K.}~\bibnamefont {Medjanik}},
  \bibinfo {author} {\bibfnamefont {C.}~\bibnamefont {Tusche}}, \bibinfo
  {author} {\bibfnamefont {D.}~\bibnamefont {Kutnyakhov}}, \bibinfo {author}
  {\bibfnamefont {S.}~\bibnamefont {Nepijko}}, \bibinfo {author} {\bibfnamefont
  {A.}~\bibnamefont {Oelsner}}, \bibinfo {author} {\bibfnamefont
  {J.}~\bibnamefont {Braun}}, \bibinfo {author} {\bibfnamefont
  {J.}~\bibnamefont {Min{\'{a}}r}}, \bibinfo {author} {\bibfnamefont
  {S.}~\bibnamefont {Borek}}, \bibinfo {author} {\bibfnamefont
  {H.}~\bibnamefont {Ebert}}, \bibinfo {author} {\bibfnamefont
  {H.}~\bibnamefont {Elmers}}, \bibinfo {author} {\bibfnamefont
  {J.}~\bibnamefont {Kirschner}},\ and\ \bibinfo {author} {\bibfnamefont
  {G.}~\bibnamefont {Sch{\"o}nhense}},\ }\href
  {https://doi.org/10.1016/j.ultramic.2015.07.008} {\bibfield  {journal}
  {\bibinfo  {journal} {Ultramicroscopy}\ }\textbf {\bibinfo {volume} {159}},\
  \bibinfo {pages} {453} (\bibinfo {year} {2015})}\BibitemShut {NoStop}%
\bibitem [{\citenamefont {Medjanik}\ \emph {et~al.}(2017)\citenamefont
  {Medjanik}, \citenamefont {Fedchenko}, \citenamefont {Chernov}, \citenamefont
  {Kutnyakhov}, \citenamefont {Ellguth}, \citenamefont {Oelsner}, \citenamefont
  {Sch\"onhense}, \citenamefont {Peixoto}, \citenamefont {Lutz}, \citenamefont
  {Min}, \citenamefont {Reinert}, \citenamefont {D\"aster}, \citenamefont
  {Acremann}, \citenamefont {Viefhaus}, \citenamefont {Wurth}, \citenamefont
  {Elmers},\ and\ \citenamefont {Sch\"onhense}}]{Medjanik2017}%
  \BibitemOpen
  \bibfield  {author} {\bibinfo {author} {\bibfnamefont {K.}~\bibnamefont
  {Medjanik}}, \bibinfo {author} {\bibfnamefont {O.}~\bibnamefont {Fedchenko}},
  \bibinfo {author} {\bibfnamefont {S.}~\bibnamefont {Chernov}}, \bibinfo
  {author} {\bibfnamefont {D.}~\bibnamefont {Kutnyakhov}}, \bibinfo {author}
  {\bibfnamefont {M.}~\bibnamefont {Ellguth}}, \bibinfo {author} {\bibfnamefont
  {A.}~\bibnamefont {Oelsner}}, \bibinfo {author} {\bibfnamefont
  {B.}~\bibnamefont {Sch\"onhense}}, \bibinfo {author} {\bibfnamefont
  {T.~R.~F.}\ \bibnamefont {Peixoto}}, \bibinfo {author} {\bibfnamefont
  {P.}~\bibnamefont {Lutz}}, \bibinfo {author} {\bibfnamefont {C.-H.}\
  \bibnamefont {Min}}, \bibinfo {author} {\bibfnamefont {F.}~\bibnamefont
  {Reinert}}, \bibinfo {author} {\bibfnamefont {S.}~\bibnamefont {D\"aster}},
  \bibinfo {author} {\bibfnamefont {Y.}~\bibnamefont {Acremann}}, \bibinfo
  {author} {\bibfnamefont {J.}~\bibnamefont {Viefhaus}}, \bibinfo {author}
  {\bibfnamefont {W.}~\bibnamefont {Wurth}}, \bibinfo {author} {\bibfnamefont
  {H.~J.}\ \bibnamefont {Elmers}},\ and\ \bibinfo {author} {\bibfnamefont
  {G.}~\bibnamefont {Sch\"onhense}},\ }\href {https://doi.org/10.1038/nmat4875}
  {\bibfield  {journal} {\bibinfo  {journal} {Nat. Mater.}\ }\textbf {\bibinfo
  {volume} {16}},\ \bibinfo {pages} {615} (\bibinfo {year} {2017})}\BibitemShut
  {NoStop}%
\bibitem [{\citenamefont {Mad{\'{e}}o}\ \emph {et~al.}(2020)\citenamefont
  {Mad{\'{e}}o}, \citenamefont {Man}, \citenamefont {Sahoo}, \citenamefont
  {Campbell}, \citenamefont {Pareek}, \citenamefont {Wong}, \citenamefont
  {Al-Mahboob}, \citenamefont {Chan}, \citenamefont {Karmakar}, \citenamefont
  {Mariserla}, \citenamefont {Li}, \citenamefont {Heinz}, \citenamefont {Cao},\
  and\ \citenamefont {Dani}}]{Madeo2020}%
  \BibitemOpen
  \bibfield  {author} {\bibinfo {author} {\bibfnamefont {J.}~\bibnamefont
  {Mad{\'{e}}o}}, \bibinfo {author} {\bibfnamefont {M.~K.~L.}\ \bibnamefont
  {Man}}, \bibinfo {author} {\bibfnamefont {C.}~\bibnamefont {Sahoo}}, \bibinfo
  {author} {\bibfnamefont {M.}~\bibnamefont {Campbell}}, \bibinfo {author}
  {\bibfnamefont {V.}~\bibnamefont {Pareek}}, \bibinfo {author} {\bibfnamefont
  {E.~L.}\ \bibnamefont {Wong}}, \bibinfo {author} {\bibfnamefont
  {A.}~\bibnamefont {Al-Mahboob}}, \bibinfo {author} {\bibfnamefont {N.~S.}\
  \bibnamefont {Chan}}, \bibinfo {author} {\bibfnamefont {A.}~\bibnamefont
  {Karmakar}}, \bibinfo {author} {\bibfnamefont {B.~M.~K.}\ \bibnamefont
  {Mariserla}}, \bibinfo {author} {\bibfnamefont {X.}~\bibnamefont {Li}},
  \bibinfo {author} {\bibfnamefont {T.~F.}\ \bibnamefont {Heinz}}, \bibinfo
  {author} {\bibfnamefont {T.}~\bibnamefont {Cao}},\ and\ \bibinfo {author}
  {\bibfnamefont {K.~M.}\ \bibnamefont {Dani}},\ }\href
  {https://doi.org/10.1126/science.aba1029} {\bibfield  {journal} {\bibinfo
  {journal} {Science}\ }\textbf {\bibinfo {volume} {370}},\ \bibinfo {pages}
  {1199} (\bibinfo {year} {2020})}\BibitemShut {NoStop}%
\bibitem [{\citenamefont {Wallauer}\ \emph {et~al.}(2021)\citenamefont
  {Wallauer}, \citenamefont {Perea-Causin}, \citenamefont {M\"unster},
  \citenamefont {Zajusch}, \citenamefont {Brem}, \citenamefont {G\"udde},
  \citenamefont {Tanimura}, \citenamefont {Lin}, \citenamefont {Huber},
  \citenamefont {Malic},\ and\ \citenamefont {H\"ofer}}]{Wallauer2021a}%
  \BibitemOpen
  \bibfield  {author} {\bibinfo {author} {\bibfnamefont {R.}~\bibnamefont
  {Wallauer}}, \bibinfo {author} {\bibfnamefont {R.}~\bibnamefont
  {Perea-Causin}}, \bibinfo {author} {\bibfnamefont {L.}~\bibnamefont
  {M\"unster}}, \bibinfo {author} {\bibfnamefont {S.}~\bibnamefont {Zajusch}},
  \bibinfo {author} {\bibfnamefont {S.}~\bibnamefont {Brem}}, \bibinfo {author}
  {\bibfnamefont {J.}~\bibnamefont {G\"udde}}, \bibinfo {author} {\bibfnamefont
  {K.}~\bibnamefont {Tanimura}}, \bibinfo {author} {\bibfnamefont {K.-Q.}\
  \bibnamefont {Lin}}, \bibinfo {author} {\bibfnamefont {R.}~\bibnamefont
  {Huber}}, \bibinfo {author} {\bibfnamefont {E.}~\bibnamefont {Malic}},\ and\
  \bibinfo {author} {\bibfnamefont {U.}~\bibnamefont {H\"ofer}},\ }\href
  {https://doi.org/10.1021/acs.nanolett.1c01839} {\bibfield  {journal}
  {\bibinfo  {journal} {Nano Letters}\ }\textbf {\bibinfo {volume} {21}},\
  \bibinfo {pages} {5867} (\bibinfo {year} {2021})}\BibitemShut {NoStop}%
\bibitem [{\citenamefont {Guo}\ \emph {et~al.}(2018)\citenamefont {Guo},
  \citenamefont {Wu}, \citenamefont {Cao}, \citenamefont {Monahan},
  \citenamefont {Lee}, \citenamefont {Louie},\ and\ \citenamefont
  {Fleming}}]{Guo2018}%
  \BibitemOpen
  \bibfield  {author} {\bibinfo {author} {\bibfnamefont {L.}~\bibnamefont
  {Guo}}, \bibinfo {author} {\bibfnamefont {M.}~\bibnamefont {Wu}}, \bibinfo
  {author} {\bibfnamefont {T.}~\bibnamefont {Cao}}, \bibinfo {author}
  {\bibfnamefont {D.~M.}\ \bibnamefont {Monahan}}, \bibinfo {author}
  {\bibfnamefont {Y.-H.}\ \bibnamefont {Lee}}, \bibinfo {author} {\bibfnamefont
  {S.~G.}\ \bibnamefont {Louie}},\ and\ \bibinfo {author} {\bibfnamefont
  {G.~R.}\ \bibnamefont {Fleming}},\ }\href
  {https://doi.org/10.1038/s41567-018-0362-y} {\bibfield  {journal} {\bibinfo
  {journal} {Nature Physics}\ }\textbf {\bibinfo {volume} {15}},\ \bibinfo
  {pages} {228} (\bibinfo {year} {2018})}\BibitemShut {NoStop}%
\bibitem [{\citenamefont {Li}\ \emph {et~al.}(2014)\citenamefont {Li},
  \citenamefont {Chernikov}, \citenamefont {Zhang}, \citenamefont {Rigosi},
  \citenamefont {Hill}, \citenamefont {van~der Zande}, \citenamefont {Chenet},
  \citenamefont {Shih}, \citenamefont {Hone},\ and\ \citenamefont
  {Heinz}}]{Li2014a}%
  \BibitemOpen
  \bibfield  {author} {\bibinfo {author} {\bibfnamefont {Y.}~\bibnamefont
  {Li}}, \bibinfo {author} {\bibfnamefont {A.}~\bibnamefont {Chernikov}},
  \bibinfo {author} {\bibfnamefont {X.}~\bibnamefont {Zhang}}, \bibinfo
  {author} {\bibfnamefont {A.}~\bibnamefont {Rigosi}}, \bibinfo {author}
  {\bibfnamefont {H.~M.}\ \bibnamefont {Hill}}, \bibinfo {author}
  {\bibfnamefont {A.~M.}\ \bibnamefont {van~der Zande}}, \bibinfo {author}
  {\bibfnamefont {D.~A.}\ \bibnamefont {Chenet}}, \bibinfo {author}
  {\bibfnamefont {E.-M.}\ \bibnamefont {Shih}}, \bibinfo {author}
  {\bibfnamefont {J.}~\bibnamefont {Hone}},\ and\ \bibinfo {author}
  {\bibfnamefont {T.~F.}\ \bibnamefont {Heinz}},\ }\href
  {https://doi.org/10.1103/physrevb.90.205422} {\bibfield  {journal} {\bibinfo
  {journal} {Phys. Rev. B}\ }\textbf {\bibinfo {volume} {90}},\ \bibinfo
  {pages} {205422} (\bibinfo {year} {2014})}\BibitemShut {NoStop}%
\bibitem [{\citenamefont {Steinhoff}\ \emph {et~al.}(2017)\citenamefont
  {Steinhoff}, \citenamefont {Florian}, \citenamefont {R\"osner}, \citenamefont
  {Sch\"onhoff}, \citenamefont {Wehling},\ and\ \citenamefont
  {Jahnke}}]{Steinhoff2017}%
  \BibitemOpen
  \bibfield  {author} {\bibinfo {author} {\bibfnamefont {A.}~\bibnamefont
  {Steinhoff}}, \bibinfo {author} {\bibfnamefont {M.}~\bibnamefont {Florian}},
  \bibinfo {author} {\bibfnamefont {M.}~\bibnamefont {R\"osner}}, \bibinfo
  {author} {\bibfnamefont {G.}~\bibnamefont {Sch\"onhoff}}, \bibinfo {author}
  {\bibfnamefont {T.~O.}\ \bibnamefont {Wehling}},\ and\ \bibinfo {author}
  {\bibfnamefont {F.}~\bibnamefont {Jahnke}},\ }\bibfield  {journal} {\bibinfo
  {journal} {Nature Communications}\ }\textbf {\bibinfo {volume} {8}},\ \href
  {https://doi.org/10.1038/s41467-017-01298-6} {10.1038/s41467-017-01298-6}
  (\bibinfo {year} {2017})\BibitemShut {NoStop}%
\bibitem [{\citenamefont {Wang}\ \emph {et~al.}(2017)\citenamefont {Wang},
  \citenamefont {Robert}, \citenamefont {Glazov}, \citenamefont {Cadiz},
  \citenamefont {Courtade}, \citenamefont {Amand}, \citenamefont {Lagarde},
  \citenamefont {Taniguchi}, \citenamefont {Watanabe}, \citenamefont
  {Urbaszek},\ and\ \citenamefont {Marie}}]{Wang2017a}%
  \BibitemOpen
  \bibfield  {author} {\bibinfo {author} {\bibfnamefont {G.}~\bibnamefont
  {Wang}}, \bibinfo {author} {\bibfnamefont {C.}~\bibnamefont {Robert}},
  \bibinfo {author} {\bibfnamefont {M.}~\bibnamefont {Glazov}}, \bibinfo
  {author} {\bibfnamefont {F.}~\bibnamefont {Cadiz}}, \bibinfo {author}
  {\bibfnamefont {E.}~\bibnamefont {Courtade}}, \bibinfo {author}
  {\bibfnamefont {T.}~\bibnamefont {Amand}}, \bibinfo {author} {\bibfnamefont
  {D.}~\bibnamefont {Lagarde}}, \bibinfo {author} {\bibfnamefont
  {T.}~\bibnamefont {Taniguchi}}, \bibinfo {author} {\bibfnamefont
  {K.}~\bibnamefont {Watanabe}}, \bibinfo {author} {\bibfnamefont
  {B.}~\bibnamefont {Urbaszek}},\ and\ \bibinfo {author} {\bibfnamefont
  {X.}~\bibnamefont {Marie}},\ }\href
  {https://doi.org/10.1103/physrevlett.119.047401} {\bibfield  {journal}
  {\bibinfo  {journal} {Physical Review Letters}\ }\textbf {\bibinfo {volume}
  {119}},\ \bibinfo {pages} {047401} (\bibinfo {year} {2017})}\BibitemShut
  {NoStop}%
\bibitem [{\citenamefont {Echeverry}\ \emph {et~al.}(2016)\citenamefont
  {Echeverry}, \citenamefont {Urbaszek}, \citenamefont {Amand}, \citenamefont
  {Marie},\ and\ \citenamefont {Gerber}}]{Echeverry2016}%
  \BibitemOpen
  \bibfield  {author} {\bibinfo {author} {\bibfnamefont {J.~P.}\ \bibnamefont
  {Echeverry}}, \bibinfo {author} {\bibfnamefont {B.}~\bibnamefont {Urbaszek}},
  \bibinfo {author} {\bibfnamefont {T.}~\bibnamefont {Amand}}, \bibinfo
  {author} {\bibfnamefont {X.}~\bibnamefont {Marie}},\ and\ \bibinfo {author}
  {\bibfnamefont {I.~C.}\ \bibnamefont {Gerber}},\ }\href
  {https://doi.org/10.1103/physrevb.93.121107} {\bibfield  {journal} {\bibinfo
  {journal} {Physical Review B}\ }\textbf {\bibinfo {volume} {93}},\ \bibinfo
  {pages} {121107} (\bibinfo {year} {2016})}\BibitemShut {NoStop}%
\bibitem [{\citenamefont {van Stokkum}\ \emph {et~al.}(2004)\citenamefont {van
  Stokkum}, \citenamefont {Larsen},\ and\ \citenamefont {van
  Grondelle}}]{Stokkum2004}%
  \BibitemOpen
  \bibfield  {author} {\bibinfo {author} {\bibfnamefont {I.~H.}\ \bibnamefont
  {van Stokkum}}, \bibinfo {author} {\bibfnamefont {D.~S.}\ \bibnamefont
  {Larsen}},\ and\ \bibinfo {author} {\bibfnamefont {R.}~\bibnamefont {van
  Grondelle}},\ }\href {https://doi.org/10.1016/j.bbabio.2004.04.011}
  {\bibfield  {journal} {\bibinfo  {journal} {Biochimica et Biophysica Acta
  ({BBA}) - Bioenergetics}\ }\textbf {\bibinfo {volume} {1657}},\ \bibinfo
  {pages} {82} (\bibinfo {year} {2004})}\BibitemShut {NoStop}%
\bibitem [{\citenamefont {Beckwith}\ \emph {et~al.}(2020)\citenamefont
  {Beckwith}, \citenamefont {Rumble},\ and\ \citenamefont
  {Vauthey}}]{Beckwith2020}%
  \BibitemOpen
  \bibfield  {author} {\bibinfo {author} {\bibfnamefont {J.~S.}\ \bibnamefont
  {Beckwith}}, \bibinfo {author} {\bibfnamefont {C.~A.}\ \bibnamefont
  {Rumble}},\ and\ \bibinfo {author} {\bibfnamefont {E.}~\bibnamefont
  {Vauthey}},\ }\href {https://doi.org/10.1080/0144235x.2020.1757942}
  {\bibfield  {journal} {\bibinfo  {journal} {International Reviews in Physical
  Chemistry}\ }\textbf {\bibinfo {volume} {39}},\ \bibinfo {pages} {135}
  (\bibinfo {year} {2020})}\BibitemShut {NoStop}%
\bibitem [{\citenamefont {Bisgaard}\ \emph {et~al.}(2009)\citenamefont
  {Bisgaard}, \citenamefont {Clarkin}, \citenamefont {Wu}, \citenamefont {Lee},
  \citenamefont {Ge{\ss}ner}, \citenamefont {Hayden},\ and\ \citenamefont
  {Stolow}}]{Bisgaard2009}%
  \BibitemOpen
  \bibfield  {author} {\bibinfo {author} {\bibfnamefont {C.~Z.}\ \bibnamefont
  {Bisgaard}}, \bibinfo {author} {\bibfnamefont {O.~J.}\ \bibnamefont
  {Clarkin}}, \bibinfo {author} {\bibfnamefont {G.}~\bibnamefont {Wu}},
  \bibinfo {author} {\bibfnamefont {A.~M.~D.}\ \bibnamefont {Lee}}, \bibinfo
  {author} {\bibfnamefont {O.}~\bibnamefont {Ge{\ss}ner}}, \bibinfo {author}
  {\bibfnamefont {C.~C.}\ \bibnamefont {Hayden}},\ and\ \bibinfo {author}
  {\bibfnamefont {A.}~\bibnamefont {Stolow}},\ }\href
  {https://doi.org/10.1126/science.1169183} {\bibfield  {journal} {\bibinfo
  {journal} {Science}\ }\textbf {\bibinfo {volume} {323}},\ \bibinfo {pages}
  {1464} (\bibinfo {year} {2009})}\BibitemShut {NoStop}%
\bibitem [{\citenamefont {Jin}\ \emph {et~al.}(2012)\citenamefont {Jin},
  \citenamefont {Rupp}, \citenamefont {Chevalier}, \citenamefont {Wolf},
  \citenamefont {Rojas}, \citenamefont {Lefkidis}, \citenamefont {Kr{\"u}ger},
  \citenamefont {Diller},\ and\ \citenamefont {H{\"u}bner}}]{Jin2012}%
  \BibitemOpen
  \bibfield  {author} {\bibinfo {author} {\bibfnamefont {W.}~\bibnamefont
  {Jin}}, \bibinfo {author} {\bibfnamefont {F.}~\bibnamefont {Rupp}}, \bibinfo
  {author} {\bibfnamefont {K.}~\bibnamefont {Chevalier}}, \bibinfo {author}
  {\bibfnamefont {M.~M.~N.}\ \bibnamefont {Wolf}}, \bibinfo {author}
  {\bibfnamefont {M.~C.}\ \bibnamefont {Rojas}}, \bibinfo {author}
  {\bibfnamefont {G.}~\bibnamefont {Lefkidis}}, \bibinfo {author}
  {\bibfnamefont {H.-J.}\ \bibnamefont {Kr{\"u}ger}}, \bibinfo {author}
  {\bibfnamefont {R.}~\bibnamefont {Diller}},\ and\ \bibinfo {author}
  {\bibfnamefont {W.}~\bibnamefont {H{\"u}bner}},\ }\href
  {https://doi.org/10.1103/physrevlett.109.267209} {\bibfield  {journal}
  {\bibinfo  {journal} {Physical Review Letters}\ }\textbf {\bibinfo {volume}
  {109}},\ \bibinfo {pages} {267209} (\bibinfo {year} {2012})}\BibitemShut
  {NoStop}%
\bibitem [{\citenamefont {Rebholz}\ \emph {et~al.}(2021)\citenamefont
  {Rebholz}, \citenamefont {Ding}, \citenamefont {Despr{\'{e}}}, \citenamefont
  {Aufleger}, \citenamefont {Hartmann}, \citenamefont {Meyer}, \citenamefont
  {Stoo{\ss}}, \citenamefont {Magunia}, \citenamefont {Wachs}, \citenamefont
  {Birk}, \citenamefont {Mi}, \citenamefont {Borisova}, \citenamefont
  {da~Costa~Castanheira}, \citenamefont {Rupprecht}, \citenamefont {Schmid},
  \citenamefont {Schnorr}, \citenamefont {Schr{\"o}ter}, \citenamefont
  {Moshammer}, \citenamefont {Loh}, \citenamefont {Attar}, \citenamefont
  {Leone}, \citenamefont {Gaumnitz}, \citenamefont {W{\"o}rner}, \citenamefont
  {Roling}, \citenamefont {Butz}, \citenamefont {Zacharias}, \citenamefont
  {D{\"u}sterer}, \citenamefont {Treusch}, \citenamefont {Brenner},
  \citenamefont {Vester}, \citenamefont {Kuleff}, \citenamefont {Ott},\ and\
  \citenamefont {Pfeifer}}]{Rebholz2021}%
  \BibitemOpen
  \bibfield  {author} {\bibinfo {author} {\bibfnamefont {M.}~\bibnamefont
  {Rebholz}}, \bibinfo {author} {\bibfnamefont {T.}~\bibnamefont {Ding}},
  \bibinfo {author} {\bibfnamefont {V.}~\bibnamefont {Despr{\'{e}}}}, \bibinfo
  {author} {\bibfnamefont {L.}~\bibnamefont {Aufleger}}, \bibinfo {author}
  {\bibfnamefont {M.}~\bibnamefont {Hartmann}}, \bibinfo {author}
  {\bibfnamefont {K.}~\bibnamefont {Meyer}}, \bibinfo {author} {\bibfnamefont
  {V.}~\bibnamefont {Stoo{\ss}}}, \bibinfo {author} {\bibfnamefont
  {A.}~\bibnamefont {Magunia}}, \bibinfo {author} {\bibfnamefont
  {D.}~\bibnamefont {Wachs}}, \bibinfo {author} {\bibfnamefont
  {P.}~\bibnamefont {Birk}}, \bibinfo {author} {\bibfnamefont {Y.}~\bibnamefont
  {Mi}}, \bibinfo {author} {\bibfnamefont {G.~D.}\ \bibnamefont {Borisova}},
  \bibinfo {author} {\bibfnamefont {C.}~\bibnamefont {da~Costa~Castanheira}},
  \bibinfo {author} {\bibfnamefont {P.}~\bibnamefont {Rupprecht}}, \bibinfo
  {author} {\bibfnamefont {G.}~\bibnamefont {Schmid}}, \bibinfo {author}
  {\bibfnamefont {K.}~\bibnamefont {Schnorr}}, \bibinfo {author} {\bibfnamefont
  {C.~D.}\ \bibnamefont {Schr{\"o}ter}}, \bibinfo {author} {\bibfnamefont
  {R.}~\bibnamefont {Moshammer}}, \bibinfo {author} {\bibfnamefont {Z.-H.}\
  \bibnamefont {Loh}}, \bibinfo {author} {\bibfnamefont {A.~R.}\ \bibnamefont
  {Attar}}, \bibinfo {author} {\bibfnamefont {S.~R.}\ \bibnamefont {Leone}},
  \bibinfo {author} {\bibfnamefont {T.}~\bibnamefont {Gaumnitz}}, \bibinfo
  {author} {\bibfnamefont {H.~J.}\ \bibnamefont {W{\"o}rner}}, \bibinfo
  {author} {\bibfnamefont {S.}~\bibnamefont {Roling}}, \bibinfo {author}
  {\bibfnamefont {M.}~\bibnamefont {Butz}}, \bibinfo {author} {\bibfnamefont
  {H.}~\bibnamefont {Zacharias}}, \bibinfo {author} {\bibfnamefont
  {S.}~\bibnamefont {D{\"u}sterer}}, \bibinfo {author} {\bibfnamefont
  {R.}~\bibnamefont {Treusch}}, \bibinfo {author} {\bibfnamefont
  {G.}~\bibnamefont {Brenner}}, \bibinfo {author} {\bibfnamefont
  {J.}~\bibnamefont {Vester}}, \bibinfo {author} {\bibfnamefont {A.~I.}\
  \bibnamefont {Kuleff}}, \bibinfo {author} {\bibfnamefont {C.}~\bibnamefont
  {Ott}},\ and\ \bibinfo {author} {\bibfnamefont {T.}~\bibnamefont {Pfeifer}},\
  }\href {https://doi.org/10.1103/physrevx.11.031001} {\bibfield  {journal}
  {\bibinfo  {journal} {Physical Review X}\ }\textbf {\bibinfo {volume} {11}},\
  \bibinfo {pages} {031001} (\bibinfo {year} {2021})}\BibitemShut {NoStop}%
\end{thebibliography}%

\end{document}


\title{Supplemental Material for\\ Momentum-Resolved Exciton Coupling and Valley Polarization Dynamics in Monolayer WS$_2$}

\author{Alice Kunin}
\author{Sergey Chernov}
\author{Jin Bakalis}
\affiliation{Department of Chemistry, Stony Brook University, Stony Brook, New York 11794, USA.}

\author{Ziling Li}
\author{Shuyu Cheng}
\affiliation{Department of Physics, The Ohio State University, Columbus, Ohio 43210, USA.}

\author{Zachary H. Withers}
\affiliation{Department of Physics and Astronomy, Stony Brook University, Stony Brook, New York 11794, USA.}

\author{Michael G. White}
\affiliation{Department of Chemistry, Stony Brook University, Stony Brook, New York 11794, USA.}
\affiliation{Chemistry Division, Brookhaven National Laboratory, Upton 11973 New York, USA.}

\author{Gerd Sch\"onhense}
\affiliation{Johannes Gutenberg-Universit\"at, Institut f\"ur Physik, D-55099 Mainz, Germany.}

\author{Xu Du}
\affiliation{Department of Physics and Astronomy, Stony Brook University, Stony Brook, New York 11794, USA.}

\author{Roland K. Kawakami}
\affiliation{Department of Physics, The Ohio State University, Columbus, Ohio 43210, USA.}

\author{Thomas K. Allison}
\affiliation{Department of Chemistry, Stony Brook University, Stony Brook, New York 11794, USA.}
\affiliation{Department of Physics and Astronomy, Stony Brook University, Stony Brook, New York 11794, USA.}

 \email{thomas.allison@stonybrook.edu}

\maketitle

\renewcommand{\thepage}{S\arabic{page}} 
\renewcommand{\thesection}{S\arabic{section}}  
\renewcommand{\thesubsection}{S\arabic{subsection}}  
\renewcommand{\thetable}{S\arabic{table}}  
\renewcommand{\thefigure}{S\arabic{figure}}
\renewcommand{\theequation}{S\arabic{equation}}

\onecolumngrid

\subsection{Sample fabrication and characterization} \label{sample}

\begin{figure}[h]
    \centering
    \includegraphics{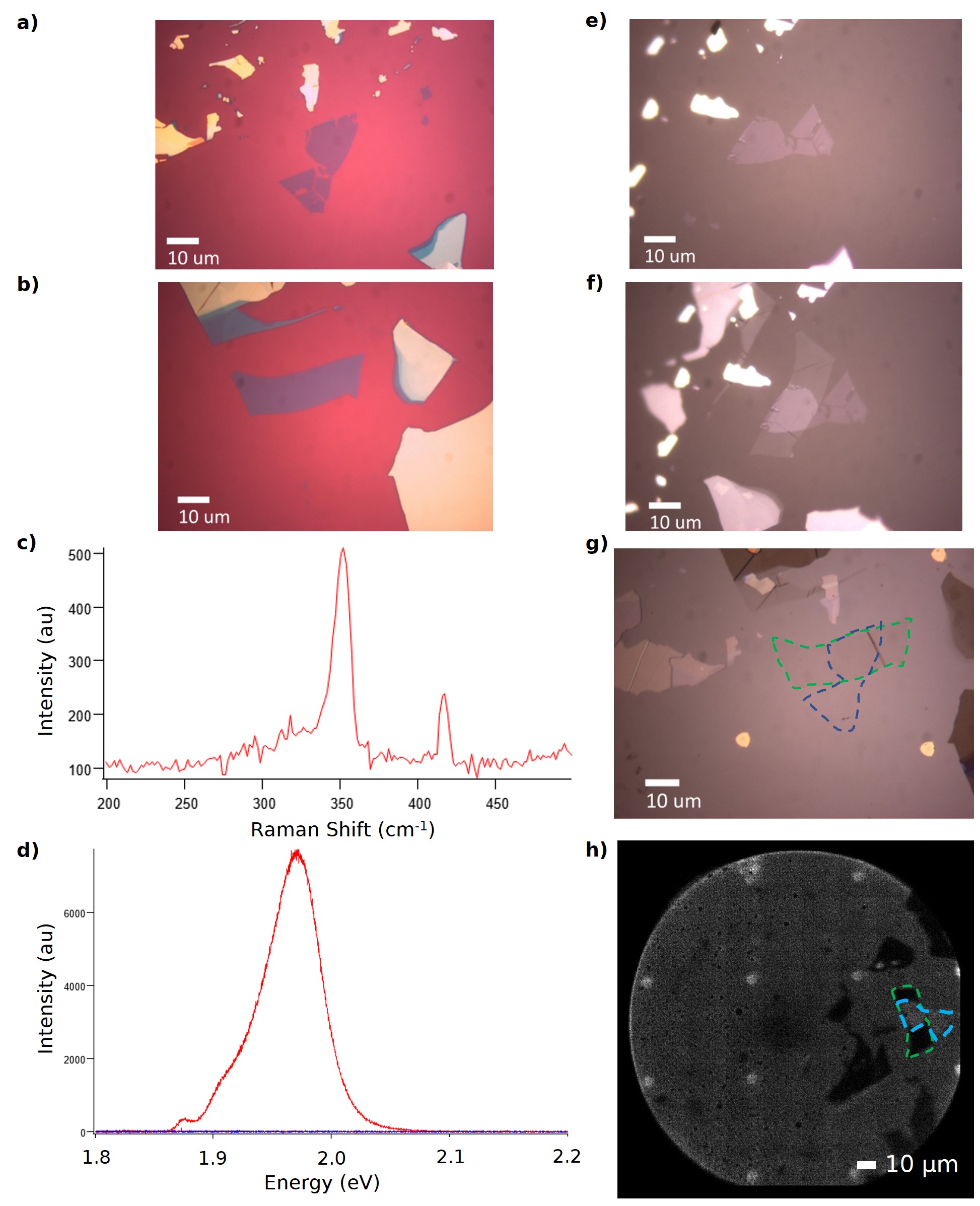}
    \caption{\textbf{Sample characterization.} Optical microscope images of \textbf{a)} the exfoliated monolayer WS$_2$ flake and \textbf{b)} the hBN.
    \textbf{c)} Raman spectrum of the exfoliated WS$_2$, $h\nu_{\textss{excitation}} = 2.41$ eV. 
    The intensity ratio of the WS$_2$ 2LA mode at the M point ($\sim$350 cm$^{-1}$) and the A$_{1g}$ mode ($\sim$415 cm$^{-1}$) of $>$2 is indicative of a monolayer sample \cite{Berkdemir2013, Saito2016}. 
    \textbf{d)} Photoluminescence emission of the exfoliated WS$_2$ shows strong emission intensity arising from the monolayer. $h\nu_{\textss{excitation}} = 2.41$ eV.
    \textbf{e)} Optical microscope image of the WS$_2$ picked up on the PDMS/PC stamp and \textbf{f)} the PDMS/PC/WS$_2$ with the picked up bottom hBN flake. 
    \textbf{g)} Image of the WS$_2$/hBN on the target Si substrate. The blue and green dashed lines indicate the outline of the monolayer WS$_2$ and hBN, respectively. The inclusion of the hBN buffer layer is essential to preserve the electronic structure of the monolayer TMDs \cite{Man2016}.  
      \textbf{h)} Image of the sample in the real-space imaging mode of the momentum microscope, taken with $h\nu_{\textss{probe}} = 4.75$ eV. 
      The WS$_2$ flake extends onto the Si substrate to prevent sample charging during ARPES measurements. 
    }
    \label{fig:sample}
\end{figure}

To assemble the monolayer WS$_2$/hBN/Si heterostructure, WS$_2$ (HQ Graphene, n-type) and hexagonal boron nitride (hBN) flakes are first exfoliated onto separate SiO$_2$(300 nm)/Si substrates (Fig. \ref{fig:sample}a) and b)).  
Raman spectroscopy (Fig. \ref{fig:sample}c)) and photoluminescence (PL) measurements (Fig. \ref{fig:sample}d)) are used to distinguish monolayer WS$_2$ flakes.
The spectra are measured at room temperature with 514 nm (2.41 eV) excitation in a backscattering configuration using a Renishaw Raman microscope.
The power of the excitation beam is $\sim$100 $\upmu$W, and a 100$\times$ objective lens focused the beam to a spot size of $\sim$1 $\upmu$m on the target flake. 
The collected signal is dispersed by a grating with a groove density of 1800/mm. 
The integration time is set to 120 s for Raman measurements and 5 s for PL measurements.
The strong PL signal and the obvious longitudinal acoustic mode ($\sim$350 cm$^{-1}$) in the Raman spectrum show that the target WS$_2$ transferred to the sample stack is a monolayer \cite{Berkdemir2013}.

Next, a dry transfer method is used to stack the WS$_2$/hBN heterostructure. 
A polydimethylsiloxane (PDMS) hemisphere is first made on a clean glass slide and then covered by a thin film of polycarbonate (PC). 
This PDMS/PC stamp is then used to pick up the monolayer WS$_2$ flake from the SiO$_2$/Si substrate (Fig. \ref{fig:sample}e)). 
The pick-up procedure is to lower the PDMS/PC stamp and heat the sample stage to 70 $^\circ$C, and when the target flake is fully covered by PC film, shut down the heating and slowly detach the PDMS/PC from the sample stage; the WS$_2$ flake is picked up by the PDMS/PC stamp after separation. 
Then, the PDMS/PC/WS$_2$ is used to further pick up the bottom hBN flake ($\sim$10-20 nm thickness) by the same procedure (Fig. \ref{fig:sample}f)).
The PDMS/PC/WS$_2$/hBN is then transferred onto a pre-patterned gold-grid-marked Si substrate with good alignment by heating the sample stage to 130 $^\circ$C and slowly lifting up the PDMS stamp; the PC/WS$_2$/hBN remains on the Si substrate. 
The PC film is then dissolved in chloroform. 
Afterwards, the WS$_2$/hBN heterostructure (Fig. \ref{fig:sample}g)) is annealed at 300 $^\circ$C in ultra-high vacuum (UHV) for 1 hour to clean up any polymer residue. 
For ARPES measurements, the finished sample was annealed to 150 $^\circ$C in UHV daily for 30-60 minutes and allowed to cool completely.
The sample can be clearly identified using the real-space imaging mode of the momentum microscope (Fig. \ref{fig:sample}h)).

\subsection{tr-ARPES experimental apparatus and data analysis} \label{instr}

The experiments presented here are driven by a home-built Yb:fiber frequency comb laser \cite{Li_RSI2016} producing 1 $\upmu$J, 185 fs pulses centered at 1035 nm, at 61.3 MHz repetition rate.
The laser pulses are resonantly enhanced in an optical cavity to produce $\sim$10-10.5 kW of intracavity average power.
At the cavity focus, the laser reaches a peak intensity of $\sim$10$^{14}$ W/cm$^2$ and high-order harmonics are generated in a jet of argon gas.
Harmonics from 10 to 40 eV are separated by a time-preserving grating monochromator, and the desired, isolated harmonic is refocused onto the sample \cite{Corder2018a, Corder2018}.
Pump pulses are generated by doubling a portion of the fundamental IR in a BBO crystal to 517.5 nm (2.4 eV). 
The polarization of the pump pulses is controlled by a quarter-wave plate.
Both the pump and probe pulses impinge on the sample at approximately 48$^\circ$ incidence angle.
The sizes of the focused 2.4 eV pump and XUV probe beams on the sample are 80 x 200 $\upmu$m$^2$ and 24 x 16 $\upmu$m$^2$, respectively.
The instrument response function (IRF, $200\pm 20$ fs Gaussian FWHM) was measured independently with auxiliary experiments on short-lived hot carriers in metallic samples and graphene.

For tr-ARPES measurements, we employ a custom time-of-flight momentum microscope to measure the real- and momentum-space images of the sample \cite{Chernov2015, Medjanik2017}.
In the real-space imaging mode, we use a broadband Xe-Hg lamp to uniformly illuminate the full sample.
The work function contrast between monolayer WS$_2$, hBN, and silicon allows the WS$_2$/hBN overlapping region of interest to be clearly identified for the tr-ARPES experiments (Fig. \ref{fig:sample}h)). 
For momentum-space measurements, we implement an electron high-pass filter using two grids in front of the detector to pass only a $\sim$4.4 eV wide energy region of the photoemission distribution to the detector. 
This allows us to detect only the desired photoemission signal near the Fermi level and mitigate saturation of the detector by suppressing the strong photoemission signal from fully occupied states below the relevant regions of the valence band.
The energy cut-off is apparent at the bottom of the band structure at $-$3 eV in Fig. 1e) of the main text.
The high-pass filter is tunable and is adjusted to pass approximately 1$-$1.5 eV below the valence band maximum (VBM) for pump-probe experiments. 
All measurements are performed in vacuum better than 4 x 10$^{-10}$ Torr.

The measured tr-ARPES 4D ($k_x, k_y, E, t$) photoelectron distributions are normalized at each individual pump-probe delay to the maximum intensity in the image at that delay.  
The VBM energy is determined by fitting the upper and lower spin-orbit split band intensities in the K valleys to a double Gaussian and extracting the center of the upper fit band.
At our low excitation fluence, we do not observe any time-resolved shifts of the VBM in this work.
We also do not observe any band gap renormalization or any laser-assisted photoelectric effect (LAPE) signal.
The presented background-subtracted signals are produced by averaging the 3D ($k_x, k_y, E$) images of the five most negative pump-probe delays and subtracting this average from the 4D ($k_x, k_y, E, t$) distribution.
To produce the valley asymmetry ratios presented in Fig. 3 of the main text and Fig. \ref{fig:temp_dep}, the intensities of the K$^+$ and K$^-$ valleys are scaled to the intensity at the longest pump-probe delay for the corresponding valleys in the linearly polarized excitation data. 
This allows for normalization of ARPES matrix element effects that cause unequal intensities in the different valleys across the momentum space image, in particular due to the direction of the probe electric field.

\FloatBarrier

\begin{figure}[b!ht]
    \centering
    \includegraphics{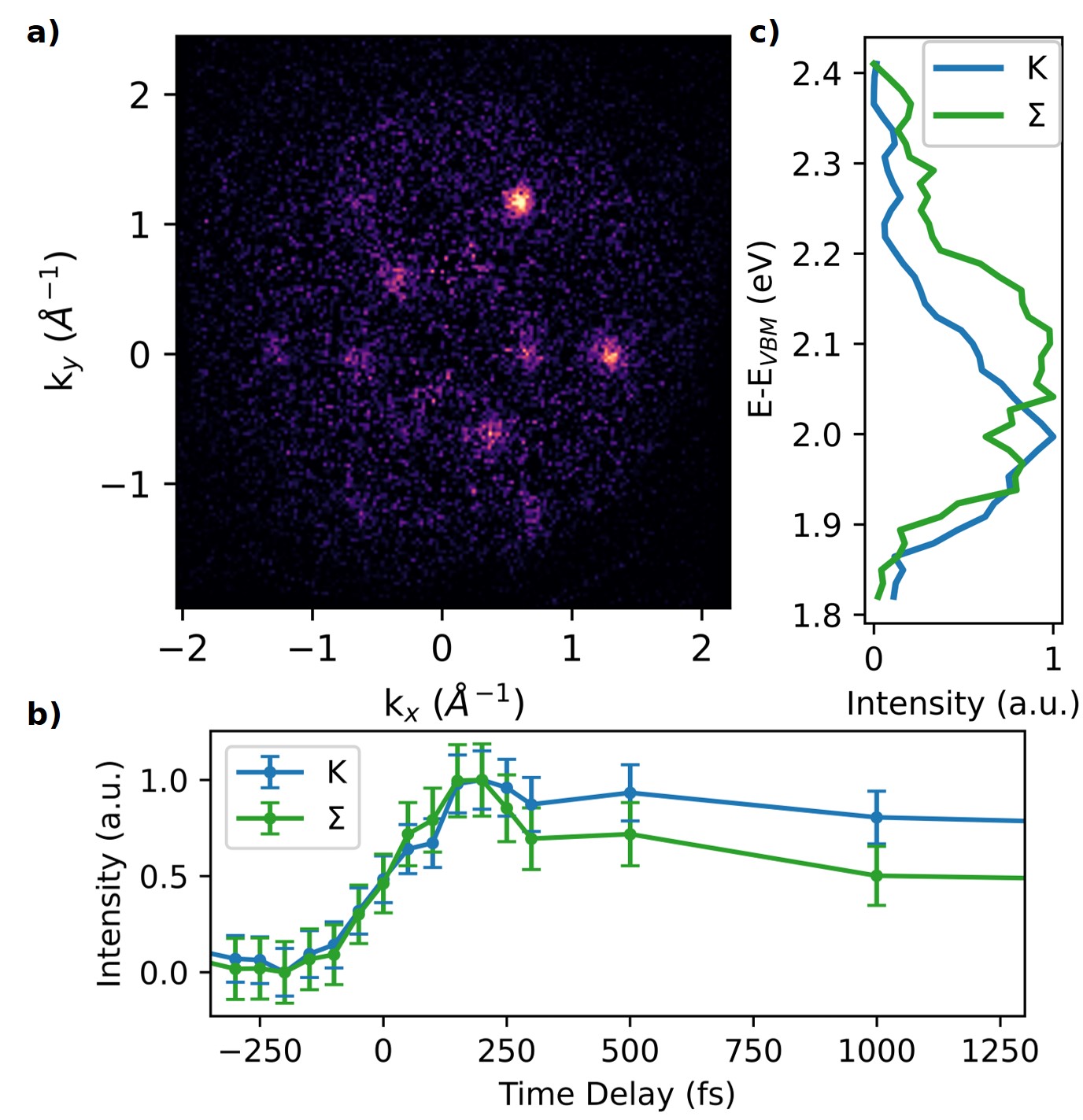}
    \caption{\textbf{Comparison of K and $\Sigma$ valley dynamics. a)} Raw exciton signal at 300 fs time delay, taken with 5 $\upmu$J/cm$^2$ pump fluence ($h\nu_{\textss{probe}} = 20.4$ eV). 
    \textbf{b)} The integrated intensity in each valley shows that the K and $\Sigma$ valley signals appear at the same time.
     \textbf{c)} The energy distribution curves for the K and $\Sigma$ valley populations at 250 fs time delay. The $\Sigma$ valley population is centered approximately 100 meV energetically above the K valley signal. 
    }
    \label{fig:sigma}
\end{figure}

\subsection{Sigma valley dynamics} \label{sigma}

With our tunable XUV probe, we observe strong dependence of the $\Sigma$ valley photoemission intensity on the probe photon energy.
Above approximately 23 eV, the intensity of the photoemission signal in the $\Sigma$ valleys is extremely weak and nearly undetectable. 
With 22.8 eV and 20.4 eV probe energies, we observe the ratio K/$\Sigma$ for the maximum total integrated valley intensities is approximately 
1.73 and 0.96, respectively.
An example of the exciton signals observed with $h\nu_{\textss{probe}} = 20.4$ eV is shown in Fig. \ref{fig:sigma}a).

Upon photoexcitation, we observe that the appearance of signal in the $\Sigma$ valleys rises at the same time as that of the K valleys (Fig. \ref{fig:sigma}b)). 
This $\Sigma$ valley signal appears centered near 2.1 eV above the VBM, approximately 100 meV higher in energy than the K valley signal, as evidenced by the energy distribution curves presented in Fig. \ref{fig:sigma}c).  
This is in contrast to recent tr-ARPES measurements at higher excitation densities for monolayer WSe$_2$/hBN and monolayer WS$_2$ on bare silicon \cite{Madeo2020, Wallauer2021a}, where the $\Sigma$ valley was observed to be roughly isoenergetic with the K valley.  
The prompt appearance of signal in the $\Sigma$ valleys at early times could result from the pump excitation of the B exciton resonance which may be situated energetically near the electronic band gap. 
Photoexcitation above the band gap has previously shown prompt and strong appearance of excitons with electrons in the $\Sigma$ valleys of monolayer WSe$_2$/hBN \cite{Madeo2020}.
Interestingly, calculations for monolayer MoS$_2$ have shown that the B$_{1s}$ exciton, which includes a mixture of the A$_{1s}$ exciton due to intravalley Coulomb exchange coupling, may show small, but nonzero, amplitude for the wavefunction in the interior of the Brillouin zone towards the $\Sigma$ valley \cite{Guo2018}.

\FloatBarrier
\subsection{Pump fluence dependence} \label{fluence_dep}
\begin{figure}[h]
    \centering
    \includegraphics[width=5in]{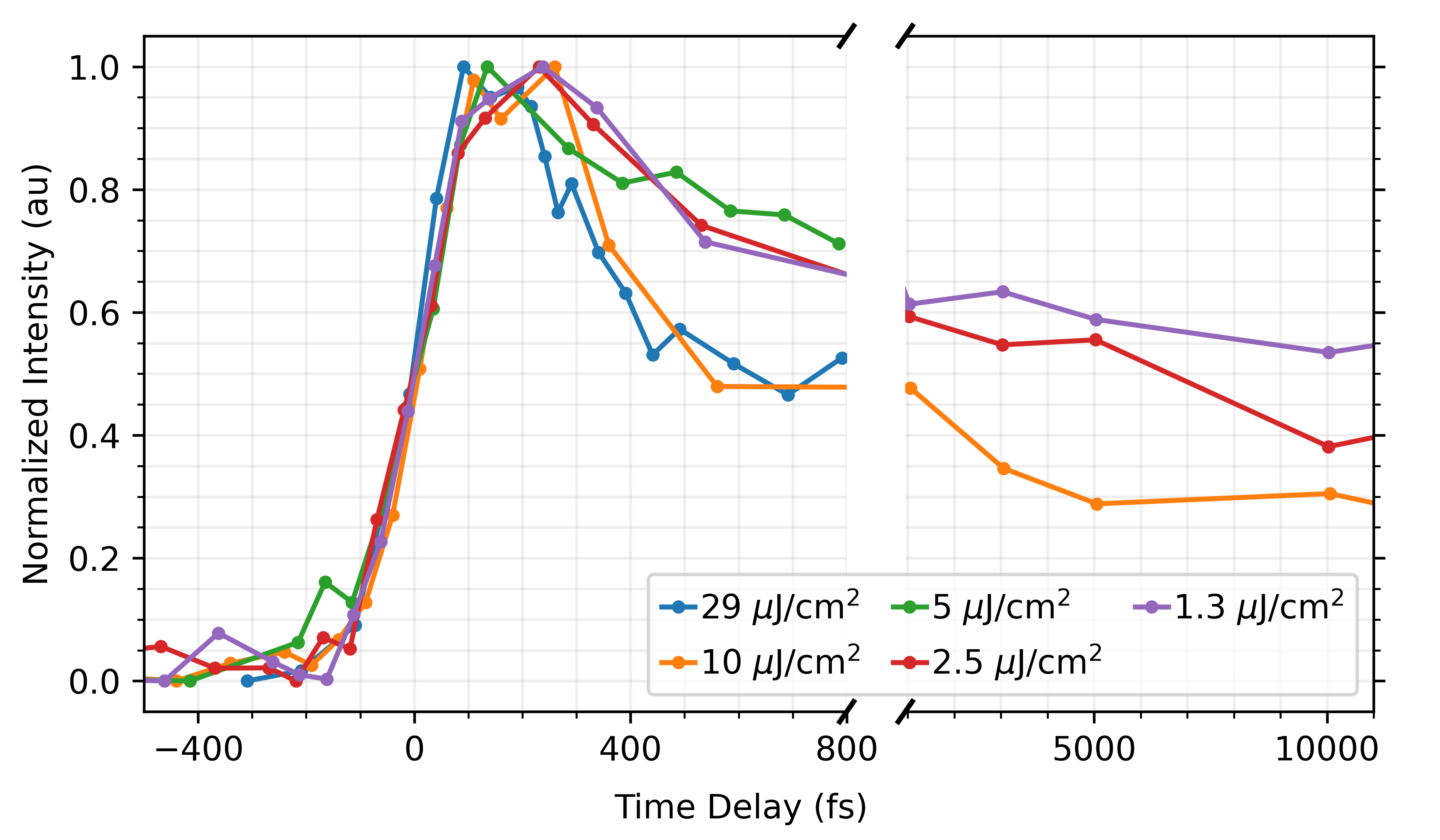}
    \caption{\textbf{Pump fluence dependence.} All data presented was taken with \textit{p}-polarized pump excitation and $h\nu_{\textss{probe}}$ = 27.6 eV. 
    }
    \label{fig:fluence_dep}
\end{figure}

The integrated K valley signal intensities for various pump pulse fluences are shown in Fig. \ref{fig:fluence_dep}.
We observe that, with pump fluences $\leq$ 5 $\upmu$J/cm$^2$, the time dynamics are invariant to the fluence.
Based on these results, all of the data presented in this work was taken with an incident pump fluence of 5 $\upmu$J/cm$^2$.
Accounting for the 5.5\% absorption of monolayer WS$_2$ at 2.4 eV \cite{Li2014a}, we estimate that this fluence corresponds to an excited carrier density of approximately 7 x 10$^{11}$ carriers/cm$^2$. 
This excitation density is well below the $\sim$3 x 10$^{12}$ carriers/cm$^2$ limit of the Mott transition \cite{Steinhoff2017}.
In this pump fluence regime, the sample maintains a temperature of 302-303 K during the experiment.

\FloatBarrier
\subsection{Nature of the circularly polarized photoexcitation} \label{AOI}

\begin{figure}[h]
    \centering
    \includegraphics[width=3in]{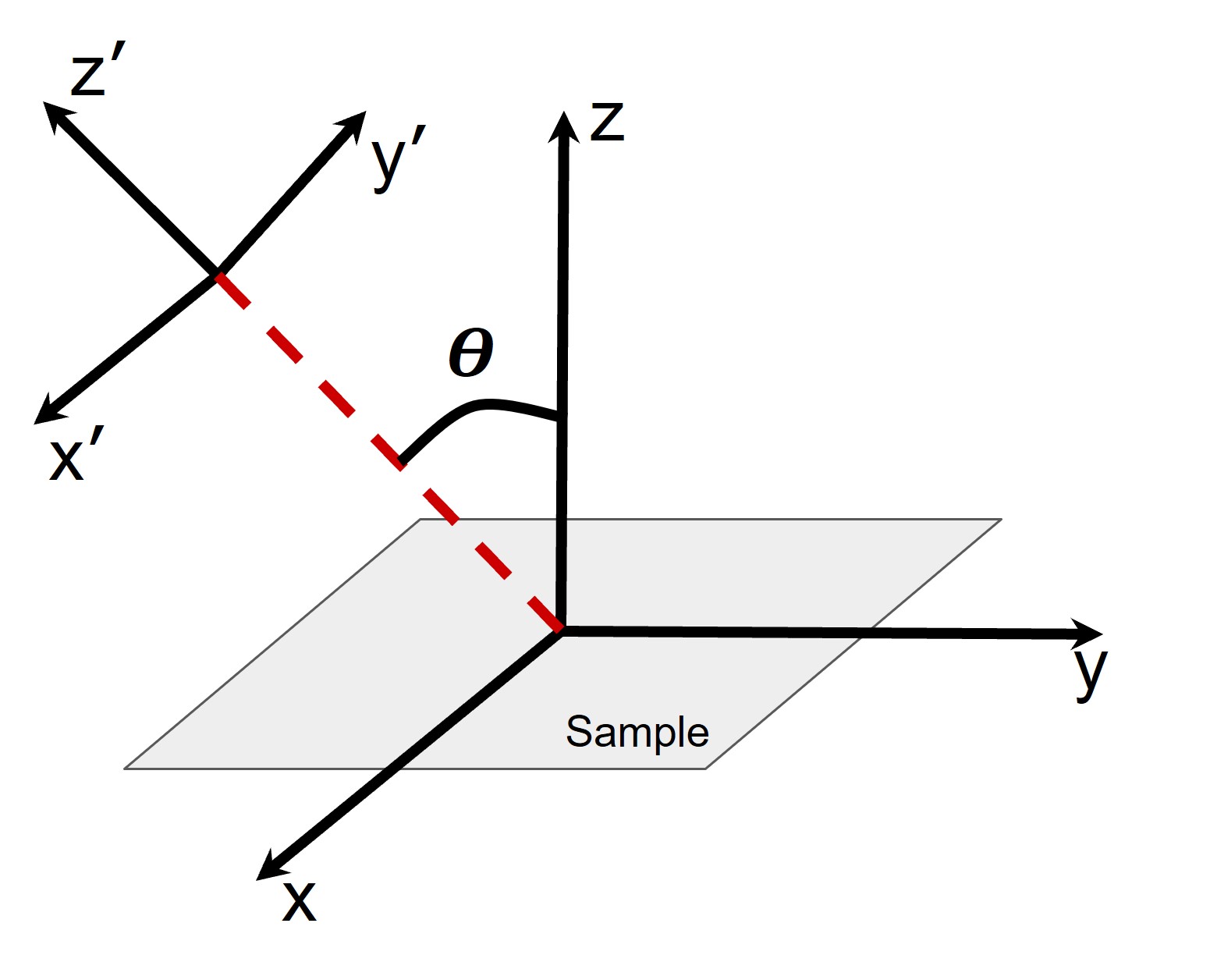}
    \caption{\textbf{Incident photoexcitation geometry.} $\theta$ denotes the angle of incidence and $\{\hat x',\hat y',\hat z'\}$ and $\{\hat x,\hat y,\hat z\}$ denote the bases of the incident light propagation and the sample, respectively.
    }
    \label{fig:aoi}
\end{figure}

Here, we consider the effect of the 48$^\circ$ incidence angle of the pump pulse on the nature of the pump polarization in the plane of the sample.
For light incident on the sample at a polar angle $\theta$ from the sample normal, we define $\{\hat x',\hat y',\hat z'\}$ as the basis of the incident light such that $-\hat z'$ is the propagation direction (Fig. \ref{fig:aoi}).
The incident circularly polarized electric field in this basis, $\vec E_{\pm}'$, can be expressed as:
\begin{align}
\vec E_{\pm}'&= E_0\hat\sigma_{\pm}'\\
&= E_0(\cos\omega t\,\hat x'\pm\sin\omega t\,\hat y') , 
\end{align}
\noindent where E$_0$ denotes the amplitude, $\omega$ is the angular frequency, and $\hat\sigma_{\pm}'$ denotes the right- and left-circular polarization states.
In the sample basis $\{\hat x,\hat y,\hat z\}$, $\hat x'=\hat x$ and $\hat y'=\cos\theta\,\hat y+\sin\theta\,\hat z$.
Thus, the electric field in the sample basis is given by:
\begin{align} 
\vec E_\pm&=E_0(\cos\omega t\,\hat x\pm\sin\omega t(\cos\theta\,\hat y+\sin\theta\,\hat z))\\
&=E_0(\cos\omega t\,\hat x\pm\sin\omega t\cos\theta\,\hat y\pm\sin\omega t\sin\theta\,\hat z) .
\end{align}

\noindent In the sample basis, the circularly polarized light can be parameterized as:
\begin{equation}
\hat\sigma_{\pm}=\cos\omega t\,\hat x\pm\sin\omega t\,\hat y ,
\end{equation}

\noindent and the $\hat x$ and $\hat y$ terms can be reexpressed as:

\begin{align}
\cos\omega t\,\hat x&=\frac{1}{2}(\hat\sigma_++\hat\sigma_-)\\
\sin\omega t\,\hat y&=\frac{1}{2}(\hat\sigma_+-\hat\sigma_-) .
\end{align}

\noindent The electric field can then be written as:
\begin{align}
\vec E_\pm&=E_0\Big(\frac{1}{2}(\hat\sigma_++\hat\sigma_-)\pm\cos\theta\frac{1}{2}(\hat\sigma_+-\hat\sigma_-)\pm\sin\omega t\sin\theta\,\hat z\Big)\\
&=E_0\bigg(\Big(\frac{1\pm\cos\theta}{2}\Big)\hat\sigma_++\Big(\frac{1\mp\cos\theta}{2}\Big)\hat\sigma_-\pm\sin\omega t\sin\theta\,\hat z\bigg) .
\end{align}

\noindent In our experimental geometry, $\theta\cong$ 48$^\circ$. 
For photoexcitation of the sample by incident left circularly polarized light ($\hat\sigma_-'$), the electric field in the sample plane is given by:
 
\begin{equation}
\vec E_-=E_0(0.165\hat\sigma_++0.834\hat\sigma_-+0.743\sin\omega t\hat z) ,
\end{equation}

\noindent indicating that in the plane of sample, the amplitude ratio between the $\hat\sigma_-$ component and the $\hat\sigma_+$ component is approximately 5:1 and the intensity ratio is 25:1. 
This ratio is reversed in the case of right circularly polarized photoexcitation.
Thus, our circularly polarized photoexcitation of the sample is predominantly of the desired helicity, but contains a small contribution from the opposite helicity as well as an out-of-plane component.
In WS$_2$, excitation by the out-of-plane $\hat z$ component of the electric field has a very small transition dipole moment \cite{Wang2017a}, particularly for the B exciton resonance \cite{Echeverry2016}, and thus we do not expect any appreciable contribution from this $\hat z$ component. 
This is confirmed experimentally by comparing the results of $s$- and $p$-polarized excitation, with no differences observed within our measurement uncertainty.

\subsection{Global analysis} \label{GA}

The global analysis (GA) algorithm employed here is similar to that previously applied to a variety of time-resolved spectroscopies \cite{Stokkum2004, Beckwith2020}.
GA is widely used to decompose conjested time-resolved spectra into individual spectral components described by exponential time dynamics \cite{Bisgaard2009, Jin2012, Rebholz2021}.
This approach assumes that the spectral and temporal components of a time-resolved spectrum $I(E, t)$ can be separated (i.e., that the spectral components do not shift in time). 

Here, the momentum-integrated signal of the desired valleys, $I(E, t)$, is decomposed into two principal spectral components, $S_1(E)$ and $S_2(E)$, each described by exponential time dynamics $f_1(t)$ and $f_2(t)$ convolved with the Gaussian IRF:
\begin{align}
I(E, t) &= S_1(E) f_1(t) + S_2(E) f_2(t)\\
         &= S_1(E)[c_1(e^{-t/\tau_1}\ast IRF)] + S_2(E)[c_2(-e^{-t/\tau_1}\ast IRF + e^{-t/\tau_2}\ast IRF)] .\label{eq:GA}
\end{align}

\noindent The two exponential decay lifetimes are denoted by $\tau_1$ and $\tau_2$, and $c_1$ and $c_2$ are amplitude constants.
An intensity offset factor, $y_0$, is included as a fit parameter as the initial, constant value of $S_1(E)$ and $S_2(E)$.
The time dynamics for the spectral component $S_2(E)$ in Eqn. \ref{eq:GA} arise from the assumption that component 2 is formed by the decaying population of component 1 rather than by direct excitation by the pump pulse. 
The global fit is performed by minimizing $\chi^2 = \sum (I_{\textss{exp.}} -I_{\textss{model}})^2 / \sigma_{\textss{exp.}}^2$. 
We find reduced $\chi^2$ values of 1.2$-$1.4.
 
For the GA of the separated K$^+$ and K$^-$ circular excitation data, we employ the same method but we allow the onset time for the time dynamics, $t_0$, to be a fit parameter:

\begin{equation}
I(E, t) = S_1(E)[c_1(e^{-(t-t_0)/\tau_1}\ast IRF)] + S_2(E)[c_2(-e^{-(t-t_0)/\tau_1}\ast IRF + e^{-(t-t_0)/\tau_2}\ast IRF)] .
\end{equation}

\noindent The K$^+$ and K$^-$ valleys are fit separately, and the shift in the onset of the time dynamics between the two valleys, $\Delta t$, is given by:

\begin{equation}
\Delta t = t_\mathrm{{0, K_{pumped}}} - t_\mathrm{{0, K_{unpumped}}} ,
\end{equation}

\noindent where $t_{0, K_{\textss{pumped}}}$ and $t_{0, K_{\textss{unpumped}}}$ are the $t_0$ parameters for the pumped and unpumped K valleys, respectively. 
The fitted lifetimes $\tau_1$ and the time shifts $\Delta t$ for the K valley signals for \textit{s}-, $\sigma^-$, and $\sigma^+$ excitation are presented in Table \ref{tab:GA}. The dominant source of the error in $\tau_1$ is the systematic uncertainty in the IRF width, which is estimated by repeating the fit over the IRF confidence interval and adding the spread in the fit results in quadrature with the statistical error. 

\begin{table}[h!]
\centering

\begin{tabular}{|P{1.5cm}|P{1.5cm}|P{1.5cm}|P{1.5cm}|P{1.5cm}|P{1.5cm}|P{1.5cm}|}
\hline
\textbf{Data} & $\boldsymbol\tau\mathbf{_1}$\textbf{ (fs)} & $\boldsymbol\tau\mathbf{_2}$\textbf{ (fs)} & \textbf{y}$\mathbf{_0}$\textbf{ (a.u.)} & \textbf{c}$\mathbf{_1}$\textbf{ (a.u.)} & \textbf{c}$\mathbf{_2}$\textbf{ (a.u.)} & $\boldsymbol\Delta$\textbf{t (fs)} \\ \hline \hline
\textit{s}, all K   & 378 $\pm$ 40 & 15500 & -8E-05 & 1.29 & 2.28 &          \\ \hline
\textit{s}, K$^+$  & 382 $\pm$ 56  & 15150  & -2E-05 & 2.83 & 5.22   & \multirow{2}{*}{6 $\pm$ 5}    \\\cline{1-6} 
\textit{s}, K$^-$ & 389 $\pm$ 49 & 15700 & -7E-05 & 2.39 & 4.04 &                \\ \hline 
\hline
$\sigma^-$, K$^-$  & 382 $\pm$ 44  & 35400 & -1E-04 & 1.40 & 2.58 &  \multirow{2}{*}{43 $\pm$ 4}    \\\cline{1-6} 
$\sigma^-$, K$^+$ & 354 $\pm$ 51 & 18100 & -1E-04 & 1.95  &  3.03 &            \\   \hline 
\hline
$\sigma^+$, K$^+$  & 360 $\pm$ 59 & 21000 & 1E-05 & 2.41  & 3.98 &  \multirow{2}{*}{53 $\pm$ 6}    \\\cline{1-6}
$\sigma^+$, K$^-$ & 440 $\pm$ 53 & 15700 & -2E-04 & 2.15 & 2.46 &          \\    \hline
\end{tabular}

\caption{\textbf{Global analysis fit results.} 
In all fits, the value of $\tau_2$ was found to be substantially longer than the longest recorded pump-probe delay of the dataset. 
The FWHM of the IRF is fixed at 200 fs.
All fitted experimental data presented here was taken with $h\nu_{\textss{probe}} = 25.2$ eV.}
\label{tab:GA}
\end{table}

\subsection{Temperature dependence} \label{cold}

\begin{figure}[h]
    \centering
    \includegraphics{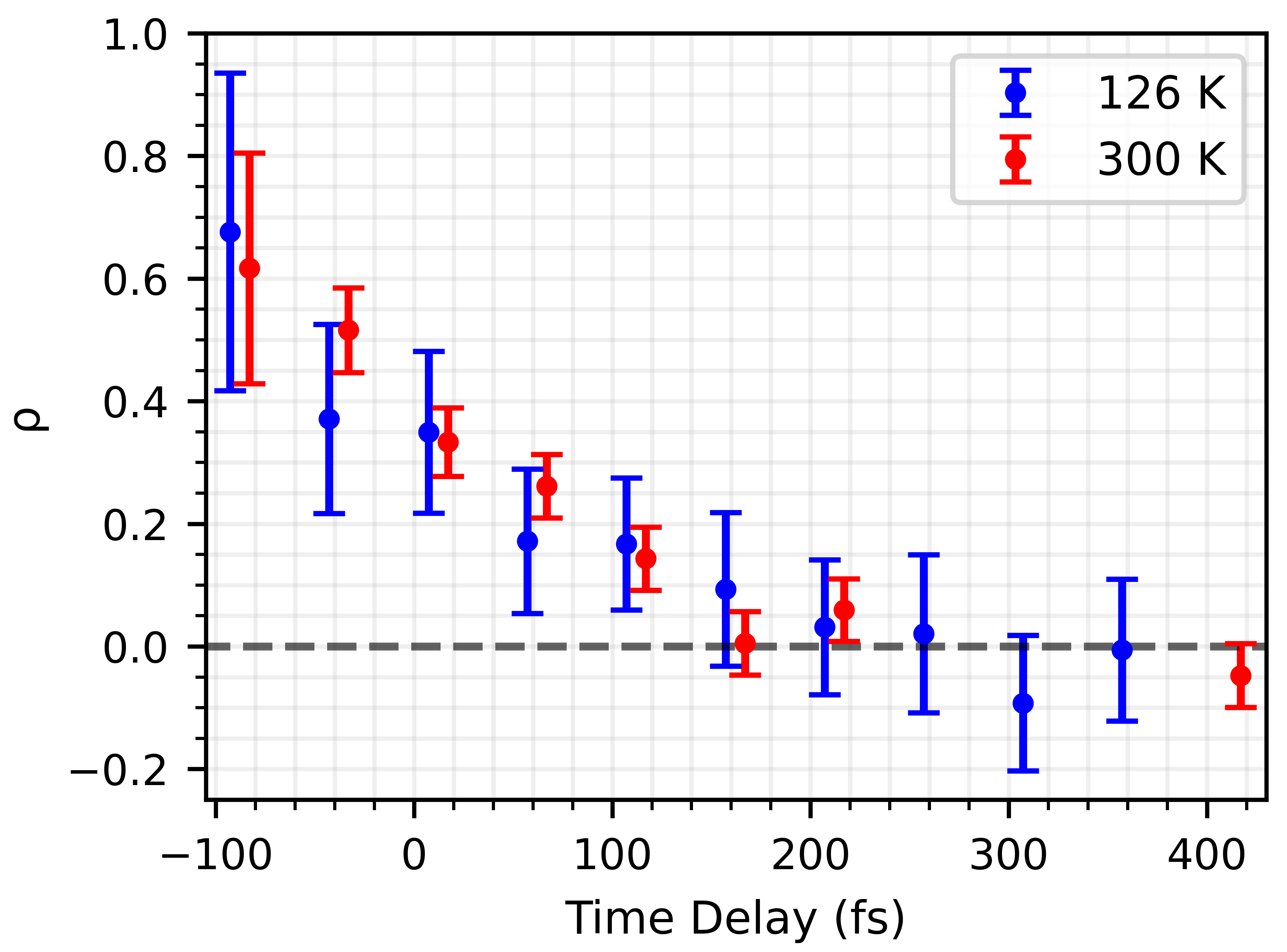}
    \caption{\textbf{Valley asymmetry temperature dependence.} The valley asymmetry ($\rho$) for $\sigma^+$ photoexcitation at room temperature and 126 K show very similar timescales for valley depolarization ($h\nu_{\textss{probe}} = 25.2$ eV). 
    }
    \label{fig:temp_dep}
\end{figure}

To examine the possible role of 
exciton-phonon coupling in the ultrafast valley depolarization, we performed additional experiments with the sample held at 126 K. 
The valley asymmetry, given by:

\begin{equation}
  \rho(t) = \frac{I_{K^+}(t) - I_{K^-}(t)}{I_{K^+}(t) + I_{K^-}(t)},
\end{equation}

\noindent for $\sigma^+$ photoexcitation at room temperature and 126 K is shown in Fig. \ref{fig:temp_dep}. 
I$_{K^+}$ and I$_{K^-}$ refer to the integrated intensity in the K$^+$ and K$^-$ valleys, respectively. 
The strong similarity between the observed timescales for the loss of valley asymmetry at each temperature indicates that exciton-phonon interactions do not play a significant role in the valley depolarization mechanism.

\subsection{Comparison of K$^+$ and K$^-$ valleys after $s$-polarized excitation}

\begin{figure}[h]
    \centering
    \includegraphics{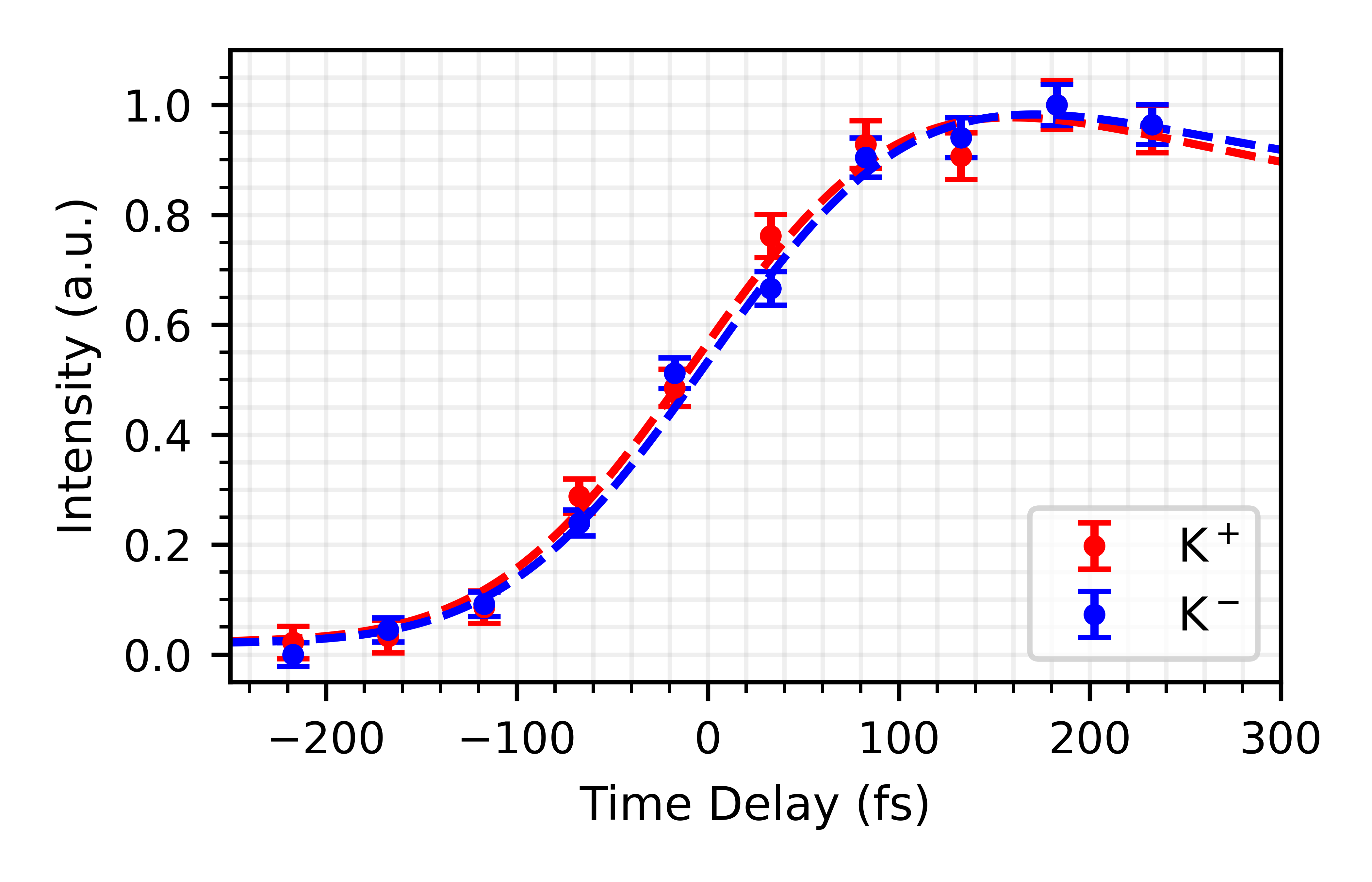}
    \caption{\textbf{Integrated intensities of the K$^+$ and K$^-$ valleys following \textit{s}-polarized excitation.} 
    Points are the experimental integrated intensities with statistical error, and dashed lines are the integrated intensities of the GA fit for each valley ($h\nu_{\textss{probe}} = 25.2$ eV). 
    }
    \label{fig:integ_int}
\end{figure}

Here, we include the integrated intensities following \textit{s}-polarized photoexcitation for the K$^+$ and K$^-$ valleys separately (Fig. \ref{fig:integ_int}). In contrast to the data recorded after circularly polarized excitation, we do not observe any notable differences between the K$^+$ and K$^-$ valleys under linearly polarized photoexcitation, as expected. 

\FloatBarrier

\bibliographystyle{apsrev4-2}
\bibliography{WS2_ML_paper}